%% file: main.tex
\documentclass[aps, pra,twocolumn,superscriptaddress, 10pt]{revtex4-2}

\usepackage{amsmath}
\usepackage{amssymb}
\usepackage{amsfonts}
\usepackage{amsthm}
\usepackage{graphicx}
\usepackage{braket}
\usepackage{algorithm}
\usepackage[noend]{algpseudocode}
\usepackage{qcircuit}
\usepackage{float}
\usepackage{color}
\usepackage{bm}
\usepackage[version=3]{mhchem}
\usepackage{multirow}
\usepackage{array}
\usepackage{tablefootnote}
\usepackage{threeparttable}
\usepackage[colorlinks=true, allcolors=blue]{hyperref}
\usepackage{pgffor}

\theoremstyle{plain}

\newcommand{\bs}[1]{\boldsymbol{#1}}

\newcommand{\lstickx}[1]{\lstick{\makebox[1.5em][l]{$#1$}}}
\newcommand{\arrep}[1]{\ar @<4pt> @/^/[#1]|-{\mbox{ $\times L$ }}}
\newcommand{\arreq}[1]{\ar @<4pt> @/^/[#1]|-{\mbox{$L=1$}}}
\newcommand{\arrer}[1]{\ar @<4pt> @/^/[#1]|-{\mbox{$L=2$}}}

\begin{document}
\title{Optimizing a parameterized controlled gate using free quaternion selection}
\author{Hiroyoshi Kurogi}
\affiliation{Department of Chemistry, Graduate School of Science, Kyushu University, 744 Motooka, Nishi-ku, Fukuoka, 819-0395, Japan}
\affiliation{Quantum Computing Center, Keio University, 3-14-1 Hiyoshi, Kohoku-ku, Yokohama, Kanagawa, 223-8522, Japan}

\author{Katsuhiro Endo}
\affiliation{Research Center for Computational Design of Advanced Functional Materials, National Institute of Advanced Industrial Science and Technology (AIST), 1-1-1 Umezono, Tsukuba, Ibaraki, 305-8568, Japan}
\affiliation{Quantum Computing Center, Keio University, 3-14-1 Hiyoshi, Kohoku-ku, Yokohama, Kanagawa, 223-8522, Japan}

\author{Yuki Sato}
\affiliation{Toyota Central R \& D Labs., Inc., Koraku Mori Building 10F, 1-4-14 Koraku, Bunkyo-ku, Tokyo 112-0004, Japan}
\affiliation{Quantum Computing Center, Keio University, 3-14-1 Hiyoshi, Kohoku-ku, Yokohama, Kanagawa, 223-8522, Japan}

\author{Michihiko Sugawara}
\affiliation{Quantum Computing Center, Keio University, 3-14-1 Hiyoshi, Kohoku-ku, Yokohama, Kanagawa, 223-8522, Japan}

\author{Kaito Wada}
\affiliation{Graduate School of Science and Technology, Keio University, 3-14-1 Hiyoshi, Kohoku, Yokohama, Kanagawa, 223-8522, Japan}

\author{Kenji Sugisaki}
\affiliation{Quantum Computing Center, Keio University, 3-14-1 Hiyoshi, Kohoku-ku, Yokohama, Kanagawa, 223-8522, Japan}
\affiliation{Graduate School of Science and Technology, Keio University, 7-1 Shinkawasaki, Saiwai-ku, Kawasaki 212-0032, Japan}
\affiliation{Keio University Sustainable Quantum Artificial Intelligence Center (KSQAIC), Keio University, 2-15-45 Mita, Minato-ku, Tokyo 108-8345, Japan}
\affiliation{Centre for Quantum Engineering, Research and Education, TCG Centres for Research and Education in Science and Technology, Sector V, Salt Lake, Kolkata 700091, India}

\author{Shu Kanno}
\affiliation{Mitsubishi Chemical Corporation, Science \& Innovation Center, Yokohama, 227-8502, Japan}
\affiliation{Quantum Computing Center, Keio University, 3-14-1 Hiyoshi, Kohoku-ku, Yokohama, Kanagawa, 223-8522, Japan}

\author{Hiroshi C. Watanabe}
\email{hcwatanabe@chem.kyushu-univ.jp}
\affiliation{Department of Chemistry, Faculty of Science, Kyushu University, 744 Motooka, Nishi-ku, Fukuoka, 819-0395, Japan}
\affiliation{Quantum Computing Center, Keio University, 3-14-1 Hiyoshi, Kohoku-ku, Yokohama, Kanagawa, 223-8522, Japan}

\author{Haruyuki Nakano}
\affiliation{Department of Chemistry, Faculty of Science, Kyushu University, 744 Motooka, Nishi-ku, Fukuoka, 819-0395, Japan}
\affiliation{Quantum Computing Center, Keio University, 3-14-1 Hiyoshi, Kohoku-ku, Yokohama, Kanagawa, 223-8522, Japan}

\begin{abstract}
In variational quantum algorithms, parameterization is typically applied to single-qubit gates.
In this study, we instead parameterize a generalized controlled gate and propose an algorithm to locally minimize the cost function by maximally optimizing these parameters.
This method extends the Free Quaternion Selection (FQS) technique, which was originally developed for single-qubit gate optimization.
To evaluate its performance, we apply the proposed method to a variety of quantum optimization tasks, including the Variational Quantum Eigensolver (VQE) for both Ising and molecular Hamiltonians, fidelity maximization in general variational quantum algorithms (VQAs), and unitary compilation of time evolution operators.
Across these applications, our method demonstrates efficient optimization, enhanced expressibility, and the ability to construct shallower circuits compared to existing techniques.
Moreover, the method can be generalized to optimize particle-number-conserving gates, which are particularly relevant for quantum chemistry.
Leveraging this capability, we further demonstrate that the method achieves superior quantum compilation of molecular time-evolution operators by approximating them with shallower circuits than standard Trotter decomposition.

\end{abstract}
\maketitle
\def\thefootnote{*}\footnotetext{These authors contributed equally to this work}
\def\thefootnote{\arabic{footnote}}

\section{Introduction}
\input{introduction}

\section{Methods}\label{sec:methods}
\input{methods}

\section{Numerical experiments}\label{sec:results}
\input{results}

\section{conclusions}\label{sec:Disciussions and conclusions}
\input{conclusions}

\section{Acknowledgement}
H.C.W. was supported by JSPS KAKENHI Grant Numbers 23K03266, Council for Science, Technology and Innovation (CSTI), Cross-ministerial Strategic Innovation Promotion Program (SIP), “Promoting the application of advanced quantum technology platforms to social issues” (Funding agency:QST), and CSTIP grant from National Research Council Canada (AQC203). 
K.W. was supported by JSPS KAKENHI Grant Number JP 24KJ1963.
K.S. acknowledges support from Quantum Leap Flagship Program (Grant No. JPMXS0120319794) from the MEXT, Japan, Center of Innovations for Sustainable Quantum AI (JPMJPF2221) from JST, Japan, and Grants-in-Aid for Scientific Research C (21K03407) and for Transformative Research Area B (23H03819) from JSPS, Japan. 
H.N. was supported by JSPS  KAKENHI Grant Number 21K04980.
\clearpage

\bibliography{references}

\newpage
\onecolumngrid

\appendix
\input{appendix}

\end{document}

%% file: introduction.tex
The variational quantum algorithm (VQA) is a hybrid algorithm between classical and quantum computers where the expected value of observables is evaluated through measurements using a trial wave function (ansatz) reproduced on a parameterized quantum circuit (PQC).
Then, VQA repeats the feedback cycle between measurements of the observable on a quantum device and parameter tuning performed by a classical computer. 
Since it is executable on present noisy quantum devices, VQA has been very widely applied to demonstrate the performance of quantum devices. 
In fact, recent studies comparing the first demonstration of VQA on a real device \cite{peruzzo2014variational, kandala2017hardware} and recent application reports have confirmed the remarkable improvement in performance of quantum devices \cite{ollitrault2020quantum,gao2021applications, stanisic2022observing, guo2024experimental,farrell2024scalable}, although they are still subject to severe noise and decoherence.
However, intensive studies on VQA have also revealed critical limitations regarding the trainability on PQC.
Then, it has become clear that, as the number of qubits increases, the mean value of the observable measurement on randomized PQCs converges exponentially to a trivial value, which is called probabilistic concentration \cite{mcclean2018barren, arrasmith2021effect,cerezo2021cost}.
The probabilistic concentration is also known as barren plateau, meaning that as the number of qubits increases, the number of measurements required for a proper parameter update also exponentially grows.
Note that a barren plateau is closely related to expressibility, i.e. versatility of a trial wave function that is termed ansatz \cite{holmes2022connecting}. 
That is, although an increase in the circuit depth with variational parameters makes it possible to express various types of solutions, it also leads to a loss of trainability because of the exponential extension of search space.
The other problem arises from accumulated hardware errors that cause the expected value to converge to a trivial value, which is called deterministic concentration or noise-induced barren plateau~\cite{wang2021noise}.
This is an ironic consequence in contrast to the attention that VQA has received for being an effective algorithm for the noisy intermediate-scale quantum (NISQ) era. 
As prerequisites to avoid these two types of concentration, one should use either a tailored ansatz for the target system or as shallow a quantum circuit as possible in combination with local observables.
However, it has recently become clear that these two approaches alone are not sufficient to avoid the barren plateau.
For example, UCCSD ansatz~\cite{peruzzo2014variational}, the most widely known physics-inspired circuit structure for molecular Hamiltonian, can also induce a barren plateau~\cite{mao2023barren}.
In addition, a barren plateau can be induced even with shallow circuits if the input state is a quantum entangled state that obeys volume law~\cite{ortiz2021entanglement, leone2024practical}.
Because of these two problems, hopes for the quantum advantage of VQA are fading fast.
These series of studies have revealed many limitations of VQA that must be avoided in order to realize its quantum advantage, but they have not necessarily denied the existence of problems and settings in which VQA shows quantum advantage in certain conditions.

Given that most barren plateau proofs have been based on fixed ansatz structures, a possible alternative is to use variable structure circuits, such as Variable ansatz (VAns) \cite{bilkis2023semi} and its analogs
\cite{chivilikhin2020mog, cincio2021machine},
and Adaptive Derivative-Assembled Psuedo-Trotter ansatz (ADAPT) VQE \cite{grimsley2019adaptive} including its derivatives~\cite{tang2021qubit,yordanov2021qubit}.
Unlike the conventional UCCSD approach where the ansatz is determined according to the Hamiltonian before optimization, ADAPT-VQE selects the operator that can lower the energy the most at each iteration. 
This allows the optimization to proceed efficiently in a relatively shallow circuit, and it is suggested that it can actually mitigate the barren plateau
\cite{grimsley2023adaptive}.
In contrast to the ADAPT-VQE where the gate group is repeatedly appended to the main circuit,
in VAns approach, the circuit structure drastically varies during the optimization through an insertion and simplification protocol, which saves the circuit depth expressing the target state with a shallower circuit.
However, this method involves a lot of heuristic protocols which may cause a lot of computational overhead.

The concept of circuit structure optimization is also seen in the Rotoselect \cite{ostaszewski2021structure} method, where an analytical optimization method by Nakanishi, Fujii, and Todo (NFT) \cite{nakanishi2020sequential} (also termed Rotosolve \cite{ostaszewski2021structure}) for rotation angle around a fixed-axis is used to reduce the initial ansatz dependence by providing an axis-selective degree of freedom during cost optimization.
This concept was extended through Free-axis selection (Fraxis) \cite{watanabe2021optimizing,watanabe2023optimizing} and Free Quaternion selection (FQS) \cite{wada2022simulating, sato2023variational,wada2024sequential}, where multi-parameters with respect to a single-qubit gate are simultaneously optimized using matrix diagonalization.
Since these approaches can incorporate the parameter correlations elevating single-qubit gate expressibility, it has been confirmed that Fraxis and FQS achieved greater efficiency than both other local optimizers and the conventional optimizers such as COBYLA and ADAM \cite{wada2024sequential}.
Moreover, the advantage over other optimizers holds even on noisy simulators and real quantum devices.
Furthermore, it has a high affinity with circuit structure optimization because it allows one to insert the ideal single-qubit gate in the optimal form.

In this study, we extend FQS to a controlled unitary gate to guarantee finding the optimal parameters for maximal cost reduction.
Here, we term this method controlled FQS. 
Note that this concept may resemble the unitary block optimization scheme (UBOS) and single-gate tomography~\cite{ben2024quantum}. However, those methods use classical optimizers to find the optimal parameters for a gate of interest, which are heuristic and offer no guarantee of optimality. In contrast, our method yields the truly optimal cost value and parameter set for the target gate, although it is limited to controlled unitaries. 
This contrasts with the above methods, which aim to optimize a more general local gate.
We confirm that controlled FQS achieves not only efficient optimization but also higher resolution to describe target quantum states of interest with a shallower circuit.
In other words, our method can enhance both expressibility and trainability in a good balance.
In this paper, to confirm the performance, we apply the controlled FQS method to VQE for the Ising model and molecular Hamiltonian, VQA for fidelity maximization.
Also, we demonstrate applications of controlled FQS to the unitary compilation of time evolution operators and the reproduced dynamics by molecular Hamiltonian, where it achieves a highly accurate approximation of target unitary with a remarkable compression rate.
Note that although tensor-network-based classical unitary compilation algorithms for shallow time evolution circuits have been studied in recent years~\cite{Anselme_Martin2024-dw,causer2024scalable,Lin2021-km,Mc_Keever2023-bu,Mc_Keever2024-ji,Kanno2024-iw}, the unitary compilation with VQA has the advantage of greater flexibility in ansatz.

%% file: methods.tex
\subsection{Overview of Free quaternion selection}
A general single-qubit gate $R_{\bm{n}} (\theta) \in \mathrm{SU}(2)$ is conventionally represented using a rotational axis $\bm{n}$ and a rotation angle $\theta$ as
\begin{align}\label{eqn:su2}
    R_{\bm{n}}(\theta)=e^{-i\frac{\theta}{2}\bm{n}\cdot \vec{\sigma}},
\end{align}
where $\vec{\sigma}=(X, Y, Z)$.

We define an extended Pauli matrix as $\vec{\varsigma} \equiv (\varsigma_i,\varsigma_\mathrm{x},\varsigma_\mathrm{y},\varsigma_\mathrm{z})=(I, -iX, -iY, -iZ)$.
Then, a general single-qubit gate can be mapped to a unit quaternion $\bm{q}$ as,  
\begin{align}\label{eq:fqs_gate}
    R(\bm{q})
    =\bm{q}\cdot\vec{\varsigma},
\end{align}
where $\bm{q}\in \mathbb{R}^4$.
Consider a quantum circuit $\prod_k W_k R_k $ where $R_k$ and $W_k$ denote the $k$-th single-qubit gate and fixed-parameter gate, respectively.
When applying $R(\bm{q})$ to an $n$-qubit system, it is embedded as $I^{\otimes m} \otimes R(\bm{q}) \otimes I^{\otimes n - m - 1}$ for some $m \leq n$.  
For notational simplicity, we omit the identity operators hereafter.
Focusing on a specific gate $R_j $, the entire circuit can be rewritten as
\begin{align}\label{eqn:QuantumCircuit}
\prod_k W_k R_k = V_2 R_j(\bm{q}) V_1,
\end{align}
where $V_1$ and $V_2$ are partial circuits before and after $R_j(\bm{q})$, respectively.
Then, the expectation value of an observable $H$ is then given by
\begin{align}\label{eqn:MeanValue_fqs}
\braket{H} 
&= \mathrm{tr}[H V_2 R_d(\bm{q}) V_1 \rho_0 V_1^{\dagger} R_d(\bm{q})^{\dagger} V_2^{\dagger} ] \notag \\
&= \mathrm{tr}[H' R(\bm{q})  \rho'  R(\bm{q})^{\dagger}  ]\notag \\
&= \bm{q}^{T} J \bm{q}, 
\end{align}
where $\rho_0$ is the initial state's density matrix, $H'=V_2^{\dagger}HV_2$, $\rho_0'=V_1 \rho_0 V_1^{\dagger}$,  $J\in \mathbb{R}^{4\times4}$ and $J_{\mu\nu}=\frac{1}{2} \mathrm{tr}[(\varsigma_\mu^{\dagger} H' \varsigma_\nu+ \varsigma_\nu^{\dagger} H' \varsigma_\mu) {\rho'_0}]$.
Since Eq.~\eqref{eqn:MeanValue_fqs} is a quadratic form, $\braket{H}$ can be minimized (or maximized) by solving the associated eigenvalue problem.

\subsection{Free quaternion selection for a controlled unitary gate (cFQS)}
\subsubsection{Cost landscape tomography}
This section extends the cost landscape tomography framework developed for single-qubit gates in~\cite{endo2023optimal} to the case of controlled-unitary gates.
We begin by expressing a controlled-unitary gate $U(\bm{q}) \in \mathrm{SU}(4)$ as
\begin{align}\label{eqn:su4}
U(\bm{q})=\ket{0}\bra{0}_\mathrm{c} \otimes I +\ket{1}\bra{1}_\mathrm{c} \otimes R(\bm{q}),
\end{align}
where $R(\bm{q}) \in \mathrm{SU}(2)$ is defined as in Eq.~\eqref{eqn:su2}, and the subscript $c$ denotes the control qubit.
An arbitrary quantum state $\ket{\Phi}$ can be expressed as
\begin{align}\label{eqn:SchmidtDecomposition}
\ket{\Phi}
&=\lambda_0\ket{0_\mathrm{c}\varphi_0}+\lambda_1\ket{1_\mathrm{c} \varphi_1},
\end{align}
where $\lambda_0, \lambda_1 \in \mathbb{C}$ and $|\lambda_0|^2+|\lambda_1|^2=1$. 
For simplicity, the control-qubit subscript $c$ is omitted hereafter. 
The states $\ket{\varphi_0}$ and $\ket{\varphi_1}$ belong to the remaining qubits.
Applying the controlled gate $U(\bm{q})$ to $\ket{\Phi}$ yields
\begin{align}\label{eqn:AppliedState}
\ket{\Psi}
=U\ket{\Phi}
&=\lambda_0 \ket{0\varphi_0}
+\lambda_1  R(\bm{q})\ket{1\varphi_1}\notag \\
&=\lambda_0 \ket{0\varphi_0}
+\lambda_1 \sum_{\mu\in \{\mathrm{i,x,y,z}\}} q_\mu \varsigma_\mu\ket{1\varphi_1} .
\end{align}
where we used the expansion $R(\bm{q}) = \bm{q} \cdot \vec{\varsigma}$.
Suppose a quantum circuit composed of $D$ controlled-unitary gates $\{U_d(\bm{q}_d)\}_{d=1}^D$ interleaved with other gates. 
Focusing on the $d$-th controlled gate, the circuit can be written as $V_2 U_d(\bm{q}_d) V_1$, where $V_1$ and $V_2$ are the circuit segments before and after $U_d$, respectively.
The expectation value $m$ of the observable $H$ is then given as,
\begin{align}\label{eqn:MeanValue}
\braket{H}\equiv m 
&= \mathrm{tr}[H V_2 U_d V_1 \rho_0 V_1^{\dagger} U^{\dagger}_d V_2^{\dagger} ] \notag \\
&= \mathrm{tr}[H' U_d  \rho'  U^{\dagger}_d  ]\notag \\
&= \bra{\Phi}U_d^{\dagger} H' U_d \ket{\Phi}, 
\end{align}
where $\rho_0$ is the initial state, $\rho'=\ket{\Phi}\bra{\Phi}=V_1 \rho_0 V_1^{\dagger}$ and $H'=V_2^{\dagger} H V_2$.
Substituting Eqs.~\eqref{eqn:su2}, \eqref{eqn:su4}, and \eqref{eqn:AppliedState} into Eq.~\eqref{eqn:MeanValue}, we obtain
\begin{align}\label{eqn:quadratic}
m &= |\lambda_0|^2 \bra{0\varphi_0} H' \ket{0\varphi_0}
+ 2 \mathrm{Re}\left[\lambda_0^* \lambda_1 \bra{0\varphi_0} H' R \ket{1\varphi_1} \right] \notag \\
&\quad + |\lambda_1|^2 \bra{1\varphi_1} R^{\dagger} H' R \ket{1\varphi_1} \notag \\
&= |\lambda_0|^2 \bra{0\varphi_0} H' \ket{0\varphi_0}
+ 2 \sum_{\mu} q_\mu \mathrm{Re}\left[\lambda_0^* \lambda_1 \bra{0\varphi_0} H' \varsigma_\mu \ket{1\varphi_1} \right] \notag \\
&\quad + |\lambda_1|^2 \sum_{\mu,\nu} q_\mu q_\nu \bra{1\varphi_1} \varsigma_\mu^\dagger H' \varsigma_\nu \ket{1\varphi_1}.
\end{align}
If the parameter $\lambda_0$ is real and controllable, Eq.~\ref{eqn:quadratic} can be rewritten in quadratic form by expressing $q_0= \lambda_0$, solvable in the same way as FQS.
However, this assumption does not hold in general. 
Consequently, the eigenvector of $E$ corresponding to the smallest eigenvalue cannot be directly used for optimization.  
Instead, we fix $q_0=1$, and express the cost function as follows,
\begin{align}\label{eqn:quadratic_form}
m =\sum_{\mu,\nu\in\{\mathrm{0,i,x,y,z}\}} q_\mu q_\nu E_{\mu\nu},
\end{align}
where $E$ is a symmetric matrix whose elements are as follows.
\begin{align}\label{eqn:matrix_elemenets}
E_{00} &= |\lambda_0|^2 \bra{0\varphi_0}H'\ket{0\varphi_0}\notag\\
E_{0\mu} 
&= {\rm Re}[\lambda_0^\ast\lambda_1 \bra{0\varphi_0}H'\varsigma_\mu\ket{1\varphi_1}]~~~\mu \in \{\mathrm{i,x,y,z}\} \\
E_{\mu\nu} 
&=|\lambda_1|^2 \bra{1\varphi_1}\varsigma_\mu^\dagger H'\varsigma_\nu\ket{1\varphi_1}~~~\mu,\nu \in \{\mathrm{i,x,y,z}\}\notag.
\end{align}
Then, we define an extended vector $\tilde{\bm{q}}$ mapped by function $h$ as
\begin{align}
\label{eqn:extended_vec}
\tilde{\bm{q}}
=&h(\bm{q})\notag\\
= &(q_\mathrm{i}^2, 
q_\mathrm{x}^2, 
q_\mathrm{y}^2, 
q_\mathrm{z}^2, 
2q_\mathrm{i} q_\mathrm{x}, 
2q_\mathrm{i} q_\mathrm{y}, 
2q_\mathrm{i} q_\mathrm{z}, 
2q_\mathrm{x} q_\mathrm{y}, 
2q_\mathrm{x} q_\mathrm{z}, 
2q_\mathrm{y} q_\mathrm{z})^T\notag\\
&\oplus
(2q_{\mathrm{i}}, 
2q_{\mathrm{x}}, 
2q_{\mathrm{y}}, 
2q_\mathrm{z})^T\notag\\
&\oplus
1.
\end{align}
We also define a corresponding coefficient vector\begin{align}
 \bm{e} =& (E_\mathrm{ii}, E_\mathrm{xx}, E_\mathrm{yy}, E_\mathrm{zz}, E_\mathrm{ix}, E_\mathrm{iy}, E_\mathrm{iz}, E_\mathrm{xy}, E_\mathrm{xz}, E_\mathrm{yz})^T \notag\\
 &\oplus  (E_\mathrm{0i}, E_\mathrm{0x}, E_\mathrm{0y}, E_\mathrm{0z})^T \notag\\
 &\oplus E_{00}.
\end{align}
With these definitions, the expectation value can be written as a linear function:
\begin{align}\label{eqn:meanvalue}
m =  \tilde{\bm{q}}^T ~ \bm{e}.
\end{align}
This implies that once $\bm{e}$ is known, $\braket{H}$ can be predicted for any $\bm{q}$ without additional measurement.

To estimate $\bm{e}$, suppose we perform measurements using $N$ different parameters $\{\tilde{\bm{q}}_i\}_{i=1}^N$, forming the parameter configuration matrix $Q$ and measurement vector $\bm{m}$:
\begin{align}\label{eqn:Qmatrix}
Q=
\begin{bmatrix}
\tilde{\bm{q}}_1^T\\
\tilde{\bm{q}}_2^T\\
 \vdots \\
\tilde{\bm{q}}_N^T\\
\end{bmatrix}\in \mathbb{R}^{N \times 15}, \quad
\bm{m}=
\begin{bmatrix}
m_1\\
m_2\\
\vdots\\
m_N
\end{bmatrix}, 
\end{align}
where $m_i = \mathrm{tr}[H' U(\bm{q}_i) \rho' U(\bm{q}_i)^\dagger]$.
Then, $\bm{e}$ is determined from the linear system
\begin{align}\label{eqn:LinearEquation}
Q \bm{e} = \bm{m}.
\end{align}

If $N < 15$, $Q$ is rank-deficient and $\bm{e}$ cannot be uniquely determined. 
Even for $N \geq 15$, $\mathrm{rank}(Q)$ is at most 14 in general. 
In such cases, we append a null-space vector $\tilde{\bm{q}}_{15} \in \mathrm{Ker}(\tilde{Q})$, defined as
\begin{align}\label{eqn:p15}
\tilde{\bm{q}}_{15} = (-1, -1, -1, -1, 0, 0, 0, 0, 0, 0, 0, 0, 0, 0, 1)^T,
\end{align}
and set the corresponding $m_{15} = 0$, yielding a full-rank augmented matrix.

Once $Q$ is full rank by choosing $\{\bm{q}\}$ such that $\{\tilde{\bm{q}}_n\}$ are independent of each other, $\bm{e}$ can be estimated via:
\begin{align}
\bm{e} = Q^{-1} \bm{m}.
\end{align}
After obtaining $\bm{e}$, the expectation value can be recast in Eq.~\eqref{eqn:quadratic}, then we obtain
\begin{align}\label{eqn:mean_value}
\braket{H}=\bm{q}^T J \bm{q} + 2 \bm{a}^T \bm{q} + b,\quad \text{with } |\bm{q}| = 1,
\end{align}
where $\bm{a}=(E_{0i},E_{0x},E_{0y},E_{0z})\in \mathbb{R}^4$, $b=E_{00}\in \mathbb{R}$, and $J = (E_{kl})_{k,l \in \{\mathrm{i,x,y,z\}} } \in \mathbb{R}^{4 \times 4}$ is a symmetric matrix.
Thus, with only 14 measurements (plus one kernel constraint), we can reconstruct $\braket{H}$ for arbitrary $\bm{q}$ without further sampling. 
For the state vector simulations in this study, we employed the configuration given in Eq.~\eqref{eqn:PC_cFQS_sv} in Appendix~\ref{apdx:parameter configuration}. 
The choice of configuration, however, significantly influences optimization performance under shot noise. 
To improve robustness, we adopted the C-cost strategy introduced in~\cite{endo2023optimal}, which minimizes statistical errors in estimating the relevant eigenvalues. 
Unlike standard FQS, controlled-FQS does not extract the eigenvalue explicitly, but instead uses it to guide parameter search (as detailed in the next section).
The configuration obtained through the protocol is shown in Appendix~\ref{apdx:parameter configuration}.
The final parameter configuration is presented in Appendix~\ref{apdx:parameter configuration}. 
It demonstrated significantly improved convergence in noisy (QASM) simulations and was thus employed throughout this work.

\subsubsection{Optimal parameter search}
We consider the following optimization problem:
\begin{align}\label{eqn:min_problem}
\underset{\bm{q}}{\mathrm{min}}~ \braket{H}(\bm{q}) = \bm{q}^T J \bm{q} + 2 \bm{a}^T \bm{q} + b,
\end{align}
where $|\bm{q}| = 1$.
Let $r_i$ and $\bm{n}_i$ denote the $i$-th eigenvalue and corresponding unit eigenvector of $J$, ordered such that $r_i < r_{i+1}$.  
For simplicity, we assume all eigenvalues are non-degenerate, but the derivation can be extended to the degenerate case.
Any vector $\bm{q} \in \mathbb{R}^4$ with unit norm can be expressed as a linear combination of $\{\bm{n}_i\}$:
\begin{align}\label{eqn:linear_combination}
\bm{q}=\sum_{i=1}^4 c_i \bm{n}_i,
\end{align}
where the coefficients satisfy the normalization constraint:
\begin{align}\label{eqn:normalization}
\sum_{i=1}^4 c_i^2 = 1.
\end{align}
Substituting Eq.~\eqref{eqn:linear_combination} into Eq.~\eqref{eqn:min_problem}, we obtain:
\begin{align}
\braket{H}=\sum_{i}^4 r_i c_i^2 + 2  \sum_{i=1}^4  c_i \bm{a}^T \bm{n}_i + b.
\end{align}
To minimize (or maximize) $\braket{H}$ under the constraint~\eqref{eqn:normalization}, we define the Lagrangian:
\begin{align}\label{eqn:lagrangan}
    \mathcal{L}\equiv \sum_{i=1}^4 r_i c_i^2 + 2 \sum_{i}^4 c_i 
 \bm{a}^T \bm{n}_i + b - \Lambda(\sum_i^4 c_i^2 -1) ,
    \end{align}
where $\Lambda$ is a Lagrange multiplier. 
The stationary condition gives:
\begin{align}\label{eqn:stationary}
\frac{\partial \mathcal{L}}{\partial c_i} = 2 (r_i  -\Lambda) c_i + 2 \bm{a}^T \bm{n}_i =0,
\end{align}
which yields
\begin{align}\label{eqn:coefficient}
c_i = \frac{\bm{a}^T\bm{n}_i}{\Lambda - r_i}.
\end{align}
Substituting Eq.~\eqref{eqn:coefficient} into the normalization condition~\eqref{eqn:normalization} results in:
$\sum_i c_i^2=1$, we obtain 
\begin{align}\label{eqn:constraint}
\sum_{i=1}^4 (\frac{\bm{a}^T\bm{n}_i}{\Lambda - r_i})^2 =1.
\end{align}
We define the function
\begin{align}
f(\Lambda)=\sum_{i=1}^4 (\frac{\bm{a}^T\bm{n}_i}{\Lambda - r_i})^2-1.
\end{align}
and seek its roots, which correspond to valid solutions of the constrained optimization problem.
To solve Eq.~\eqref{eqn:min_problem}, we search for roots of $f(\Lambda) = 0$.  
Note that $f(\Lambda)$ is a convex function in a range of $(r_{i-1}, r_{i})$ and  
$\lim_{\Lambda \to r_i}f(\Lambda)=+\infty$, if $\bm{a}^T \bm{n}_i \neq 0$ and $\bm{a}^T \bm{n}_{i+1} \neq 0$.
Let $s_i \in (r_i, r_{i+1})$ denote the unique local minimum of the function $f(\Lambda)$ in that interval, i.e., the point where $\partial f / \partial \Lambda = 0$. 
Since $f(\Lambda)$ is convex in each interval, the local minimum $s_i$ can be efficiently found using ternary search algorithm. 
We define $s_0 = s_4 = -\infty$. 
Then, for each $i = 1, 2, 3, 4$, the function $f(\Lambda)$ is monotonic in the intervals $(s_{i-1}, r_i)$ and $(r_i, s_i)$.

\begin{algorithm}[H]
\caption{Free quaternion selection for controlled-gate} \label{algo:CFQS}
\begin{algorithmic}[1]
\Require PQC structure $V_1$ and $V_2$, target gate index $d$, observable $H$, parameter configuration $\{\bm{q}\}$.
\Ensure Optimized cost $\braket{H}_{\mathrm{min}}$ and parameters $\bs{q}^\ast$.

\Procedure{MainProcedure}{}
 \State $\bm{e} \gets$ \Call{SubRoutine1}{$V_1, V_2, H, \{\bm{q}\}$}
\State Extract $J$, $\bm{a}$, and $b$ from $\bm{e}$ using Eq.~\eqref{eqn:mean_value}
 \State $\braket{H}_{min}, \bm{q}^* \gets$ \Call{SubRoutine2}{$J, \bm{a}, b$}
 \State \Return{$\braket{H}_{min}, \bm{q}^* $}
\EndProcedure
\State  
\Procedure{Subroutine1}{$V_1, V_2, H, \{\bm{q}\}$}
\For {$\bm{q}$ in $ \{{\bm q_1}, {\bm q_2},..., \bm q_{14}\}$}
\State Append $h({\bm{q}})$ to $Q$.
\State Append $\mathrm{tr}[H V_2 R_d(\bm{q}) V_1 \rho_0 V_1^{\dagger} R_d(\bm{q})^{\dagger} V_2^{\dagger} ]$ to ${\bm m}$.
\EndFor
\State Append $\tilde{\bm{q}}_{15}\in \mathrm{Ker}(Q)$ in Eq~\eqref{eqn:p15} to $Q$.
\State Append $0$ to $\bm{m}$
\State \Return{$Q^{-1} \bm{m}$}
\EndProcedure
\State  
\Procedure{Subroutine2}{$J, \bm{a}, b$}
\State Compute eigenvalues $r_i$ and eigenvectors $\bm{n}_i$ of $J$
\State Find local minima of $f(\Lambda)$ via ternary search
\State Identify all $\Lambda$ such that $f(\Lambda) = 0$ using binary search\For {$\{\Lambda\}$}
  \State Compute $c_i$ from $\Lambda$ using Eq.~\eqref{eqn:coefficient}
  \State Reconstruct $\bm{q}$ from $\{c_i\}$ via Eq.~\eqref{eqn:linear_combination}
  \State Evaluate $\braket{H}(\bm{q})$ using Eq.~\eqref{eqn:mean_value}
\EndFor
\State Select the lowest $\braket{H}(\bm{q})$ and corresponding $\bm{q}$ 
\State \Return{$\braket{H}_\mathrm{min}, \bm{q}^* $}
  \EndProcedure
\end{algorithmic}
\end{algorithm}
If $f(s_i) < 0$, each valid interval contains at most two roots as:
\begin{align}
\exists! \Lambda \in (s_{i-1}, r_i) : f(\Lambda) = 0, \\
\exists! \Lambda \in (r_i, s_i) : f(\Lambda) = 0.
\end{align}

By applying binary search across all such intervals, we obtain the root set
\[
L = \left\{ \tilde{\Lambda} \in \mathbb{R} \mid f(\tilde{\Lambda}) = 0 \right\}, \quad \text{with } |L| \leq 8.\]

For each $\tilde{\Lambda} \in L$, compute the corresponding coefficients $\{c_i\}$, reconstruct the quaternion $\bm{q}^{(m)}$ via Eq.~\eqref{eqn:linear_combination}, and evaluate the cost $\braket{H(\bm{q}^{(m)})}$ using Eq.~\eqref{eqn:meanvalue}.  
Finally, select the $\bm{q}^*$ yielding the lowest cost $\braket{H}_{\min}$.
This procedure is easily extendable to multi-controlled gates such as the Toffoli gate.

\subsubsection{Implementation}
The present popular quantum devices such as superconducting quantum devices do not natively support general controlled-unitary gates.
However, such gates can be executed, when decomposed into fixed-axis single qubit rotations and controlled-NOT gates as shown in Fig.~\ref{fig:decomposition} \cite{barenco1995elementary}.
Since the decomposition involves two controlled operations, the CNOT depth of the circuit doubles.
When the controlled unitary operator belongs to the special unitary group $\mathrm{SU}(2)$, the relation between the quaternion parameters and the rotation angles is given by
\begin{gather}
\begin{gathered}
q_\mathrm{i} = \cos{\frac{\beta+\delta}{2}}\cos{\frac{\gamma}{2}},~~~  q_\mathrm{x} = \sin{\frac{\beta-\delta}{2}}\sin{\frac{\gamma}{2}}, \\
q_\mathrm{y} = \cos{\frac{\beta-\delta}{2}}\sin{\frac{\gamma}{2}},~~~  q_\mathrm{z} = \sin{\frac{\beta+\delta}{2}}\cos{\frac{\gamma}{2}}.
\end{gathered}
\end{gather}
In Fig.~\ref{fig:decomposition}, the single-qubit gate $R_z(\theta)$ acting on the first qubit at the end of the circuit corresponds to a phase gate.  
This gate becomes redundant when the target unitary $U$ belongs to $\mathrm{SU}(4)$.  
However, in our setup, we assume that the controlled unitary gates are initialized as either controlled-NOT or controlled-Z gates.  
In such cases, the final phase gate acquires a fixed non-zero angle, specifically $\theta = \pi/2$, and remains constant during the application of the controlled-FQS procedure.
\begin{figure}[t]
 \begin{tabular}{c}
   \Qcircuit @C=1.0em @R=.8em {
       \lstickx{} &\qw &\ctrl{1} &\qw \\
       \lstickx{} &\qw &\gate{R(\bm{q})}&\qw   
       }
   \\
   \Qcircuit @C=.24em @R=.32em 
   {\Updownarrow }
   \\
 \Qcircuit @C=0.24em @R=.32em {
    \lstickx{} & \qw & \qw & \qw &\ctrl{1} 
       &\qw   &\qw &\ctrl{1} &\gate{R_z(\theta)} &\qw  \\ 
    \lstickx{} &\qw &\gate{R_z(\beta)}&\gate{R_y(\frac{\gamma}{2})}&\targ   &\gate{R_y(-\frac{\gamma}{2})}& \gate{R_z(-\frac{\delta+\beta}{2})}&\targ & \gate{R_z(\frac{\delta-\beta}{2})}&\qw}
      \\
\end{tabular}
\caption{A decomposition of a general controlled unitary gate into ordinary fixed rotational single qubit gates and two controlled-NOT gates.}
\label{fig:decomposition}
\end{figure}

\subsubsection{Optimization of a Generalized Particle-number preserving gate}
In quantum chemical calculations, the number of electrons is typically known in advance, depending on the system under study.
In the Jordan–Wigner mapping, qubits encode orbital occupancies, and only a limited subset of computational basis states with the correct particle number contribute to the ground state.
This prior information can be exploited to restrict the variational search space, thereby improving optimization efficiency.
To achieve this, the quantum gates must also be parameterized in a manner that preserves particle number.

Protocols for constructing such gates have been proposed in prior works~\cite{gard2020efficient,kerenidis2021classical,eddins2022doubling}, where the gate is decomposed into the fixed-axis gates.
These gates are typically represented using one or two parameters, assuming that each parameter is optimized independently.
Although simultaneous optimization of two parameters has been proposed for excitation-preserving gates, those approaches are primarily tailored for time-evolution simulations~\cite{wada2022simulating}.

Here, we propose a particle number-preserving gate based on a quaternion representation, providing three degrees of freedom.
In other words, this approach generalizes particle-number preserving gates by leveraging  the controlled FQS method. 
The matrix representation of the proposed gate is given by
\begin{align}
U_\mathrm{NP}(\bm{q})=
\begin{bmatrix}
1 & 0 & 0 & 0\\
0 & q_\mathrm{i}-i q_\mathrm{z} & -q_\mathrm{y} -i q_\mathrm{x} & 0\\
0 & q_\mathrm{y} -i q_\mathrm{x} & q_\mathrm{i} + i q_\mathrm{z}  & 0\\
0 & 0 & 0 & 1\\
\end{bmatrix}
~~ \mathrm{for} ~|\bm{q}| =1  .
\label{eqn:ParticlePreserving}
\end{align}
A generalized particle-number preserving gate is represented with a complex number.
In quantum chemistry, however, in the absence of external magnetic fields and spin-orbit interactions, the ground state is generally represented in real space. 
Thus, although the quaternion representation may appear excessive at first glance, it has been reported to effectively avoid local minima and enable efficient optimization toward the target state~\cite{sato2023variational,wada2022simulating}
The generalized particle-number conserving gate is implemented using a controlled-FQS scheme as illustrated in Fig.~\ref{fig:PPgate}. 

\subsection{Self-consistent field optimization between a controlled and a negative controlled gates}
Suppose a pair of sequential controlled and negative controlled gates that shares the same control bit as in Fig.~\ref{fig:sequential_cgates}.
\begin{align}\label{eqn:sequence_cfqs}
U(\bm{q})=\ket{0}\bra{0}_\mathrm{c} \otimes R(\bm{p})
 +\ket{1}\bra{1}_\mathrm{c} \otimes R(\bm{q}).
\end{align}
Then, substituting this into Eq.~\eqref{eqn:MeanValue}, the expectation value of $m$ of the Hamiltonian $H$ becomes
\begin{align}\label{eqn:cost_function}
m
&=|\lambda_0|^2 \bra{0 \phi_0} R^{\dagger}(\bm{p}) H' R(\bm{p}) \ket{0 \phi_0} \notag \\
&\quad +|\lambda_1|^2  \bra{1 \phi_1} R^{\dagger}(\bm{q}) H' R(\bm{q})  \ket{1 \phi_1} \notag \\ 
&\quad +2  \mathrm{Re}[\lambda_0 \lambda_1  \bra{0 \phi_0} R(\bm{p})^{\dagger} H' R(\bm{q})  \ket{1 \phi_1} ] \notag  \\
&= \sum_{i,j} p_i p_j |\lambda_0|^2 \bra{0\phi_0} \varsigma_i^{\dagger} H' \varsigma_j \ket{0 \phi_0}  \notag \\
&\quad + \sum_{i,j}  q_i q_j |\lambda_1|^2 \bra{1\phi_1}  \varsigma_i^{\dagger} H' \varsigma_j \ket{1 \phi_1} \notag \\
&\quad + \sum_{i,j} p_i q_j  \lambda_0^\ast \lambda_1 \bra{0\phi_0} \varsigma_i^{\dagger} H' \varsigma_j \ket{1 \phi_1} \notag\\
&\quad + \sum_{i,j} q_j p_i  \lambda_0 \lambda_1^\ast \bra{1\phi_1} \varsigma_i^{\dagger} H' \varsigma_j \ket{0 \phi_0}\notag \\
&= \bm{p}^T J \bm{p} + \bm{p}^T K \bm{q} + \bm{q}^T K^\dagger \bm{p} + \bm{q}^T L \bm{q} \notag \\
&= 
\begin{bmatrix}
\bm{p}^T & \bm{q}^T
\end{bmatrix}
\begin{bmatrix}
J&K  \\
K^\dagger &L
\end{bmatrix}
\begin{bmatrix}
\bm{p} \\
\bm{q}
\end{bmatrix}
\end{align}
where $|\bm{p}|=1$ and $|\bm{q}|=1$, and the matrix elements are defined as follows: $J_{ij}\equiv |\lambda_0|^2 \bra{0 \phi_0} \varsigma_i^\dagger H' \varsigma_j \ket{0 \phi_0},~
 L_{ij} \equiv |\lambda_1|^2  \bra{1 \phi_1} \varsigma_i^\dagger H' \varsigma_j \ket{1 \phi_1}$,  and 
 $K_{ij} \equiv \lambda_0^\ast \lambda_1  \bra{0 \phi_0} \varsigma_i^\dagger H' \varsigma_j \ket{1 \phi_1}$.
We define an extended vector $\tilde{\bm{q}}'$ and coefficient vector $\bm{e}'$ as
\begin{align}
\label{eqn:extended_vec2}
\tilde{\bm{q}}'
=&h(\bm{p},\bm{q})\notag\\
=&(p_\mathrm{i}^2, 
p_\mathrm{x}^2, 
p_\mathrm{y}^2, 
p_\mathrm{z}^2, 
2p_\mathrm{i} p_\mathrm{x}, 
2p_\mathrm{i} p_\mathrm{y}, 
2p_\mathrm{i} p_\mathrm{z}, 
2p_\mathrm{x} p_\mathrm{y}, 
2p_\mathrm{x} p_\mathrm{z}, 
2p_\mathrm{y} p_\mathrm{z}\notag\\
&~q_\mathrm{i}^2, 
q_\mathrm{x}^2, 
q_\mathrm{y}^2, 
q_\mathrm{z}^2, 
2q_\mathrm{i} q_\mathrm{x}, 
2q_\mathrm{i} q_\mathrm{y}, 
2q_\mathrm{i} q_\mathrm{z}, 
2q_\mathrm{x} q_\mathrm{y}, 
2q_\mathrm{x} q_\mathrm{z}, 
2q_\mathrm{y} q_\mathrm{z},\notag\\
&~2p_\mathrm{i}q_\mathrm{i}, 
2p_\mathrm{i}q_\mathrm{x}, 
2p_\mathrm{i}q_\mathrm{y}, 
2p_\mathrm{i}q_\mathrm{z},\notag\\
&~2p_x q_\mathrm{i}, 
2p_\mathrm{x} q_\mathrm{x}, 
2p_\mathrm{x} q_\mathrm{y}, 
2p_\mathrm{x} q_\mathrm{y}, \notag\\
&~2p_\mathrm{y} q_\mathrm{i}, 
2p_\mathrm{y} q_\mathrm{x}, 
2p_\mathrm{y} q_\mathrm{y}, 
2p_\mathrm{y} q_\mathrm{z}, \notag\\
&~2p_\mathrm{z} q_\mathrm{i}, 
2p_\mathrm{z} q_\mathrm{x}, 
2p_\mathrm{z} q_\mathrm{y}, 
2p_\mathrm{z} q_\mathrm{z})^T
\end{align}
and 
\begin{align}
 \bm{e}' =
 &( J_\mathrm{ii}, J_\mathrm{xx}, J_\mathrm{yy}, J_\mathrm{zz}, J_\mathrm{ix}, J_\mathrm{iy}, J_\mathrm{iz}, J_\mathrm{xy}, J_\mathrm{xz}, J_\mathrm{yz} \notag\\
&~ L_\mathrm{ii}, L_\mathrm{xx}, L_\mathrm{yy}, L_\mathrm{zz}, L_\mathrm{ix}, L_\mathrm{iy}, L_\mathrm{iz}, L_\mathrm{xy}, L_\mathrm{xz}, L_\mathrm{yz} \notag\\
&~ K_\mathrm{ii}, K_\mathrm{ix}, K_\mathrm{iy}, K_\mathrm{iz}
 K_\mathrm{xi}, K_\mathrm{xx},  K_\mathrm{xy},  K_\mathrm{xz},\notag\\
& K_\mathrm{yi}, K_\mathrm{yx},  K_\mathrm{yy},  K_\mathrm{yz},
 K_\mathrm{zi}, K_\mathrm{zx},  K_\mathrm{zy},  K_\mathrm{zz})^T
\notag\\
\end{align}
\begin{figure}[t]
\begin{tabular}{ccc}   
\begin{tabular}{c}   
 \Qcircuit @C=1.0em @R=1.28 em{
   \lstickx{~} &\qw & \multigate{1}{U_\mathrm{NP}(\bm{q})} &\qw \\ 
    \lstickx{~}&\qw &\ghost{U_\mathrm{NP}(\bm{q})} &\qw \\
 }
\end{tabular}
&$\Longleftrightarrow$ &
\begin{tabular}{c}   
\Qcircuit @C=1.0em @R=.64em {
    \lstickx{~} &\qw    &\targ 
       &\ctrl{1}   &\targ  &\qw  \\ 
    \lstickx{~} &\qw   &\ctrl{-1}   & \gate{R(\bm{q})}&\ctrl{-1} &\qw}
\end{tabular}
\end{tabular}
\caption{A generalized particle number preserving gate. $R(\bm{q})$ is a generalized unitary belonging to SU(2). }
\label{fig:PPgate}
 \begin{tabular}{c}
   \Qcircuit @C=0.96em @R=1.28em {
       & \qw &\ctrlo{1} &\ctrl{1} & \qw\\ 
       &\qw &\gate{R(\bm{p})} &\gate{R(\bm{q})}  &\qw }
\end{tabular}
\caption{Sequentially located controlled and negatively controlled gates.}
\label{fig:sequential_cgates}
\end{figure}

Then, the expectation value becomes
\begin{align}\label{eqn:meanvalue_function}
m =  \tilde{\bm{q}}'^T ~ \bm{e}'.
\end{align}
By collecting different $\tilde{\bm{q}}'$ vectors, we construct the matrix $Q'$ as in Eq.~\eqref{eqn:Qmatrix}.
\begin{align}\label{eqn:extendedQmatrix}
Q'=
\begin{bmatrix}
\tilde{\bm{q}}'^T_1\\
\tilde{\bm{q}}'^T_2\\
 \vdots \\
\tilde{\bm{q}}'^T_N\\
\end{bmatrix}, ~~~ 
\bm{m}=
\begin{bmatrix}
m_1\\
m_2\\
\vdots\\
m_N
\end{bmatrix}.
\end{align}
Let $\mathcal{Q}' = (Q'^TQ')^{-1} Q'^T$ denote the pseudoinverse of $Q'$.
If $N<36$, $\mathrm{rank}(\mathcal{Q}')$ is at most $N$, so $\bm{e}'$ cannot be uniquely determined.
Even for $N \geq 36$, $\mathrm{rank}(\mathcal{Q}')$ is 35 at most. 
To make $Q'$ full rank, we append the vector $\tilde
{\bm{q}}'_{36} \in \mathrm{Ker}(Q')$:
\begin{align}
(\tilde{\bm{q}}'_{36})_i =
\begin{cases}
-\frac{1}{4}~~&\mathrm{for}~~i\in\{1,2,3,4\}\\
~~\frac{1}{4}~~&\mathrm{for}~~i\in\{11,12,13,14\}\\
~~0~~&\mathrm{otherwise}.\\
\end{cases}
\end{align}
In this study, we employed a parameter configuration $Q'$ shown in $\{\bm{p},\bm{q}\}$ shown in Eq.~\eqref{eqn:PC_SCF-cFQS_sv} for state vector simulations and in Eq.~\eqref{eqn:PC_SCF-cFQS_qasm} for QASM simulations. 

\begin{algorithm}[H]
    \caption{Self-consistent field optimization for controlled-gates via FQS}
    \label{algo:SCF-CFQS}
    \begin{algorithmic}[1]
    \Require PQC structure $V_1$ and $V_2$, target gate index $d$, observable $H$, current cost $\braket{H}$, parameter configuration $\{(\bm{p}, \bm{q})\}$, threshold value $t$.
    \Ensure Optimized parameters $(\bm{p}^\ast, \bm{q}^\ast)$.
    \Procedure{MainProcedure}{}
    \State $\bm{e}^{\prime} \gets$ \Call{Subroutine3}{$V_1, V_2, H, \{(\bm{p},\bm{q})\}$}
    \State Extract $J,K,L$ from $\bm{e}'$ using Eq.~\eqref{eqn:cost_function}
    \State $m \gets 0$
    \State $y_1 \gets \braket{H}$
    \Repeat
    \State $y_0 \gets y_1$ 
    \If{$m$ is even} 
    \State $\bm{a}  \gets \bm{q}^T K^T$
    \State $b \gets \bm{q}^T L\bm{q}$
    \State $y_1, \bm{p} \gets$ \Call{Subroutine2}{$J, \bm{a}, b$}
    \Else 
    \State $\bm{a} \gets \bm{p}^T K$
    \State $b \gets \bm{p}^T J\bm{p}$
    \State $y_1, \bm{q} \gets$ \Call{Subroutine2}{$L, \bm{a}, b$}
    \EndIf
    \State $m\gets m+1$
    \Until{$y_1 - y_0 \leq t$}
    \State \textbf{return} $(\bm{p}, \bm{q})$
    \EndProcedure
    \State
    \Procedure{Subroutine3}{$V_1, V_2, H, \{(\bm{p},\bm{q})\}$}
    \For {$(\bm{p},\bm{q})$ in $\{(\bm{p},\bm{q})_1,...,(\bm{p},\bm{q})_{35}\}$}
    \State Append $h(\bm{p},\bm{q})$ to $Q^{\prime}$.
    \State Append $\mathrm{tr}[H V_2  R_d(\bm{p}) R_d(\bm{q}) V_1 \rho_0 V_1^{\dagger} R_d(\bm{p})^{\dagger} R_d(\bm{q})^{\dagger} V_2^{\dagger}]$ to $\bm{m}$
    \EndFor
    \State Append $\bm{\tilde{q}}^{\prime}_{36}$ to $Q^{\prime}$.
    \State Append $0$ to $\bm{m}$.
    \EndProcedure
\end{algorithmic}
\end{algorithm}

For each configuration, we measured the expectation $m$, and constructed $\tilde{Q}'$.
Appending $\tilde {\bm{q}}'_{36}$ and 0 to $Q$ and $\bm{m}$ as the 36th row respectively allows for estimation via:
\begin{equation}\label{eqn:inversion_scf-cFQS}
\tilde{Q}' \bm{e}'=\bm{m}    
\end{equation} 
Once $\bm{e}'$ is obtained, we can estimate the expectation value $m$ for any arbitrary pair of $\bm{p}$ and $\bm{q}$.
Note that when $\bm{q}$ is fixed,  Eq.~\eqref{eqn:cost_function} is reduced to the same form as Eq.~\eqref{eqn:min_problem}, which allows for estimation the optimal $\bm{p}^\ast$ using Algorithm \ref{algo:CFQS}.
Likewise, fixing $\bm{p}$ enables optimization of $\bm{q}^\ast$ without additional measurement.
By iterating this process until convergence of the cost, both $\bm{p}$ and $\bm{q}$ are updated to simultaneously satisfy Eq.~\eqref{eqn:stationary}.
Note that this algorithm does not necessarily find the global optimum with respect to $\bm{p}$ and $\bm{q}$, and may instead converge to a local minimum.
Nevertheless, as it accounts for the parameter correlation, the method is expected to exhibit efficient optimization performance. 
We refer this algorithm to self-consistent field optimization for controlled gates via Free Quaternion Selection (SCF-cFQS), summarized in Algorithm \ref{algo:SCF-CFQS}.

%% file: results.tex
\begin{figure*}[htb]
 \centering
 \begin{tabular}{ccc}
 \includegraphics[width=0.45\textwidth]
{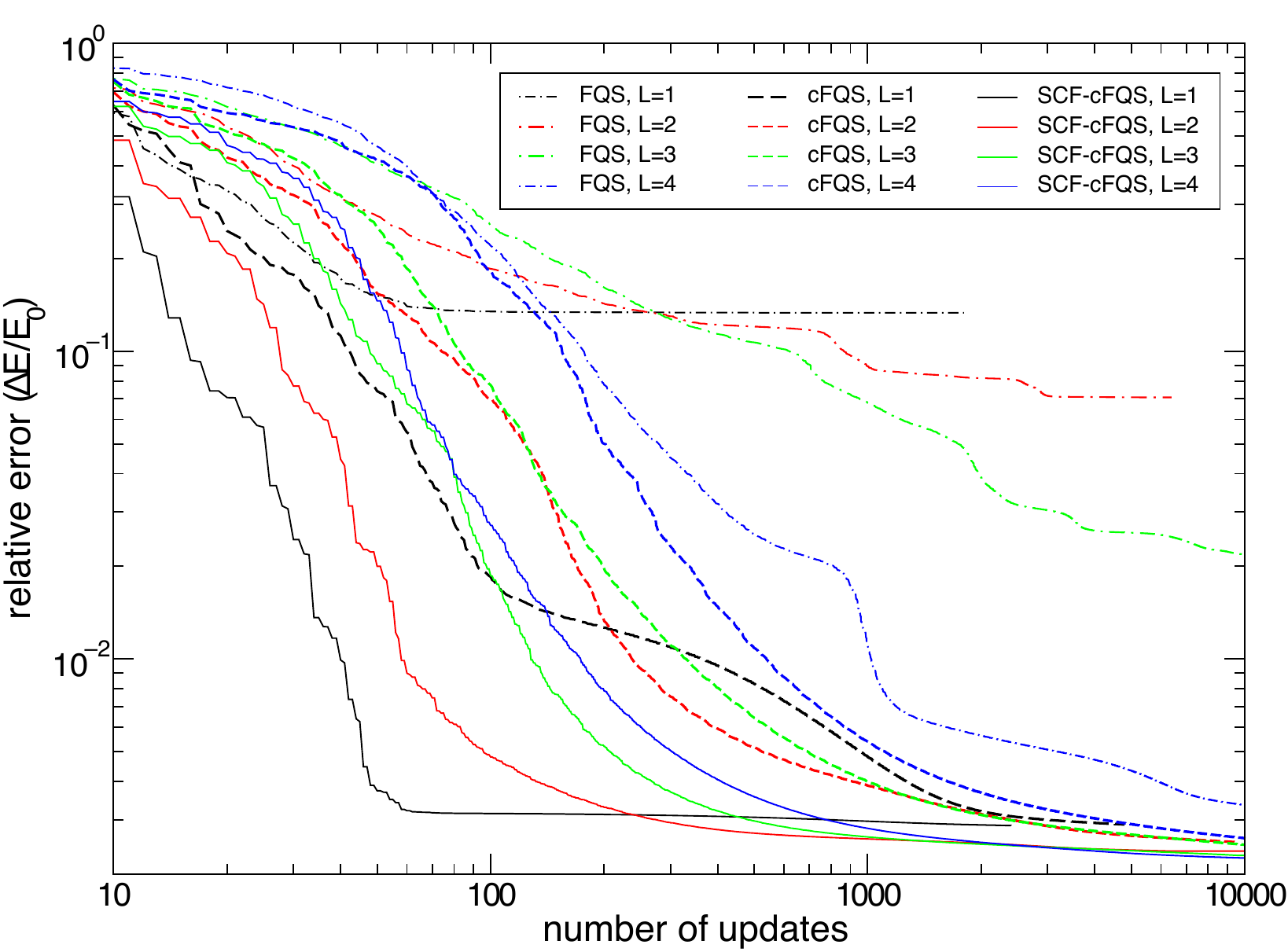} 
 &&
 \includegraphics[width=0.45\textwidth]{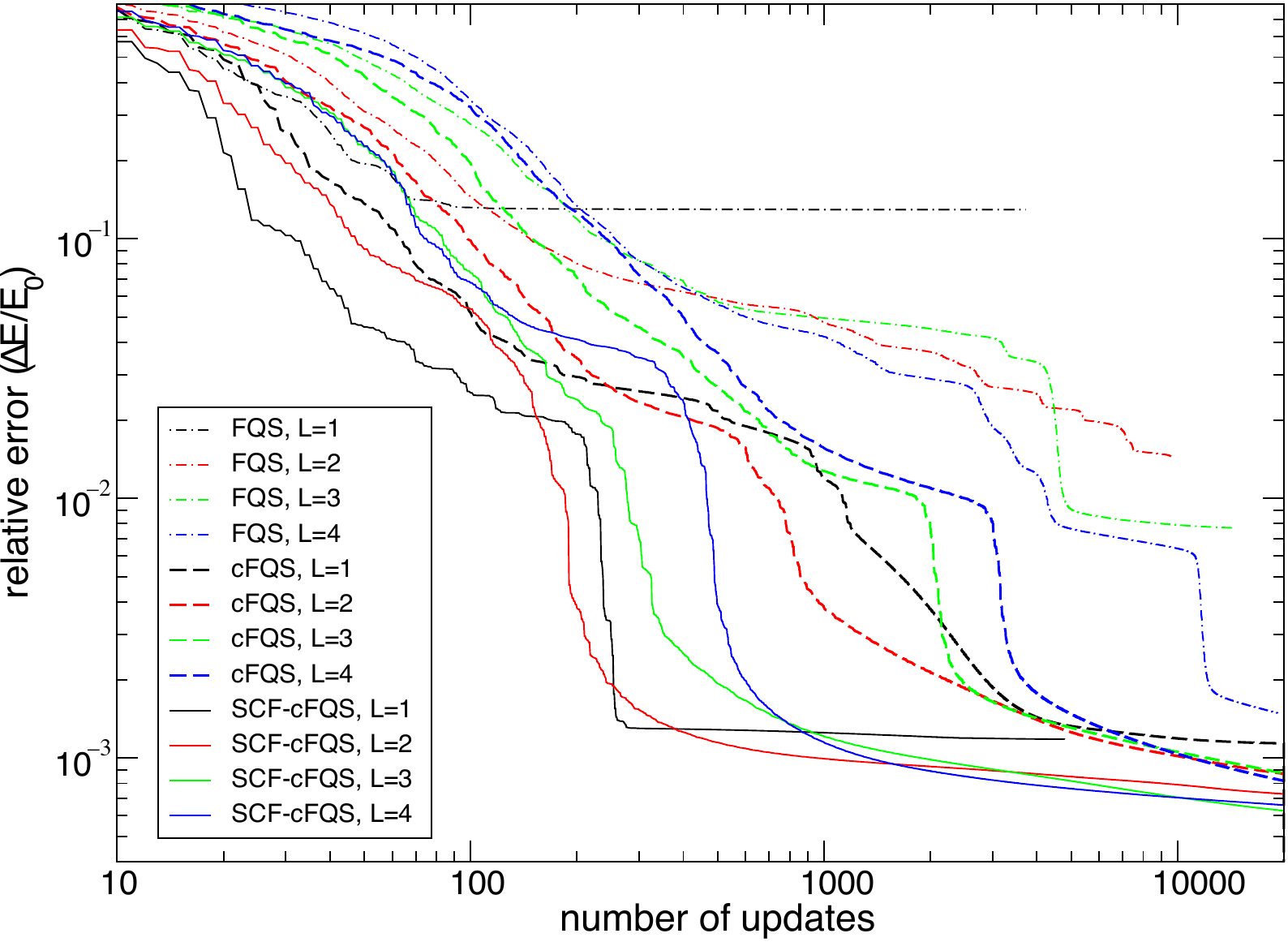} 
 \\
(a) 8 qubits   &&
(b) 12 qubits \\
 &&
 \end{tabular}
 \caption{{\bf VQE optimization trajectory for the mixed-field Ising model.} Dash-dotted, dashed, and sold lines represent the averaged trajectories of FQS, cFQS, and SCF-cFQS simulations respectively. 
 Each trajectories is averaged over independent more than ten optimizations.
 Black, red, green, and blue line colors stand for the number of layers in alternated layered ansatz employed, 
 In the FQS simulations, controlled-gates are fixed to controlled-Z gates, while cFQS and SCF-cFQS simulations, controlled gates are randomly initialized.}
 \label{fig:Ising}
 \end{figure*}

\begin{figure*}[htb]
 \centering
 \begin{tabular}{ccc} 
 \includegraphics[width=0.45\textwidth]{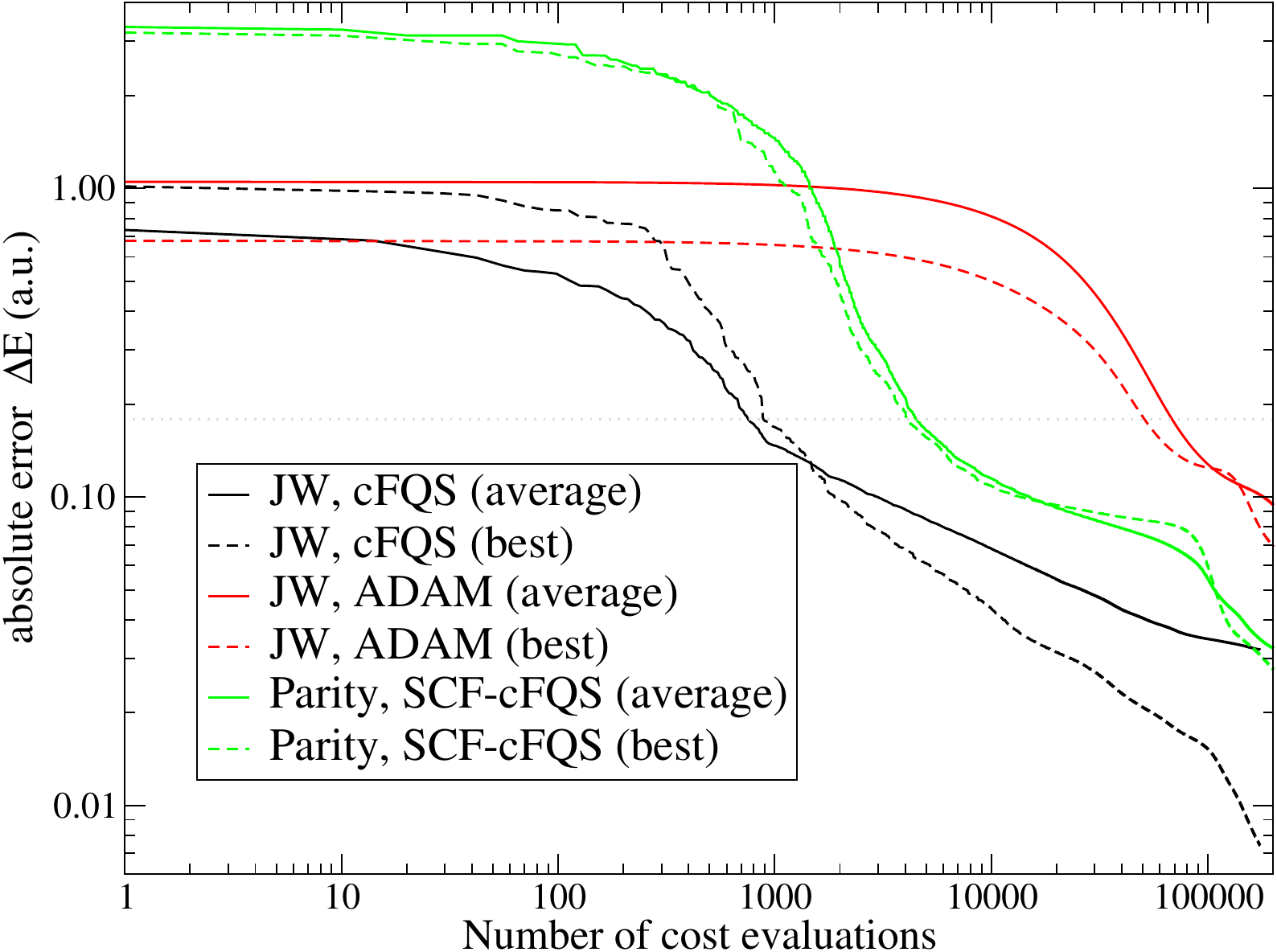}  
 &&
 \includegraphics[width=0.45\textwidth]{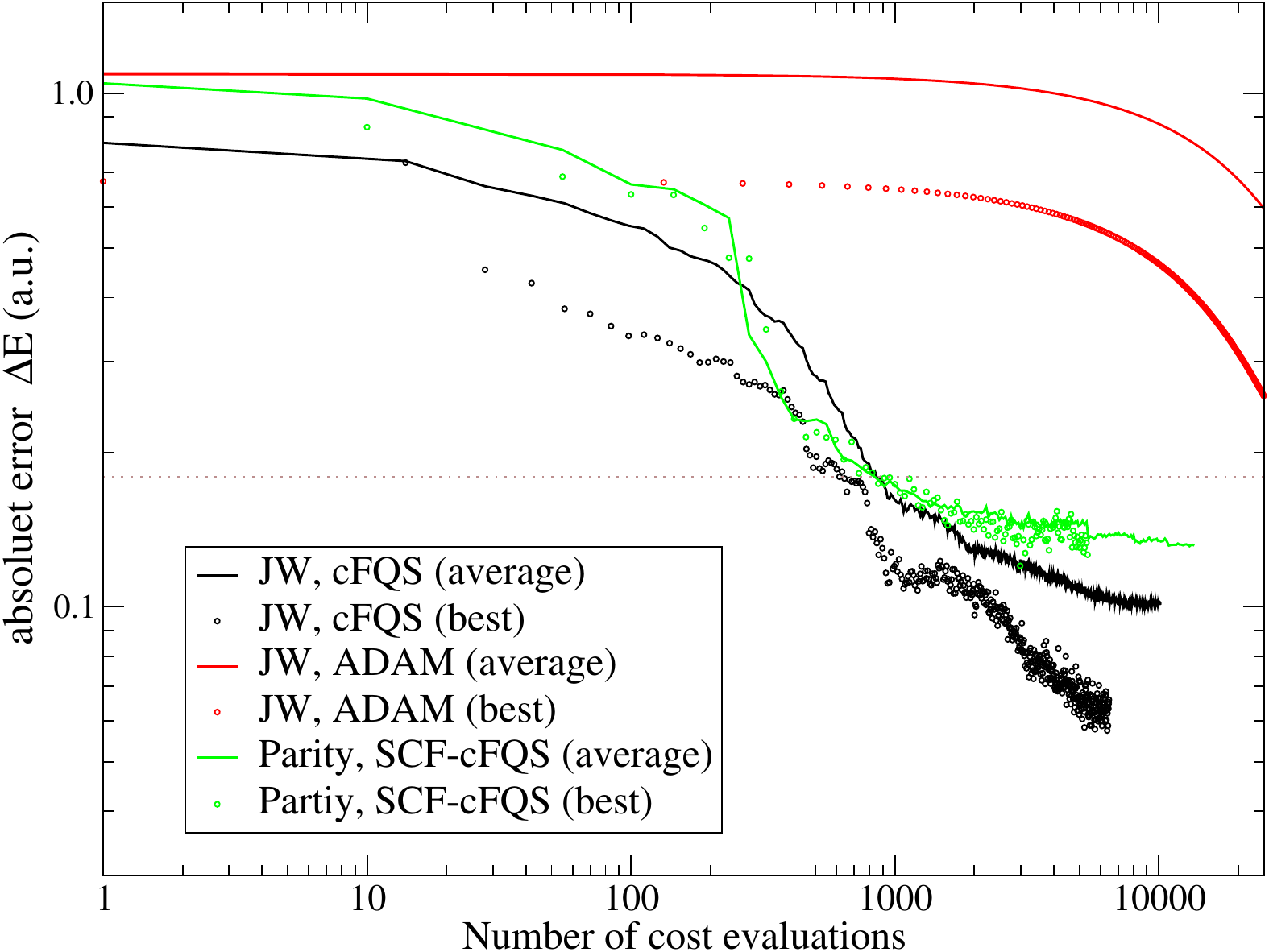}  \\
(a) Statevector simulators.
&&
(b) QASM simulators.\\
 \end{tabular}
 \caption{{\bf VQE trajectories of \ce{O2} with STO-3G basis set.}
(a), (b) Optimization trajectories under different methods and Hamiltonian mappings.
Black and red lines correspond to particle-number preserving gates with the Jordan-Wigner mapping: black for cFQS optimization, red for ADAM.
Green lines show results from SCF-cFQS combined with FQS for single-qubit gates, using the Parity mapping with two-qubit reduction.
Solid lines represent averages over 20 VQE runs.
In (a), dashed lines indicate the best-case trajectories among the 20 runs.
In (b), best-case trajectories are shown as discrete points.
The gray dotted line marks the Hartree–Fock energy.}
 \label{fig:VQE_molecules}
\end{figure*}

Throughout this paper, we term the PQC optimizations only targeting single qubit gates "FQS simulations". 
In contrast, in "cFQS" and "SCF-cFQS" simulations, controlled gates were also optimized by controlled-FQS and SCF-cFQS respectively, while single qubits gates were optimized by FQS.
We employed an alternating layered ansatz with 2-qubit blocks, where each block consists of two general single-qubit gates and one controlled gate as shown in Fig.~\ref{fig:ansatz}(a).
We applied FQS and either controlled-FQS or SCF-cFQS to the single-qubit gates and controlled-gate, respectively.
In FQS simulations, all controlled gates were initialized as controlled-Z gates, while cFQS and SCF-cFQS simulations, all single-qubit gates and controlled gates were randomly initialized.   
We remark that the update order is significant impact on the optimization performance, and employed zipping-like order [See numbers in Fig.~\ref{fig:ansatz}(a)], which performed the best of all compared as we reported in the previous studies in \cite{wada2024sequential}.
In the next section, we show VQE for Ising model and molecular Hamiltonian and VQA   
about fidelity maximization as a benchmark.
We also apply cFQS to quantum assisted quantum compilation (QAQC) of time-evolution operator of molecular Hamiltonian and discuss the performance.
In this paper, an updating cycle of all parameterized gates is referred to as a sweep.

\subsection{Variational Quantum Algorithm}
\subsubsection{VQE for a mixed field Ising model}
As the second benchmark, we carried out VQE for mixed-field one-dimensional Ising model under periodic boundary conditions whose Hamiltonian represented by
\begin{align}
H= \mathcal{J}\sum_{i=0}^{n-1} Z_i Z_{i+1} + h \sum_{i=0}^n (X_i+Z_i),
\end{align}
where $\mathcal{J}=1$ and $h=1/\sqrt{2}$.
Compared to the FQS optimization in Fig.~\ref{fig:Ising}, the resulting energy levels of cFQS and SCF-cFQS optimizations are clearly lower.
The number of controlled operations required for controlled FQS is twice that of FQS in the standard gate decomposition as in Fig.~\ref{fig:decomposition}, but the resolution of a two-qubit unitary, i.e., a gate block in Fig.~\ref{fig:gate_block}, is improved.
Note that the FQS optimization does not reach the cFQS optimization level even when compared even with doubling the number of layers so that the CNOT depth is the same.
As it has been proven that barren plateaus are induced depending on the number of layers based on alternating layered ansatz \cite{cerezo2021cost}, doubling the number of layers does not allow the FQS optimization to reach the cFQS optimization level, which implies that it is presumably an effective approach to mitigate a barren plateau at least for this system.
Note that the two-qubit unitary matrices by cFQS and SCF-cFQS both have the same number of parameters.
However, the SCF-cFQS converges faster to a better solution.
This suggests that the additional incorporation of parameter correlations by SCF-cFQS allows for efficient optimization.
The SCF-cFQS requires 35 measurements per gate update, almost twice as many as the cFQS, which requires 15 measurements, but the number of gate updates to reach the same energy for the cFQS is several to several dozen times greater than for the SCF-cFQS, which implies practical improvement of efficiency.

\subsubsection{VQE for molecular Hamiltonian}
In quantum chemistry problems, parity mapping offers a clear advantage in mitigating the barren plateau problem by enabling a two-qubit reduction through the use of symmetries.
In such cases, the SCF-cFQS method becomes particularly effective.
In contrast, when using the Jordan–Wigner mapping, it is straightforward to implement particle-number preserving gates, which makes this mapping appealing for chemistry-specific ansätze.
This raises the question: which approach is more advantageous in practice?

To address this, we performed VQE simulations for both the \ce{H2} and \ce{O2} molecules, with internuclear distances of 0.7500~\AA{} and 1.2308~\AA{}, respectively.
Molecular orbitals were obtained via self-consistent field (SCF) calculations by solving the Roothaan equations, using the 3-21G basis set for \ce{H2} and the STO-3G basis set for \ce{O2}.
For \ce{H2}, electron excitations were allowed up to LUMO+2, whereas the active space for \ce{O2} comprised six orbitals:
$\sigma_{\mathrm 2p}, \sigma^*_{\mathrm 2p}, \pi_{\mathrm 2p_x}, \pi_{\mathrm 2p_y}, \pi^*_{\mathrm 2p_x}$, and $ \pi^*_{\mathrm 2p_y} $.

The resulting Fermionic Hamiltonians were mapped to sums of Pauli tensor products using either the Jordan–Wigner or parity mapping.
In the case of parity mapping, two-qubit reduction was applied; for the Jordan–Wigner mapping, particle-number preserving ansätze were used, with the corresponding gates arranged in a causal cone structure (Appendix~\ref{apdx:particle-number-preserving-ansatz}).
Since the computational cost per optimization step differs among optimization methods, we compared performance based on the number of Hamiltonian measurements (i.e., cost evaluations), allowing for a fair assessment.

Although the barren plateau problem significantly limits the performance of variational quantum algorithms (VQAs), it primarily emerges under random initialization of parameterized quantum circuits (PQCs).
In quantum chemistry, however, the Hartree–Fock (HF) state represented by a computational basis state, often provides a good initial guess.
Nevertheless, the HF state frequently corresponds to a saddle point in the energy landscape, and using it as an initial state can cause the optimization to stagnate.
To circumvent this, a slight perturbation from the HF state is typically introduced, but the degree of perturbation is often determined heuristically.
If the perturbation is too small, gradients remain shallow and optimization progresses slowly; if it is too large, the barren plateau may reappear, nullifying the benefit of starting from the HF state.

To explore more effective initialization strategies, we introduced a metric to quantify the similarity between a given state and the HF state, and conducted VQE simulations using both HF-close and randomized initializations.
As shown in Appendix~\ref{apdx:initial_state_preparation}, we confirmed that VQE initialized from randomized states achieved lower cost values than those initialized near the HF state within a practical number of optimization steps.
Furthermore, we found that optimization performance improves when generalized single-qubit and controlled gates are initialized in complex space, rather than in real space—despite the fact that the ground states of standard molecular Hamiltonians are typically real.
This appears to be consistent with the previous reports based on FQS\cite{sato2023variational, wada2022simulating}, and we speculate that the use of complex space as a shortcut has succeeded in avoiding getting stuck in local optima.
Consequently, unless otherwise specified, we initialize particle-number preserving gates using quaternions in complex space throughout the following VQE calculations.

Figure~5 shows that, when using the local optimizers, the particle-number preserving ansatz leads to more efficient optimization than the parity-mapped ansatz, even at the cost of foregoing the two-qubit reduction.
Notably, the SCF-cFQS method still outperforms the ADAM optimizer even when both are applied to particle-number preserving ansätze.
Importantly, these trends in optimization performance were found to persist even in the presence of shot noise.

\subsubsection{Fidelity maximization} 
\begin{figure}[tb]
 \centering
 \begin{tabular}{c}
 \includegraphics[width=0.45\textwidth]{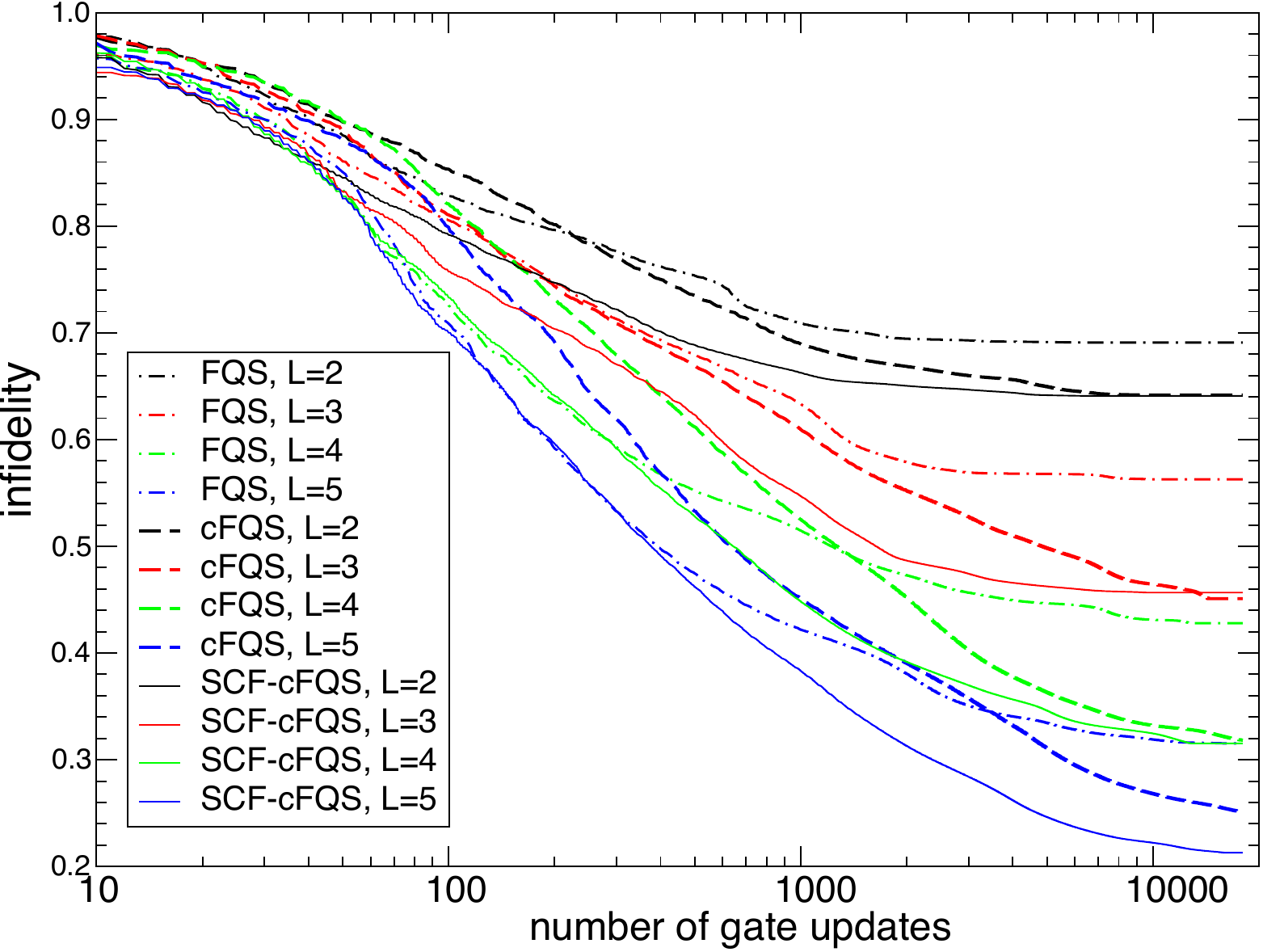} 
 \end{tabular}
 \caption{{\bf VQA for fidelity maximization}. Refer to Figs.~\ref{fig:Ising} for color and line types.}
 \label{fig:fidelity_8q}
 \end{figure}

Consistent with FQS, controlled-FQS is also applicable to optimization problems where the objective function is written as
\begin{align}\label{eqn:generalized_cost}
\sum_{k=1}^K \mathrm{tr}\left[\rho_k U^\dagger H_k U \right],
\end{align}
where $U$ is a target unitary matrix, and $\rho_k$, $H_k$ are the $k$-th input density matrix and observable, respectively. 
By setting $H_k$ as the reference quantum state, it is applicable to VQA by maximizing (or minimizing) the fidelity.
In this case, $H_k=\ket{\Phi_0}\bra{\Phi_0}$, where $\ket{\Phi_0}$ is the reference state.
Figure~\ref{fig:fidelity_8q} shows the averaged VQA trajectories for fidelity maximization over more than five independent jobs, where we employed $C(\{\bm{q}\}) \equiv 1-|\braket{\Phi_0|\Phi(\bm{q})}|^2$ as a cost function and the reference state $\ket{\Phi_0}$ were randomly generated for the respective jobs.
In line with other Hamiltonians, the pace of cFQS optimization is comparable to that of FQS, although the trajectories reach lower cost levels. 
The increased expressive capacity within a block does not compromise trainability, which is an impressive result, given that the barren plateau is known to be affected by the number of layers. 
Here again, SCF-cFQS shows much faster convergence than cFQS. 
We emphasize that this can be attributed to the parameter correlations incorporated by SCF-cFQS. 
On the other hand, the converged cost levels appear consistent between cFQS and SCF-cFQS as shown in Fig.~\ref{fig:fidelity_8q}.
It seems reasonable, given that the gate blocks in Fig.~\ref{fig:ansatz}(a) have identical expressibility in both optimizations.
In the FQS simulations, a gate block has totally six parameters and the correlation between three parameters corresponding to respective single-qubit gates is taken into account.
In the cFQS and SCF-cFQS simulations, a gate block has nine parameters in total, where three parameter sets consisting of three parameters are separately optimized in cFQS while a parameter set with three parameters regarding a single-qubit gate and the other set with six parameters related to two controlled-gates are separately optimized in the SCF-cFQS simulations, 
In SCF-cFQS, the obtained six parameters relevant to controlled gates are not necessarily optimal when viewed as a whole, but they are optimal when viewed on an individual gate considering the correlation.

\begin{figure}[tb]
 \centering
 \includegraphics[width=0.45\textwidth]{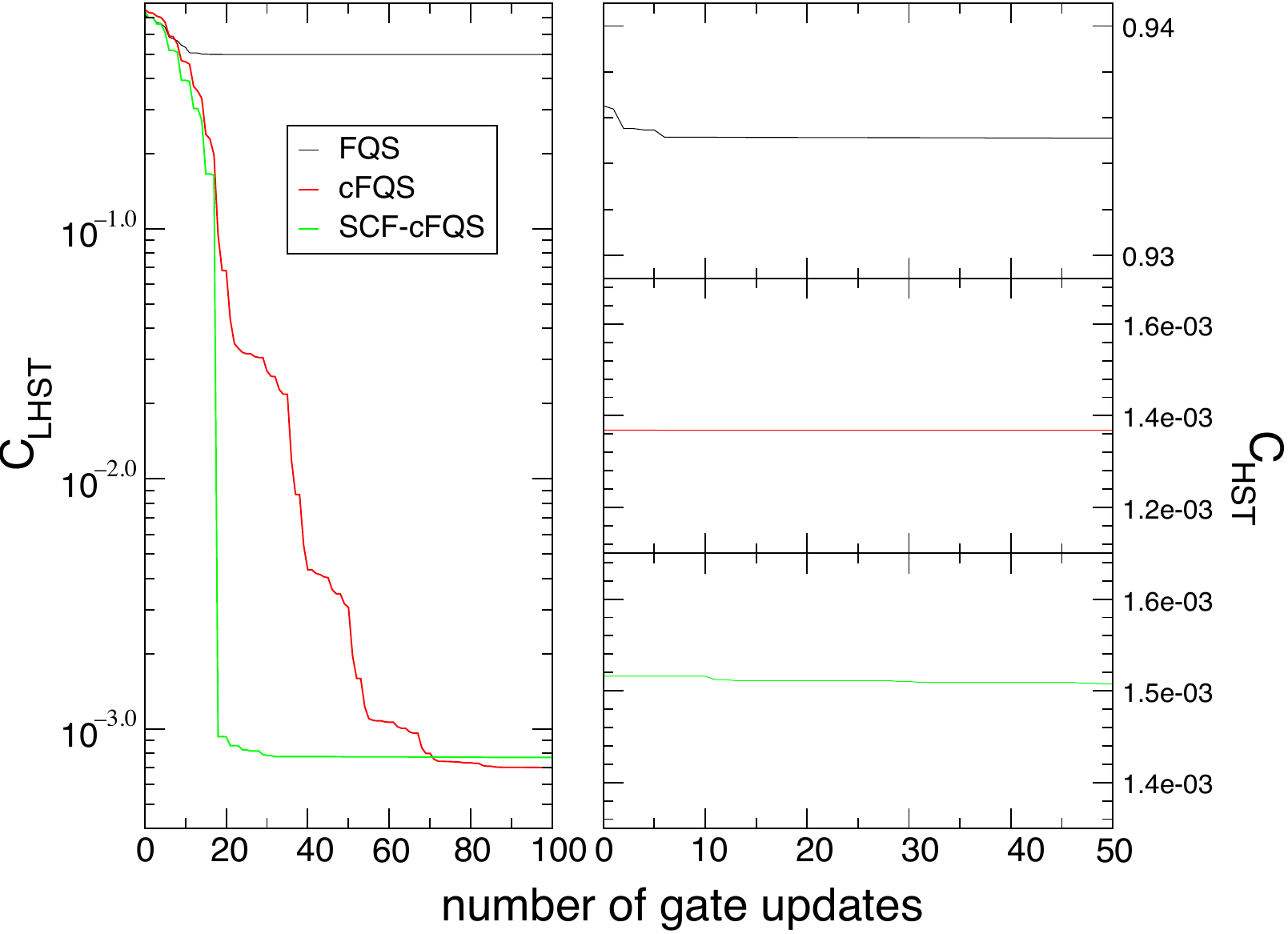} \\
 \caption{{\bf Optimization trajectories in unitary compilation of a time evolution operator for a \ce{H2}/31G based on parity mapping with two-qubit reduction.} 
 Only the best cases among five independent compilations for respective methods are explicitly presented.
 The number of circuit layer $L=1$ is employed.
 The qubit Hamiltonian is generated using parity mapping with two-qubit reduction.
 Left and right windows represent the optimizations with local and global costs, respectively.
}
 \label{fig:HST_H2}
\end{figure}
\subsection{Unitary compilation}
We approximately compile the time-evolution operator using shallow circuits.  
Here, we employ the fermionic Hamiltonian of the \ce{H2} molecule with the 6-31G basis set as the target unitary, defined as $U = \exp(-i H t / \hbar)$.  
The approximate unitary $V$ is represented by an alternating-layered ansatz with a fixed circuit depth.  
The cost function is defined based on the inner product between $U$ and $V$ as follows:
\begin{align}\label{eqn:global_HST_cost}
C_\mathrm{HST} = 1-\frac{1}{d^2} |\mathrm{tr}\left[V^{\dagger}U\right]|^2.
\end{align}
To obtain $|\mathrm{tr}\left[V^\dagger U\right]|^2$, we employed the Hilbert-Schmidt test~\cite{khatri2019quantum}, using the following relation 
\begin{align}\label{eqn:HST}
    \frac{1}{d^2}|\mathrm{tr}\left[V^\dagger U\right]|^2
    =|\bra{\Phi^+}_{\mathrm{AB}}U\otimes V^*\ket{\Phi^+}_{\mathrm{AB}}|^2,
\end{align}
where $U$ and $V^*$ act separately on systems A and B, respectively, and $\ket{\Phi^+}$ is a maximally entangled state between systems A and B.  
This entangled state can be easily prepared by applying Hadamard gates to each qubit in system A, followed by $n$ CNOT gates connecting corresponding qubit pairs between systems A and B.  
Note that Eq.~\eqref{eqn:HST} can be reformulated into the representation of Eq.~\eqref{eqn:generalized_cost}, allowing both FQS and controlled FQS to be applied to unitary compilation.  
Since the function in Eq.~\eqref{eqn:global_HST_cost} is a global cost, it can induce a barren plateau, which manifests as a concentration of spectral radius of the FQS matrix \cite{wada2022simulating}.
To alleviate this issue, we first employed the local cost function $C_\mathrm{LHST}$ during the early stages of optimization, and then switched to the global cost $C_\mathrm{HST}$ in Eq.~\eqref{eqn:global_HST_cost} once optimization had sufficiently progressed.  
As originally proposed by Khatri et al., $C_\mathrm{LHST}$ is defined as:
\begin{align}\label{eqn:local_HST_cost}
C_\mathrm{LHST}(U,V) \equiv \frac{1}{n}\sum_{j=1}^n (1-F_\mathrm{e}^{(j)}),
\end{align}
where $F_\mathrm{e}^{(j)}$ is the entanglement fidelity, which is defined as
\begin{align}
 F_\mathrm{e}^{(j)} \equiv 
 \mathrm{tr}\left[
 \ket{\Phi^+}\bra{\Phi^+}_{\mathrm{A}_j \mathrm{B}_j}
 (\mathcal{E}_j \otimes \mathcal{I}_{\mathrm{B}_j})
 \ket{\Phi^+}\bra{\Phi^+}_{\mathrm{A}_j \mathrm{B}_j}
 \right],
\end{align}
where $\mathrm{A}_j$ ($\mathrm{B}_j$) is the $j$-th qubit in system $\mathrm{A}$ ($\mathrm{B}$), $\mathcal{E}_j$ is a local channel, and $\mathcal{I}_{\mathrm{B}_j}$ is a identity channel. 

In practice, the time-evolution operator is implemented in a quantum circuit via Suzuki-Trotter decomposition. For example, the first-order Trotter decomposition can be written as 
\begin{align}
\mathrm{exp}(-i\frac{H_1+H_2}{\hbar}t)\simeq \left[\mathrm{exp}(-i\frac{H_1}{\hbar}\frac{t}{n})~\mathrm{exp}(-i\frac{H_2}{\hbar}\frac{t}{n})\right]^n, 
\end{align}
where $n$ is the number of Trotter steps. 
To suppress trotter error, the value of $n$ must be increased as $t$ increases, but the circuit length increases accordingly.
Although unitary compilation aims to compress circuit depth, the compiled time-evolution operator must still be embedded in the quantum circuit at least once.  
Thus, special attention must be paid to the time parameter $t$ to ensure that the resulting circuit remains executable on NISQ devices.  

However, if sequential compilation is allowed, it becomes possible to separately compile decomposed components and later recompile the approximated operators together.  
For instance, $\mathrm{exp}(-i\frac{H_1}{\hbar}\frac{t}{n})$ and $\mathrm{exp}(-i\frac{H_2}{\hbar}\frac{t}{n})$ can be individually approximated by $W_1$ and $W_2$, which are then $W_1$ and $W_2$ compiled together.
Therefore, in this work, we focus on $t$ in terms of compression efficiency and trainability, rather than feasibility of the reference unitary.  

To this end, we consider two time steps, $\Delta t = 1/16$ and $1.0$, and perform compilation for each.  
A small $t$ is advantageous for mitigating barren plateaus, as the target unitary is close to the identity, making warm-start strategies more effective.  
For $t=1/16$, single-qubit gates are initialized with rotation angles randomly sampled from $[0, \pi/18]$ and rotation axes randomly distributed over the Bloch sphere.  
Each controlled gate is initialized as a controlled-Z gate.  

On the other hand, a small $t$ is disadvantageous when considering downstream applications, since simulating longer time intervals would require many repetitions of the complied operators, leading to circuits too deep for NISQ hardware.  
In contrast, for $t = 1.0$, warm-starting is no loger valid, so we performed 10 independent compilations with randomly initialized parameters and selected the one yielding the lowest cost.  
While such parallel compilation is feasible on current quantum platforms, it alone is insufficient to guarantee successful compilation when barren plateaus are present.  
To improve variational optimization efficiency, we leveraged the properties of cFQS, combined with additional techniques detailed below.  

Figure~\ref{fig:HST_H2} shows the best compilation trajectory.  
The cost function was switched from $C_\mathrm{LHST}$ to $C_\mathrm{HST}$ after five parameter update sweeps.  
First, all compilations nearly converged under $C_\mathrm{LHST}$, and switching to $C_\mathrm{HST}$ led to no further improvement, indicating that parametrization of the controlled gates is crucial.  
Second, the of FQS alone showed minimal cost reduction.  
Third, while SCF-cFQS exhibited fast convergence, its final cost was comparable to that of cFQS.  
Given that SCF-cFQS requires twice as many CNOT gates as cFQS, the latter is more advantageous for circuit compression.  
Therefore, we focus on cFQS in the remainder of this study.

Next, we present the compilation results for the time-evolution operator with $\Delta t = 1.0$, where we exploit the symmetry of the operator to improve optimization efficiency.  
Specifically, by restricting the variational space to a symmetry-preserving subspace, the optimization process can be made more efficient.  
We consider two types of symmetry-based subspace constraints: restriction on the input state and restriction through the ansatz structure.  
To evaluate their respective performance, we first focus on input state restriction.  

In molecular systems, the number of spin-up and spin-down electrons, as well as the total number of electrons, is conserved during time evolution.  
In other words, the time-evolution operator $U$ acts within a subspace where both the total particle number and the total spin-$z$ component are preserved.  
This implies that simulating the evolution of a specific state does not require reproducing the entire operator $U$, but only its action within the relevant subspace.  

For simplicity, we adopt the Jordan–Wigner basis, in which each qubit represents the occupation number of a corresponding spin orbital.  
Let $\mathcal{W}$ denote the vector space spanned by a basis set $W$, and $|W| = \dim(\mathcal{W})$.

Instead of the maximally entangled state $\ket{\Phi^{+}}_\mathrm{AB}$, we employed the quantum state defined as
\begin{align}
 \ket{\Tilde{\Phi}}_\mathrm{AB} \equiv \frac{1}{\sqrt{\left| W \right|}}\sum_{\bm{w}\in W} \ket{\bm{w}}_\mathrm{A} \otimes \ket{\bm{w}}_\mathrm{B}.
 \end{align}
Here, on a quantum circuit, $\ket{\Tilde{\Phi}}_\mathrm{AB}$ is reproduced by preparing $\ket{\Tilde{\Phi}'}$ in system A, and then applying $n$ CNOT operations to connecting qubit pairs between system A and B, where $\ket{\Tilde{\Phi}'}$ is defined as
\begin{align}
 \ket{\Tilde{\Phi}'} \equiv \frac{1}{\sqrt{\left| W \right|}}\sum_{\bm{w}\in W} \ket{\bm{w}}.
\end{align}
Correspondingly, we defined the global and local cost functions as
 \begin{align}
 C_{\rm{global}}(\bm{\theta}) &\equiv 1- \frac{1}{\left| W \right|^2} | \sum_{\bm{w}\in W} \bra{\bm{w}} V^{\dagger}(\bm{\theta})U \ket{\bm{w}} |^2 \\
 C_{\rm{local}}(\bm{\theta}) &\equiv \frac{1}{n} \sum_{j=1}^n (1-\Tilde{F}_\mathrm{e}^{(j)}),
 \end{align} 
where $\Tilde{F}_\mathrm{e}^{(j)}$ denotes the entanglement fidelity when the initial state is $\ket{\Tilde{\Phi}}$ instead of the maximally entangled state $\ket{\Phi^+}$ used in $F_\mathrm{e}^{(j)}$.
We note that these cost functions reduce to the original QAQC cost functions given in Eqs.~\eqref{eqn:global_HST_cost} and \eqref{eqn:local_HST_cost} when $W$ is chosen to include the full set of computational basis states.

\begin{table*}[htb] 
     \centering
     \caption{ {\bf Input state restriction and basis set used in compilation for time evolution operator of a \ce{H2} molecule.} }
     \begin{tabular}{c|c|c} \hline
      Input state restriction & Number of states $|W|$ & Basis set $W$ \\ \hline\hline
      Compact &2    &  $\{ \ket{0101}, \ket{1010} \}$ \\ 
      $S_\mathrm{z}$ & 4    &  $\{ \ket{\bm{j}} \otimes \ket{\bm{k}} | \bm{j},\bm{k}\in\{ 01, 10\} \}$ \\ 
      Number of particle &6    &  $\{ \ket{j_1 j_2 j_3 j_4} | j_m\in\{ 0, 1\}, \sum_{m=1}^4 j_m = 2 \}$ \\ 
      No restriction & 16   &  $\{ \ket{j_1 j_2 j_3 j_4} | j_m\in\{ 0, 1\} \} $ \\ \hline
     \end{tabular}
     \label{tab:set-for-subspace-h2}
 \end{table*}

The choice of $\ket{\tilde{\Phi}}$ is arbitrary; here, we employ the Dicke state $\ket{D^n_k}$ in which all computational bases with the same Hamming weight are equally weighted, as shown below.
 \begin{align}
 \ket{D^n_k} \equiv \frac{1}{\sqrt{\binom{n}{k}}} \sum_{\mathrm{HW}(\bm{j})=k} \ket{\bm{j}},
 \end{align}
Where $\mathrm{HW}(\bm{j})$ denotes the Hamming weight of the bit string $\bm{j}$.  
Note that the particle number is consistent in the subspace spanned by the computational bases in a Dicke state. 
We adopted the Dicke state preparation protocol proposed in Ref.~\cite{bartschi2019deterministic},  
which requires no ancilla qubits and achieves circuit depth $\mathcal{O}(n)$ with $\mathcal{O}(kn)$ gates.  

Moreover, it is also possible to restrict the input states to subspaces with fixed numbers of spin-up and spin-down electrons.  
In such cases, the corresponding quantum state $\ket{\Tilde{\Phi}'}$ is given by:
 \begin{align}
 \ket{\Tilde{\Phi}'} = \ket{D^{n}_{N_{\alpha}}} \otimes \ket{D^{n}_{N_{\beta}}},
 \end{align}
where $n$ is the number of spatial orbitals, the subscript $\alpha$ ($\beta$) denotes spin-up (spin-down), and $N_\alpha$ ($N_\beta$) represents the number of $\alpha$-spin ($\beta$-spin) electrons.  
To evaluate the performance of input state restrictions, we approximated the time-evolution operator of \ce{H2}/STO-3G using several different choices of $W$, as summarized in Table~\ref{tab:set-for-subspace-h2}.  
For comparison, we used an ansatz without any conserved quantities, employing a nearest-neighbor connection with three layers.  

From quantum chemical insight, it is known that the ground state of \ce{H2} lies in the subspace spanned by the states $\{\ket{0101}, \ket{1010}\}$, which we refer to as the "compact" subspace, as shown in Table~\ref{tab:cnot-num-and-depth}.  
While there is no unique protocol for restricting input states, in this study we focus on two types of restrictions that are easily implemented using Dicke states.  
Although \ce{H2} is a special case where the compact subspace is known in advance, comparing the results obtained from the compact subspace with those from other restrictions is useful for assessing the effectiveness of the method.  

\begin{figure}[hbp]
     \centering
     \begin{tabular}{c}
     \includegraphics[width=0.45\textwidth]{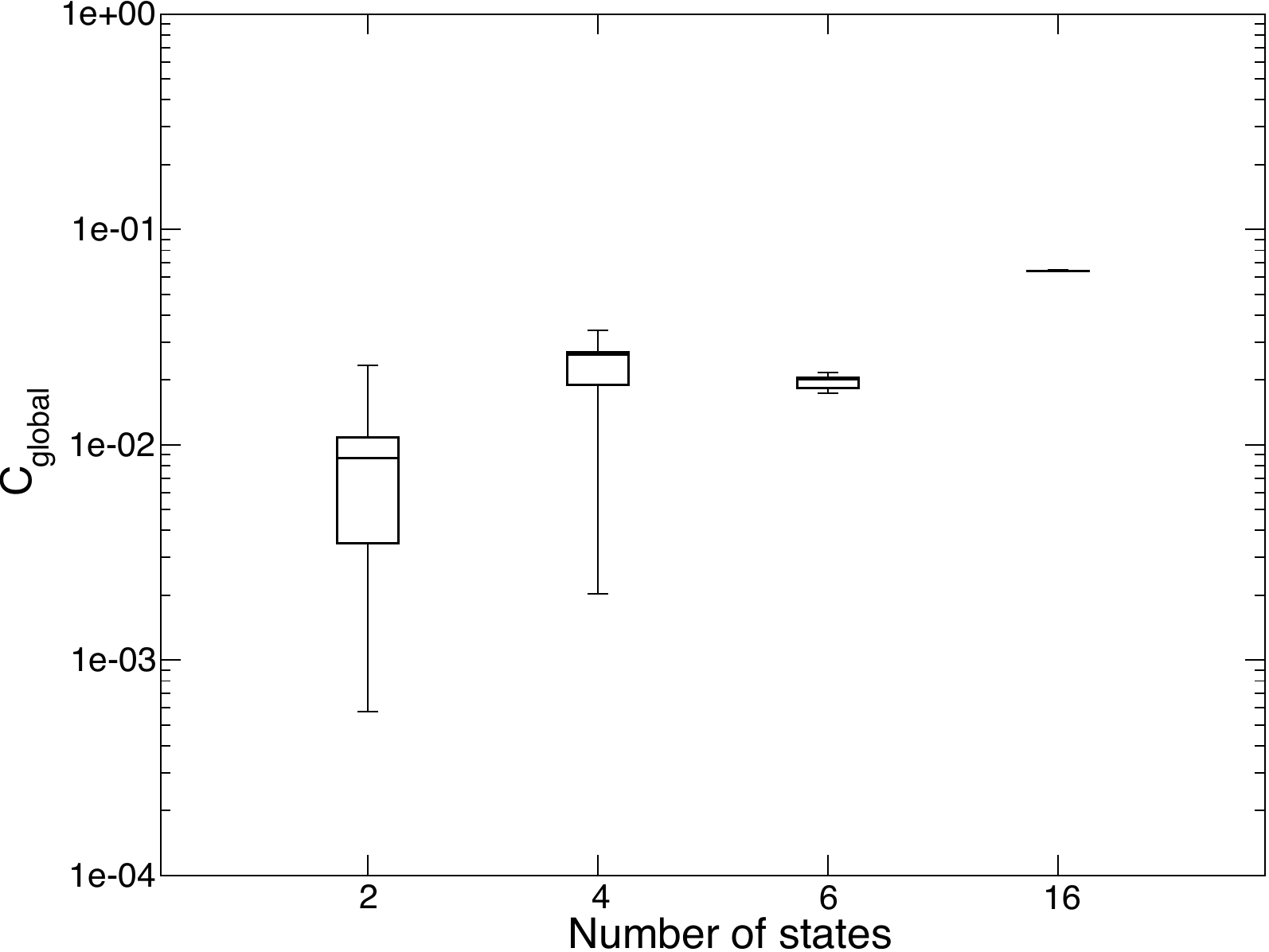} 
     \end{tabular}
     \caption{{\bf Resulting cost values of quantum compilation of time evolution operator of \ce{H2}/STO-3G with different subspace.}
     }
     \label{fig:subspace_h2}
\end{figure}

For the \ce{H2} system, restricting to the particle- and $S_z$-preserving subspace reduces the number of basis states required to four and six, respectively, as listed in Table~\ref{tab:set-for-subspace-h2}.  
The results of the compilation are summarized in Fig.~\ref{fig:subspace_h2}, where it is evident that a smaller subspace leads to a lower cost.  
The block size of the unitary operator to be compiled is given by $|W| \times |W|$, depending on the subspace size.  
As $|W|$ is effectively reduced, even shallow PQCs with limited parameter counts can reproduce the unitary action within the subspace.  
This observation suggests that optimization becomes easier, which is consistent with the results shown in Fig.~\ref{fig:subspace_h2}.
The trajectories of unitary compilation within 2- and 16-dimensional subspaces are shown in Appendix~\ref{apdx:traj_compliation_H2}.  
Most trajectories using the compact subspace exhibit faster convergence and lower final cost values compared to those using the full 16-dimensional QAQC setting.  
Note that although we initialize the PQC parameters randomly based on the warm-starting concept described above, parameter initialization in the vicinity of the identity matrix leads to better convergence than fully random initialization.  
Therefore, in the following, we employ the warm-starting strategy rather than random initialization for compiling unitary matrices with $\Delta t = 1$.
\begin{table*}[htb]
     \centering
     \caption{ {\bf Variety of the ansaetze restriction and the respective number of CNOT gates and the CNOT-depth of each ansaetz } }
     \scalebox{0.9}{
     \begin{threeparttable}         
     \begin{tabular}{c|cccccccc} \hline
      & Input state restriction & $U_\mathrm{NP}$\tnote{a}~ & Qubit connection & $L$\tnote{b}~ & Parameters & $C_{\rm{global}}$\tnote{c} & No. of CNOTs & CNOT-depth \\ \hline\hline
      \multirow{6}{*}{\ce{H2}} &\multirow{5}{*}{Spin restricted} 
      & No & Nearest neighbor    & 3 &120& 8.75E-05 & 24  & 12 \\
      &&Yes &Nearest neighbor    & 9 &108& 3.40E-03 & 144 & 72 \\ 
      &&Yes &All-to-all          & 3 &54 & 1.32E-07 & 72  & 36 \\ 
      &&Yes & $S_\mathrm{z}$-preserving    & 3 &18 & 8.08E-05 & 26  & 14 \\
      &&Yes & $S_\mathrm{z}$-preserving    & 4 &24 & 9.48E-08 & 35  & 19 \\
      \cline{2-9}
      &\multicolumn{6}{c}{The 1st-order Trotter decomposition}                & 36  & 30 \\ \hline
      \multirow{7}{*}{\ce{H3}}&\multirow{6}{*}{Spin restricted} 
      & No & Nearest neighbor & 9  &504& 1.32E-02 & 108 & 36 \\ 
      &&Yes& All-to-all       & 6  &270& 4.40E-02 & 360 & 120 \\ 
      &&Yes& All-to-all       & 10 &450& 3.22E-05 & 600 & 200 \\ 
      &&Yes& $S_\mathrm{z}$-preserving  & 5  &90 & 7.20E-03 & 124 & 64 \\
      &&Yes& $S_\mathrm{z}$-preserving  & 6  &108& 3.56E-04 & 149 & 77 \\
      &&Yes& $S_\mathrm{z}$-preserving  & 7  &126& 4.32E-05 & 174 & 90 \\   
      \cline{2-9}
      &\multicolumn{6}{c}{The 1st-order Trotter decomposition}            & 494 & - \\
\hline
      \multirow{7}{*}{\ce{LiH}}&\multirow{6}{*}{Spin restricted} 
      & No & Nearest neighbor & 9  &504& 1.33E-02 & 108 & 36 \\ 
      &&Yes& All-to-all       & 3  &135&1.23E-02 & 180 & 60 \\ 
      &&Yes& All-to-all       & 6  &270&6.85E-04 & 360 & 120 \\ 
      &&Yes& $S_\mathrm{z}$-preserving  & 5  &90 &1.08E-03 & 124 & 64 \\
      &&Yes& $S_\mathrm{z}$-preserving  & 6  &108&1.14E-03 & 149 & 77 \\
      &&Yes& $S_\mathrm{z}$-preserving  & 7  &126&4.98E-04 & 174 & 90 \\ 
    \cline{2-9}
 &\multicolumn{6}{c}{The 1st-order Trotter decomposition}  & 504 & - \\ \hline
     \end{tabular}
     \begin{tablenotes}
     \item[a] Usage of the number-preserving gate. "No" indicates the gate block in Fig.~\ref{fig:gate_block}(a) is employed instead.
     \item[b] The number of circuit layers
     \item[c] The best values of $C_\mathrm{global}$ among 10 independent optimizations.
     \end{tablenotes}
    \end{threeparttable}
     }
     \label{tab:cnot-num-and-depth}
\end{table*}

Next, we examine the effect of ansatz constraints in combination with input state restrictions—--specifically, the use of particle-number-preserving gates (see Fig.~\ref{fig:PPgate}) and their optimization via cFQS.  
Unlike in the previous sections, we vary the number of circuit layers $L$ while maintaining an effective compression rate of circuit depth, as applied to \ce{H2}, \ce{H3}, and \ce{LiH} molecules.  

To construct the molecular Hamiltonians, we employ the STO-3G basis set for all molecules.  
As a result, \ce{H2} corresponds to a 4-qubit system, while both \ce{H3} and \ce{LiH} correspond to 6-qubit systems.  
In the cases of \ce{H2} and \ce{H3}, all spin orbitals are included in the active space.  
For \ce{LiH}, we define the active space to include three spatial orbitals—i.e., six spin orbitals—associated with sigma bonding.  

To ensure a fair comparison of ansatz constraints, we consistently apply input state restrictions that preserve both the particle number and the spin-$z$ component, as represented by:
\begin{align}
    \ket{\Tilde{\Phi}'} &= \ket{D^2_1} \otimes \ket{D^2_1} \quad \rm{for\ \ce{H2}},\\
    \ket{\Tilde{\Phi}'} &= \ket{D^3_2} \otimes \ket{D^3_1} \quad \rm{for\ \ce{H3}},\\
    \ket{\Tilde{\Phi}'} &= \ket{D^3_1} \otimes \ket{D^3_1} \quad \rm{for\ \ce{LiH}}.
\end{align}
Three and four types of PQCs are used for the \ce{H2} molecule and 6-qubit systems, respectively.  
These circuits are constructed by applying gate blocks to qubit pairs as specified in Table~\ref{tab:qubit_connection} in the Appendix.  

Although the output states from ansaetze with entanglement patterns based on nearest-neighbor and all-to-all connections preserve the particle number, they do not necessarily preserve the spin-$z$ component.  
This is because the number-preserving gate blocks may connect qubits corresponding to orbitals with opposite spins (i.e., $\alpha$ and $\beta$), as illustrated in Fig.~\ref{fig:symmetric_PQC} in the Appendix~\ref{apdx:PQC_structure}.  

In contrast, both the particle number and the spin-$z$ component are strictly preserved in ansaetze where number-preserving gates only connect qubits associated with the same spin type.  
We refer to such a design as the $S_\mathrm{Z}$-preserving ansatz.  

In general, deeper circuits have higher expressibility but tend to suffer from reduced trainability.  
However, it is not straightforward to define a clear standard for comparing the expressibility and trainability among different ansätze.  

The gate block shown in Fig.~\ref{fig:gate_block}(a) contains two CNOT gates, whereas the particle-number preserving gate $U_\mathrm{NP}$ requires four CNOT gates.  
Moreover, the number of CNOT gates per circuit layer varies depending on the qubit connectivity.  

In this study, we compare the performance of the ansaetze based on the number of layers required to reach the lowest cost after 100 sweeps, as summarized in Table~\ref{tab:cnot-num-and-depth}.

First, we compare the nearest-neighbor and all-to-all qubit connectivity using $U_\mathrm{NP}$ for the \ce{H2} molecule.  
It is evident that the all-to-all connection results in a significantly lower cost compared to the nearest-neighbor configuration.  
Although the all-to-all connectivity requires more CNOT gates per layer, it achieves lower cost values with shallower circuits, and is thus more advantageous in terms of the number of controlled operations.  
Note, however, that both types of connectivity require more CNOT gates than the Trotter decomposition, suggesting that they are less efficient from the perspective of circuit compilation.  
Based on these results, we omitted experiments with the nearest-neighbor connection for the other molecules.

Second, Table~\ref{tab:cnot-num-and-depth} shows that the $S_\mathrm{z}$-preserving ansatz consistently achieved the lowest cost values across all molecules.  
These cost levels are comparable to those obtained from the number-preserving ansatz with all-to-all connectivity.  
However, the number of CNOT gates in the spin-preserving ansaetze is significantly smaller than that of the all-to-all counterpart, except in the case of the \ce{H2} molecule, indicating a more practical level of circuit compression.  

\begin{figure*}[htb]
     \centering
     \begin{tabular}{cc}
         \includegraphics[width=0.45\textwidth]{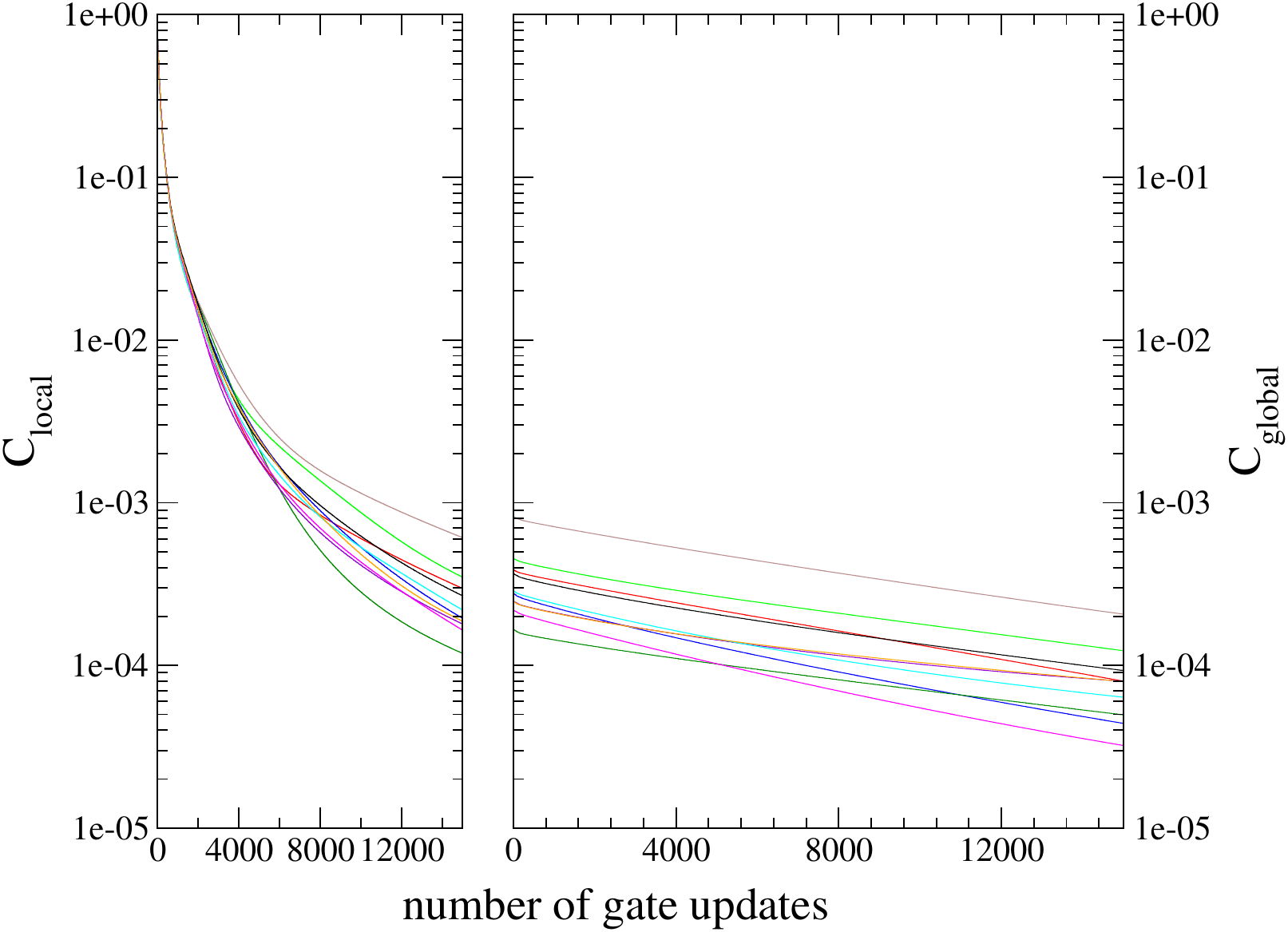}  
         &
         \includegraphics[width=0.45\textwidth]{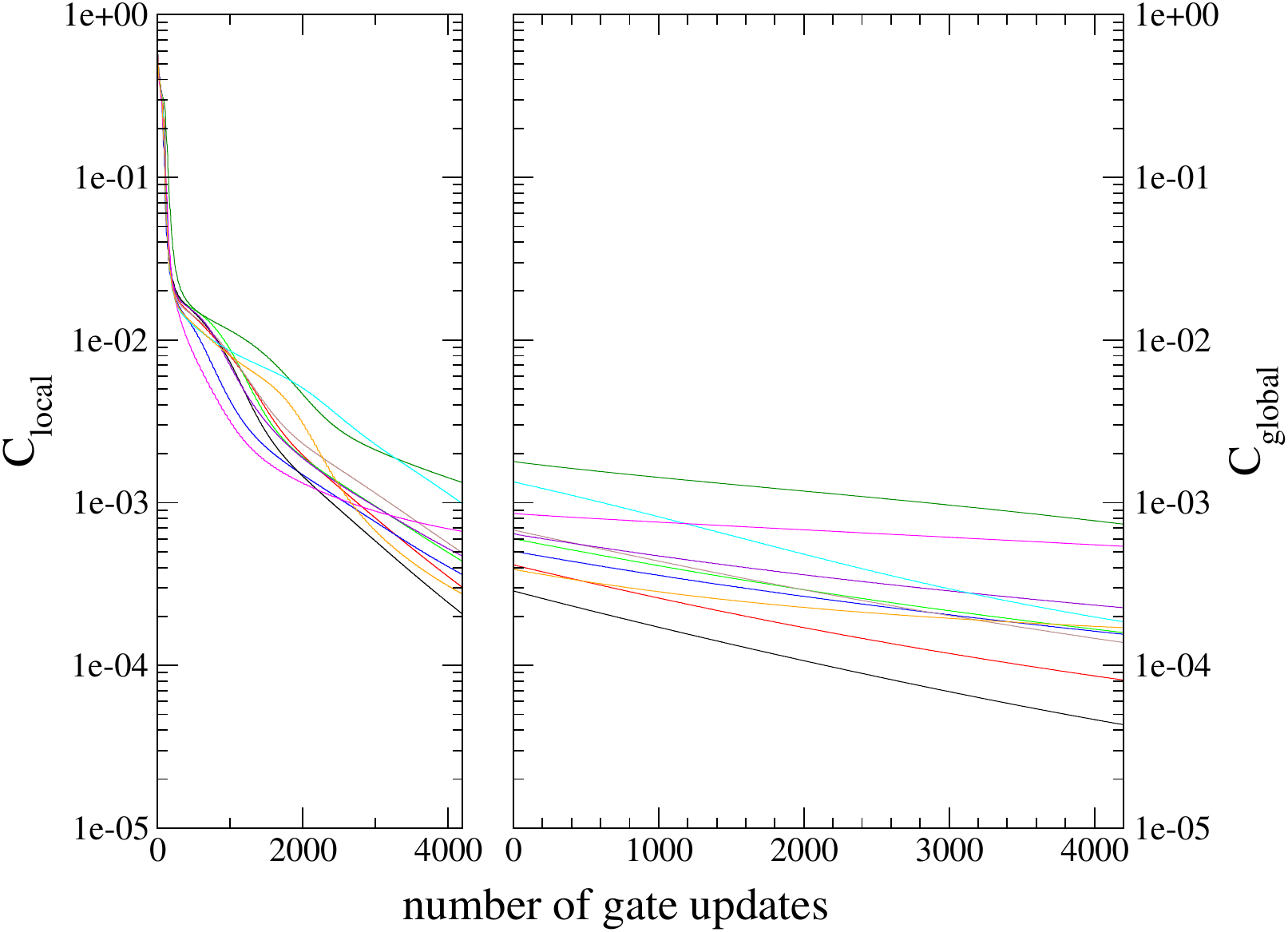}
         \\
        (a) $U_\mathrm{NP}$ with all-to-all connection and $L=10$   
        &
        (b) The spin-preserving ansatz with $L=7$
    \end{tabular}
    \\
    \caption{{\bf Optimization trajectories of QAQC for \ce{H3}/STO-3G with different ansätze.} 
    (a) Number preserving ansaetze with all-to-all connection with circuit layer $L=10$.
    (b) Z-spin preserving ansaetze with all-to-all connection with circuit layer $L=10$.
    Left and right panels show results using local and global cost functions, respectively.}
    \label{fig:traj-h3}
 \end{figure*}

Furthermore, the spin-preserving ansatz involves fewer variational parameters than the number-preserving counterparts.  
As a result, the total number of parameter updates per optimization sweep is reduced.  
Under a fixed number of sweeps, this leads to faster convergence toward the target energy levels, as confirmed in Fig.~\ref{fig:traj-h3} and Appendix~\ref{apdx:traj_compliation_H2}.  

Combined with input state restrictions, quantum compilation using symmetry-preserving ansaetze designed with generalized number-preserving gates yields a more accurate approximation of time-evolution operators, while requiring significantly fewer controlled operations.

In the original quantum compilation research, it has been shown that the cost function in unitary compilation are rigorously related to averaged fidelity between the states evolved with an exact and an approximated time evolution operators~\cite{khatri2019quantum}.
It is not the case when the input state restriction is employed although the cost is still closely related to the averaged fidelity. 
Using this relation, we can roughly estimate the fidelity error when a quantum system evolved based on a complied time evolution operator.
However, this cost function is averaged over the input states that are used for unitary compilation.
And thus, it does not necessarily provide an exact prediction of an arbitrary specific quantum state. 
Hence, we next reproduced quantum dynamics of a certain state based on the complied unitary and estimated the infidelity to verify the accuracy of approximated unitary operators.

Figure~\ref{fig:time-evolution} shows the obtained infidelities in the time evolution for a \ce{H2} and \ce{H3} molecule where the fidelity is defined as
\begin{align}
 F(t) = |\bra{\psi_{\rm{ini}}}(V(\bm{\theta})^\dagger)^{t/\Delta t} e^{-iHt}\ket{\psi_{\rm{ini}}}|^2,
\end{align}
where $V(\bm{\theta})$ is an approximated time evolution operator whose parameter $\bm{\theta}$ is extracted from the unitary compilation that provided the smallest cost value. 
For a \ce{H2} molecule, we employed the Hartree-Fock state as the initial state $\ket{\psi_{\mathrm{ini}}}$ while for \ce{H3} the state in which two spin-up orbitals and one spin-down orbital are occupied is employed.
The initial states are, in the computational basis, represented as $\ket{0101}$ and $\ket{011001}$ for \ce{H2} and \ce{H3}, respectively.
For the consistency with compilation, $\Delta t$ is chosen as 1.0.

For \ce{H2}, the operator from the spin-preserving ansatz of $L=3$ reproduced fairly accurate quantum dynamics.
We suppose that the preserved fidelity for $L=3$ is a special case that can occur in systems as small as hydrogen molecule.
For larger systems, as usual, the infidelity should monotonically increase to some magnitude in the course of time as confirmed in the \ce{H3} system.  
On the other hand, the number-preserving ansatz with all-to-all connection $L=3$ for \ce{H2} and $L=10$ for \ce{H3} for and spin-preserving ansatz 
appear to be better performance than the spin preserving one, which contradicting to the cost in unitary compilation, where the spin-preserving ansatz resulted in lower values.
As mentioned above, the cost function does not correspond rigorously to the averaged fidelity. Additionally, this can presumably be explained by the fact that lower averaged fidelity does not necessarily guarantee more accurate dynamics for an arbitrary state, which is represented by the overlap of input states.
Leaving aside the unexpected behaviors, it is important to emphasize that the time evolution operator was reproduced to maintain a infidelity smaller than 0.1 over $t=200$ while reducing the number of CNOT-gates to less than half.

%% file: conclusions.tex
It has been demonstrated that sequential local optimization of single-qubit gate using FQS is more effective than classical optimizers such as COBYLA and ADAM.
This advantage has been clearly confirmed not only on ideal statevector simulators without noise, but also on QASM simulators with shot noise, and—most notably—on real quantum devices~\cite{wada2024sequential}.
In this study, we further extended this approach by parameterizing and sequentially optimizing controlled gates, in combination with FQS for single-qubit gates, to achieve more efficient and accurate overall optimization.
Our numerical experiments also show that the benefits of controlled-FQS (cFQS) can be further enhanced by incorporating self-consistent field optimization (SCF-cFQS).
This enhancement is theoretically justified by the inclusion of parameter correlations among six variables associated with the gate.
However, two potential drawbacks of the proposed method should be noted.
First, the number of required cost evaluations, i.e., the number of observable measurements, increases for controlled gates compared to FQS.
While FQS typically requires ten measurements (reducible to nine by reusing the current cost value), cFQS requires fourteen, and SCF-cFQS requires 35.
That said, this may not pose a significant limitation if the quantum device has a sufficient number of qubits or if parallel execution across multiple devices is feasible.
Such parallelism is increasingly realistic given the capabilities of modern quantum hardware.
Accordingly, we argue that it is more appropriate to evaluate local optimizers such as FQS and cFQS in terms of the number of gate updates, rather than measurement count—as we have done in this study.
However, this criterion does not necessarily apply to gradient-based optimizers, for which the number of required measurements increases with system size—unlike local optimizers, which typically require a constant number of measurements.

The second drawback lies in the increased CNOT depth.
As illustrated in Fig.~\ref{fig:decomposition}, generalized controlled gates require at least two controlled operations, effectively doubling the CNOT depth of the circuit.
This increase in circuit depth could potentially lead to probabilistic and deterministic concentration of the expectation value, i.e., twin barren plateaus.
However, our numerical experiments indicate that in alternating layered ansätze, the use of cFQS in gate blocks does not induce a probabilistic concentration.
This can be attributed to the fact that the emergence of barren plateaus is more closely tied to the number of circuit layers rather than the expressibility of individual gate blocks.
On the contrary, cFQS and SCF-cFQS can in some cases reduce the required number of layers, thereby helping to mitigate barren plateaus, although they may not completely eliminate their exponential scaling behavior.

We emphasize that controlled-gate optimization is highly compatible with a variety of variational techniques.
It is therefore important to consider its integration with methods designed to avoid barren plateaus, such as tailored ansätze and circuit structure optimization.
As an example, we extended the expressibility of particle-number preserving gates and optimized them using cFQS.
This approach enabled efficient compilation of target unitaries using significantly shallower circuits.

Although generalized controlled unitaries require additional CNOT gates in their standard decomposition, and may thus lead to unnecessarily complex gate structures, prior work suggests that optimizations in complex space can be more successful—even when the target quantum states reside in real space~\cite{wada2022simulating, sato2023variational}.
Furthermore, the number of CNOT gates required for a particle-number preserving gate can be reduced from four to three by employing controlled-Fraxis instead of controlled-FQS.
This reduction in CNOT cost, however, comes at the expense of reduced expressibility.
Therefore, the optimal choice of method will depend on the target system and the available quantum hardware.
Lastly, we note that cFQS and SCF-cFQS are also highly compatible with adaptive optimization frameworks, such as VAns and ADAPT-VQE, as well as with advanced techniques like multi-controlled gate operations and optimized parameter configurations~\cite{endo2023optimal}, though further investigation in these directions is left for future work.

%% file: appendix.tex
\renewcommand{\theequation}{A.\arabic{equation}}
\setcounter{equation}{0}
\section{parameter configuration}
$Q_0$ and $Q_0'$ are the original parameter configurations of controlled-FQS and SCF-cFQS employed for state vector simulations, respectively.
$Q_1$ and $Q_1'$ are the improved parameter configurations of controlled-FQS and SCF-cFQS employed for QASM simulations, which was obtained by the approach in \cite{endo2023optimal}

\label{apdx:parameter configuration}
\begin{minipage}{0.3\linewidth}
\begin{align}\label{eqn:PC_cFQS_sv}
Q_0&=
\begin{bmatrix}
\bm{q}^T_1\\
\bm{q}^T_2\\
\bm{q}^T_3\\
\bm{q}^T_4\\
\bm{q}^T_5\\
\bm{q}^T_6\\
\bm{q}^T_7\\
\bm{q}^T_8\\
\bm{q}^T_9\\
\bm{q}^T_{10}\\
\bm{q}^T_{11}\\
\bm{q}^T_{12}\\
\bm{q}^T_{13}\\
\bm{q}^T_{14}
\end{bmatrix}
=
\begin{bmatrix}
1&0&0&0\\
0&1&0&0\\
0&0&1&0\\
0&0&0&1\\
1&1&0&0\\
1&0&1&0\\
1&0&0&1\\
0&1&1&0\\
0&1&0&1\\
0&0&1&1\\
1&1&1&0\\
1&1&0&1\\
1&0&1&1\\
0&1&1&1\\
\end{bmatrix}
\end{align}
\end{minipage}
\hfill
\begin{minipage}{0.7\linewidth}
\begin{align}\label{eqn:PC_cFQS_qasm}
Q_1=
\begin{bmatrix}
\bm{q}^T_1\\
\bm{q}^T_2\\
\bm{q}^T_3\\
\bm{q}^T_4\\
\bm{q}^T_5\\
\bm{q}^T_6\\
\bm{q}^T_7\\
\bm{q}^T_8\\
\bm{q}^T_9\\
\bm{q}^T_{10}\\
\bm{q}^T_{11}\\
\bm{q}^T_{12}\\
\bm{q}^T_{13}\\
\bm{q}^T_{14}
\end{bmatrix}
=
\begin{bmatrix}
+0.841464& +0.440207& -0.313043& -0.012669\\
-0.236232& +0.733872& -0.628245& +0.104569\\
-0.354510& -0.117623& +0.927570& -0.010105\\
-0.061836& -0.704459& +0.186645& +0.681966\\
+0.081448& -0.087390& -0.705379& -0.698692\\
+0.293054& +0.032108& +0.368205& -0.881767\\
+0.688379& -0.691224& -0.154162& -0.156774\\
-0.739612& +0.236695& +0.099249& -0.622173\\
-0.030371& -0.299379& -0.876030& +0.376857\\
-0.444212& -0.841304& -0.010504& -0.307852\\
+0.019976& +0.908723& +0.413518& -0.053154\\
+0.123101& +0.356778& -0.007177& +0.926015\\
+0.707437& +0.066835& +0.681001& +0.176928\\
-0.905210& -0.019329& -0.048812& +0.421708\\
\end{bmatrix}
\end{align}
\end{minipage}

\begin{align}\label{eqn:PC_SCF-cFQS_sv}
Q'_0=
\begin{bmatrix}
\bm{p}^T_1, \bm{q}^T_1\\
\bm{p}^T_2, \bm{q}^T_2\\
\bm{p}^T_3, \bm{q}^T_3\\
\bm{p}^T_4, \bm{q}^T_4\\
\bm{p}^T_5, \bm{q}^T_5\\
\bm{p}^T_6, \bm{q}^T_6\\
\bm{p}^T_7, \bm{q}^T_7\\
\bm{p}^T_8, \bm{q}^T_8\\
\bm{p}^T_9, \bm{q}^T_9\\
\bm{p}^T_{10}, \bm{q}^T_{10}\\
\bm{p}^T_{11}, \bm{q}^T_{11}\\
\bm{p}^T_{12}, \bm{q}^T_{12}\\
\bm{p}^T_{13}, \bm{q}^T_{13}\\
\bm{p}^T_{14}, \bm{q}^T_{14}\\
\bm{p}^T_{15}, \bm{q}^T_{15}\\
\bm{p}^T_{16}, \bm{q}^T_{16}\\
\bm{p}^T_{17}, \bm{q}^T_{17}\\
\bm{p}^T_{18}, \bm{q}^T_{18}\\
\bm{p}^T_{19}, \bm{q}^T_{19}\\
\bm{p}^T_{20}, \bm{q}^T_{20}\\
\bm{p}^T_{21}, \bm{q}^T_{21}\\
\bm{p}^T_{22}, \bm{q}^T_{22}\\
\bm{p}^T_{23}, \bm{q}^T_{23}\\
\bm{p}^T_{24}, \bm{q}^T_{24}\\
\bm{p}^T_{25}, \bm{q}^T_{25}\\
\bm{p}^T_{26}, \bm{q}^T_{26}\\
\bm{p}^T_{27}, \bm{q}^T_{27}\\
\bm{p}^T_{28}, \bm{q}^T_{28}\\
\bm{p}^T_{29}, \bm{q}^T_{29}\\
\bm{p}^T_{30}, \bm{q}^T_{30}\\
\bm{p}^T_{31}, \bm{q}^T_{31}\\
\bm{p}^T_{32}, \bm{q}^T_{32}\\
\bm{p}^T_{33}, \bm{q}^T_{33}\\
\bm{p}^T_{34}, \bm{q}^T_{34}\\
\bm{p}^T_{35}, \bm{q}^T_{35}
\end{bmatrix}
=
\begin{bmatrix}
\begin{pmatrix}
1&0&0&0\\
1&0&0&0\\
1&0&0&0\\
1&0&0&0\\
0&1&0&0\\
0&1&0&0\\
0&1&0&0\\
0&1&0&0\\
0&0&1&0\\
0&0&1&0\\
0&0&1&0\\
0&0&1&0\\
0&0&0&1\\
0&0&0&1\\
0&0&0&1\\
0&0&0&1\\
1&0&0&0\\
1&0&0&0\\
1&0&0&0\\
1&0&0&0\\
1&0&0&0\\
1&0&0&0\\
1&0&0&0\\
1&0&0&0\\
1&0&0&0\\
1&1&0&0\\
1&-1&0&0\\
1&0&1&0\\
1&0&-1&0\\
1&0&0&1\\
1&0&0&-1\\
0&1&1&0\\
0&1&0&1\\
0&0&1&1\\
1&1&1&1\\
\end{pmatrix}
\begin{pmatrix}
1&0&0&0\\
0&1&0&0\\
0&0&1&0\\
0&0&0&1\\
1&0&0&0\\
0&1&0&0\\
0&0&1&0\\
0&0&0&1\\
1&0&0&0\\
0&1&0&0\\
0&0&1&0\\
0&0&0&1\\
1&0&0&0\\
0&1&0&0\\
0&0&1&0\\
0&0&0&1\\
1&1&0&0\\
1&-1&0&0\\
1&0&1&0\\
1&0&-1&0\\
0&1&1&0\\
0&1&-1&0\\
0&1&0&1\\
0&0&1&1\\
0&0&1&-1\\
1&0&0&0\\
1&0&0&0\\
1&0&0&0\\
1&0&0&0\\
1&0&0&0\\
1&0&0&0\\
1&0&0&0\\
1&0&0&0\\
1&0&0&0\\
1&1&1&1
\end{pmatrix}
\end{bmatrix}
\end{align}

\begin{align}\label{eqn:PC_SCF-cFQS_qasm}
Q'_1=
\begin{bmatrix}
\bm{p}^T_1, \bm{q}^T_1\\
\bm{p}^T_2, \bm{q}^T_2\\
\bm{p}^T_3, \bm{q}^T_3\\
\bm{p}^T_4, \bm{q}^T_4\\
\bm{p}^T_5, \bm{q}^T_5\\
\bm{p}^T_6, \bm{q}^T_6\\
\bm{p}^T_7, \bm{q}^T_7\\
\bm{p}^T_8, \bm{q}^T_8\\
\bm{p}^T_9, \bm{q}^T_9\\
\bm{p}^T_{10}, \bm{q}^T_{10}\\
\bm{p}^T_{11}, \bm{q}^T_{11}\\
\bm{p}^T_{12}, \bm{q}^T_{12}\\
\bm{p}^T_{13}, \bm{q}^T_{13}\\
\bm{p}^T_{14}, \bm{q}^T_{14}\\
\bm{p}^T_{15}, \bm{q}^T_{15}\\
\bm{p}^T_{16}, \bm{q}^T_{16}\\
\bm{p}^T_{17}, \bm{q}^T_{17}\\
\bm{p}^T_{18}, \bm{q}^T_{18}\\
\bm{p}^T_{19}, \bm{q}^T_{19}\\
\bm{p}^T_{20}, \bm{q}^T_{20}\\
\bm{p}^T_{21}, \bm{q}^T_{21}\\
\bm{p}^T_{22}, \bm{q}^T_{22}\\
\bm{p}^T_{23}, \bm{q}^T_{23}\\
\bm{p}^T_{24}, \bm{q}^T_{24}\\
\bm{p}^T_{25}, \bm{q}^T_{25}\\
\bm{p}^T_{26}, \bm{q}^T_{26}\\
\bm{p}^T_{27}, \bm{q}^T_{27}\\
\bm{p}^T_{28}, \bm{q}^T_{28}\\
\bm{p}^T_{29}, \bm{q}^T_{29}\\
\bm{p}^T_{30}, \bm{q}^T_{30}\\
\bm{p}^T_{31}, \bm{q}^T_{31}\\
\bm{p}^T_{32}, \bm{q}^T_{32}\\
\bm{p}^T_{33}, \bm{q}^T_{33}\\
\bm{p}^T_{34}, \bm{q}^T_{34}\\
\bm{p}^T_{35}, \bm{q}^T_{35}
\end{bmatrix}
=
\begin{bmatrix}
\begin{pmatrix}
-0.168738& -0.377669& -0.473438& -0.325976\\
-0.047231& -0.609849& -0.320462& -0.195069\\
-0.270645& +0.642219& +0.292993& +0.370156\\
+0.288417& +0.262689& +0.204232& +0.425265\\
-0.232073& +0.160375& -0.352717& -0.340317\\
+0.465742& +0.532438& -0.161785& -0.051620\\
+0.005033& +0.663459& -0.145238& -0.166775\\
-0.049396& +0.689267& -0.039698& -0.057984\\
-0.189335& -0.374789& +0.307014& -0.284040\\
+0.297044& -0.184862& +0.621407& -0.016755\\
+0.025099& +0.015446& +0.690956& -0.170530\\
-0.008484& +0.031660& +0.722470& -0.034882\\
-0.265450& -0.360971& -0.333375& +0.172628\\
+0.236749& -0.327388& -0.132797& +0.579663\\
+0.046245& -0.017053& -0.202154& +0.679860\\
-0.037673& +0.018414& -0.015557& +0.719174\\
+0.378055& -0.204488& -0.144759& +0.068018\\
-0.492665& +0.232769& +0.288145& +0.375047\\
+0.295951& -0.027452& -0.240158& -0.285853\\
+0.436258& +0.009700& -0.045667& -0.161869\\
+0.457070& +0.105276& +0.008341& +0.188058\\
+0.356260& +0.058825& +0.269569& +0.343798\\
+0.492086& -0.102147& +0.064846& -0.020277\\
+0.442664& +0.336447& +0.093118& -0.036410\\
-0.015023& +0.185844& +0.163124& +0.564778\\
-0.000526& +0.420639& -0.505695& -0.505042\\
+0.425188& -0.589638& +0.182963& +0.030676\\
+0.011472& -0.462920& +0.472808& -0.483827\\
+0.465052& +0.151085& -0.578084& +0.128895\\
+0.021247& -0.484857& -0.469391& +0.448005\\
+0.305720& +0.210750& +0.267743& -0.619767\\
-0.373386& +0.353355& +0.499346& -0.252072\\
-0.379127& +0.360506& -0.343972& +0.425453\\
-0.253350& -0.326506& +0.460676& +0.360606\\
+0.562911& -0.008321& +0.065322& +0.028830
\end{pmatrix}
\begin{pmatrix}
-0.176545& -0.626582& -0.258796& -0.087986\\
-0.402438& +0.368249& -0.252058& -0.352148\\
-0.472026& -0.178495& +0.191706& -0.005056\\
-0.466292& -0.062608& -0.629068& -0.090412\\
+0.159521& +0.798771& -0.124395& -0.035230\\
+0.252487& +0.388189& -0.284955& -0.418468\\
+0.072618& +0.125477& +0.682434& -0.155417\\
+0.052599& +0.160433& -0.527380& +0.459236\\
+0.076053& +0.793515& +0.076680& +0.086141\\
+0.071319& +0.287377& -0.198327& -0.603455\\
-0.071561& -0.069914& +0.694660& -0.008280\\
-0.118511& -0.037576& -0.334553& +0.590223\\
+0.006697& +0.606798& +0.153507& +0.516220\\
+0.223436& +0.589225& -0.007620& -0.293179\\
-0.020900& -0.017197& +0.689158& +0.137197\\
-0.025005& -0.023223& -0.184241& +0.667592\\
+0.083045& -0.336838& -0.474636& -0.666365\\
+0.127512& -0.489994& -0.470893& +0.036296\\
-0.064015& -0.621105& +0.617871& +0.025251\\
+0.083266& -0.531518& -0.483985& +0.507555\\
-0.516304& +0.051429& +0.348933& -0.594642\\
+0.632870& -0.027783& -0.519592& +0.086498\\
-0.155714& +0.659522& -0.451482& +0.282400\\
-0.435624& +0.053767& +0.359399& +0.599190\\
+0.274839& -0.184056& +0.118540& -0.704406\\
+0.466377& -0.268501& -0.126103& -0.082250\\
+0.145396& -0.000754& +0.284435& +0.578862\\
+0.347491& -0.453299& -0.021438& -0.035322\\
+0.351055& +0.223086& +0.099730& +0.476605\\
+0.408670& -0.400898& +0.029591& +0.121736\\
+0.460832& +0.238913& +0.088536& +0.359219\\
+0.564924& +0.120497& -0.024247& -0.297641\\
+0.622612& +0.197431& +0.016464& +0.008705\\
+0.639050& +0.057386& +0.253009& +0.106098\\
+0.647186& -0.032961& +0.437916& -0.257403
\end{pmatrix}
\end{bmatrix}
\end{align}

\clearpage
\section{Circuit Structure}
\label{apdx:particle-number-preserving-ansatz}

\begin{figure}[htb]
\begin{tabular}{ccc}
\Qcircuit @C=2.0 em @R=1.8em{
  \lstickx{\ket{0}_1} 
  &\multigate{1}{1}
  &\multigate{0}{n}
  &\qw &\qw \\
  \lstickx{\ket{0}_2} 
  &\ghost{1} 
  &\multigate{1}{2} 
  &\qw &\qw \\
  \lstickx{\ket{0}_3} 
  &\multigate{1}{3}
  &\ghost{2}       
  &\qw &\qw  \\
  \lstickx{\ket{0}_4} 
  &\ghost{3} 
  &\multigate{1}{4} 
  &\qw &\qw \\
  \lstickx{\ket{0}_5} 
  &\multigate{0}{5}
  &\ghost{4}       
  &\qw &\qw \\ 
  &{\vdots}  &{\vdots}  \\
  \\
  \lstickx{\ket{0}_{n-1}} 
  &\multigate{1}{n-1} 
  &\multigate{0}{n-2} 
  &\qw &\qw \\
  \lstickx{\ket{0}_n} 
  &\ghost{n-1} 
  &\multigate{0}{n} 
  &\qw &\qw \\
 &&& \arrep{ll}
  \gategroup{1}{2}{9}{4}{1.0em}{--}}
&
~~~~~~~~~~~
&
\Qcircuit @C=1.6 em @R=1.2em{
  \lstickx{\ket{1}_0} 
  &\qw &\qw &\qw &\multigate{6}{6} & \qw &\multigate{1}{7} & \qw \\
  \lstickx{\ket{1}_1} 
  &\qw &\qw &\multigate{1}{4} &\qw & \qw &\ghost{7} & \qw\\
  \lstickx{\ket{1}_2} 
  &\qw  &\multigate{1}{2} &\ghost{4}&\qw  & \qw &\multigate{1}{8}  &\qw\\
  \lstickx{\ket{1}_3} 
  &\multigate{1}{1} &\ghost{2} &\multigate{1}{5}&\qw & \qw &\ghost{8} &\qw \\
  \lstickx{\ket{0}_4} 
  &\ghost{1} &\multigate{1}{3} &\ghost{5}  &\qw &\qw  &\multigate{1}{9} &\qw\\
  \lstickx{\ket{0}_5} 
  &\qw &\ghost{3} &\qw &\qw  &\multigate{6}{6'}& \ghost{9} &\qw\\ 
  \lstickx{\ket{0}_6} 
  &\qw &\qw &\qw  &\ghost{6} &\qw & \multigate{1}{7'}  &\qw\\ 
  \lstickx{\ket{1}_7} 
  &\qw &\qw  &\multigate{1}{4'} &\qw &\qw &\ghost{7'}  &\qw \\
  \lstickx{\ket{1}_8} 
  &\qw  &\multigate{1}{2'} &\ghost{4'} &\qw &\qw&\multigate{1}{8'} &\qw \\
  \lstickx{\ket{0}_9} 
  &\multigate{1}{1'} &\ghost{2'} &\multigate{1}{5'}&\qw  &\qw &\ghost{8'} &\qw \\
  \lstickx{\ket{1}_{10}} 
  &\ghost{1'} &\multigate{1}{3'} &\ghost{5'} &\qw  &\qw &\multigate{1}{9'} &\qw \\
  \lstickx{\ket{0}_{11}} 
  &\qw &\ghost{3'}  &\qw &\qw &\ghost{6'} &\ghost{9'} &\qw\\
 && \arreq{l} &&&&\arrer{lll}
  \gategroup{1}{2}{12}{3}{2.0em}{--}
  \gategroup{1}{4}{12}{7}{3.0em}{--}\\
}\\
(a) Alternating Ansatz 
& ~~~~~
& (b) Particle-number preserving ansatz for \ce{O2}
\end{tabular}
\caption{{\bf Ansaetze employed for optimzation.}
Each layer consists of gates in the dashed line, and the total number of layers is written as $L$.
(b) For $L = 3$ and higher, the structure repeats in the same manner as for $L = 2$.
Qubits 0 through 5 correspond to alpha-spin orbitals, while qubits 6 through 11 correspond to beta-spin orbitals.
Qubits 3,4 and 9,10 represent the HOMOs, and qubits 5 and 11 represent the LUMOs.
In optimizations, the gate blocks are sequentially updated in ascending order of the subscript.}
\label{fig:ansatz}
\end{figure}

\begin{figure*}[htb]
\begin{tabular}{ccc}
\Qcircuit @C=2.0em @R=1.0em{
 \lstickx{} & \gate{R_1(\bm{q})} & \ctrl{1} &\qw\\
 \lstickx{} & \gate{R_2(\bm{q})} & \gate{Z_3} & \qw 
 \gategroup{1}{2}{2}{3}{1.0em}{--}\\
}
&\Qcircuit @C=1.4em @R=1.0em{
 \lstickx{} & \gate{R_1(\bm{q})} & \ctrl{1} &\qw\\
 \lstickx{} & \gate{R_2(\bm{q})} & \gate{R_3(\bm{q})} & \qw 
 \gategroup{1}{2}{2}{3}{1.0em}{--}\\
}
&
\Qcircuit @C=1.4em @R=1.0em{
\lstickx{} & \gate{R_1(\bm{q})} & \ctrl{1} & \ctrlo{1} &\qw \\
\lstickx{} & \qw & \gate{R_2(\bm{q})} & \gate{R'_2(\bm{q})} &\qw 
\gategroup{1}{2}{2}{4}{1.0em}{--}}\\
(a) For FQS optimizations 
&(b) For cFQS optimizations 
&(c) For SCF-cFQS optimizations 
\end{tabular}
\caption{{\bf Two-qubit gate blocks employed in the asaetze in Fig.~\ref{fig:ansatz}.}
Each gate block consists of single-qubit gates and controlled-gates. 
In the gate block optimization, the gates are sequentially updated in ascending order of the subscript. 
In SCF-cFQS simulations, $R_2$ and $R'_2$ are self-consistently optimized until convergence.
}
\label{fig:gate_block}
\end{figure*}

\clearpage
\section{Initial State Preparation for the Molecular Hamiltonian}
\label{apdx:initial_state_preparation}
The choice of the initial state has a significant impact on the performance of VQE. To investigate effective state preparation protocols, we performed VQE simulations with initial states prepared according to three different protocols.

\textit{Protocol 1}: In the ADAM optimization of an ansatz composed of RBS gates~\cite{kerenidis2021classical}, the initial states were generated by selecting random rotation angles of the form $\theta = r \pi$, where $r$ is a random number in the interval $[-f,f]$ where $f\in\mathbb{R}$.

\textit{Protocol 2}: The initial states were prepared using multiple randomized single-qubit gates, where each gate is parameterized by a quaternion $(q_{\mathrm i}, q_x, q_y, q_z)$. In this scheme, the components $(q_x, q_y, q_z)$ were generated randomly and normalized such that their combined magnitude was a parameter of $f$. The component $q_i$ was then set to $q_i = \sqrt{1 - f^2}$.

\textit{Protocol 3}: The initial states were prepared using multiple randomized single-qubit gates, where each gate is parameterized by a quaternion $(q_i, q_x, q_y, q_z)$. In this scheme, $q_y$ was fixed to $0$, and the remaining components $(q_i, q_x, q_z)$ were generated randomly and normalized.

\begin{figure*}[h]
 \centering
 \begin{tabular}{ccc}
 \includegraphics[scale=0.27]{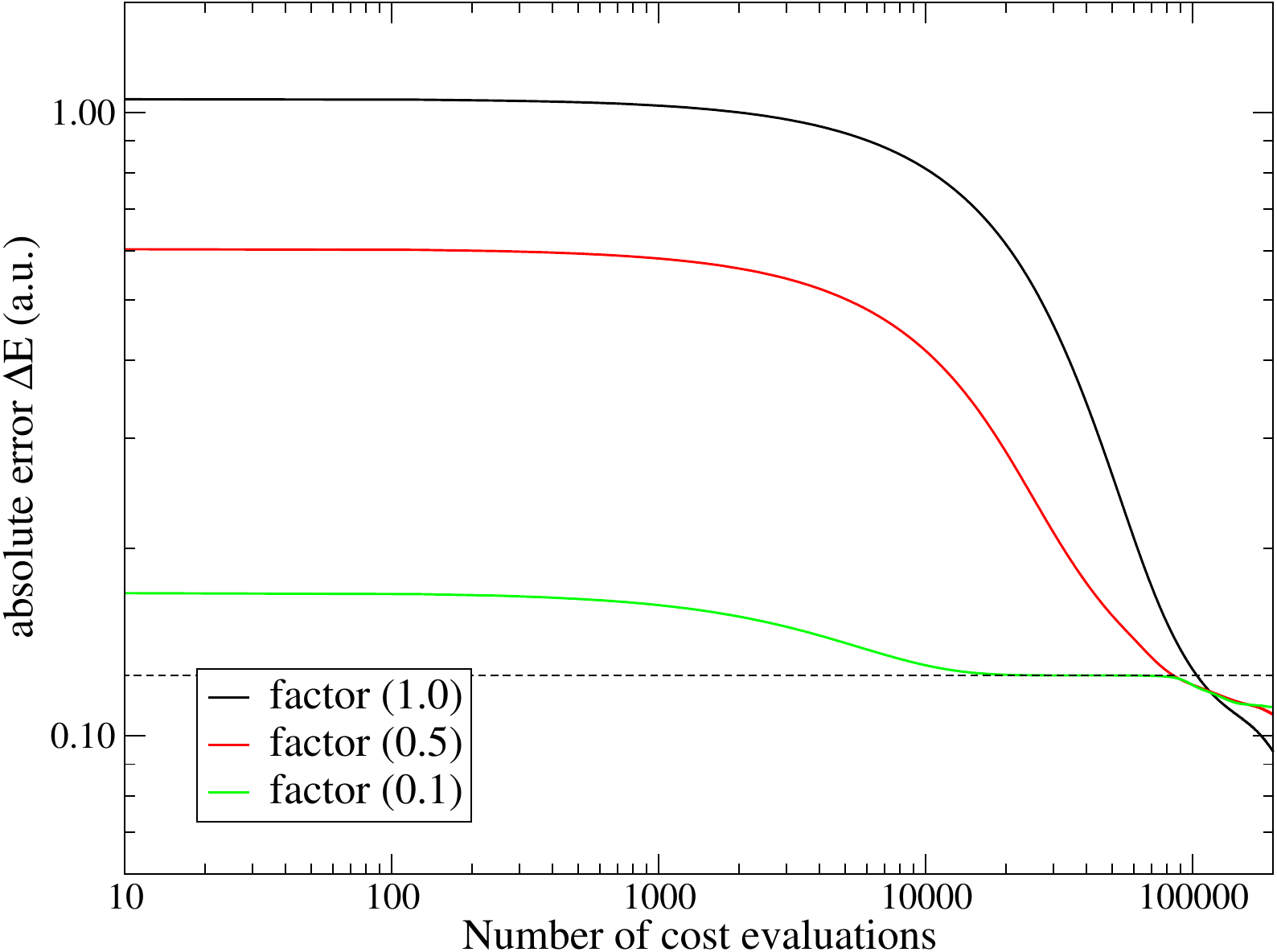}  
 &~&
 \includegraphics[scale=0.27]{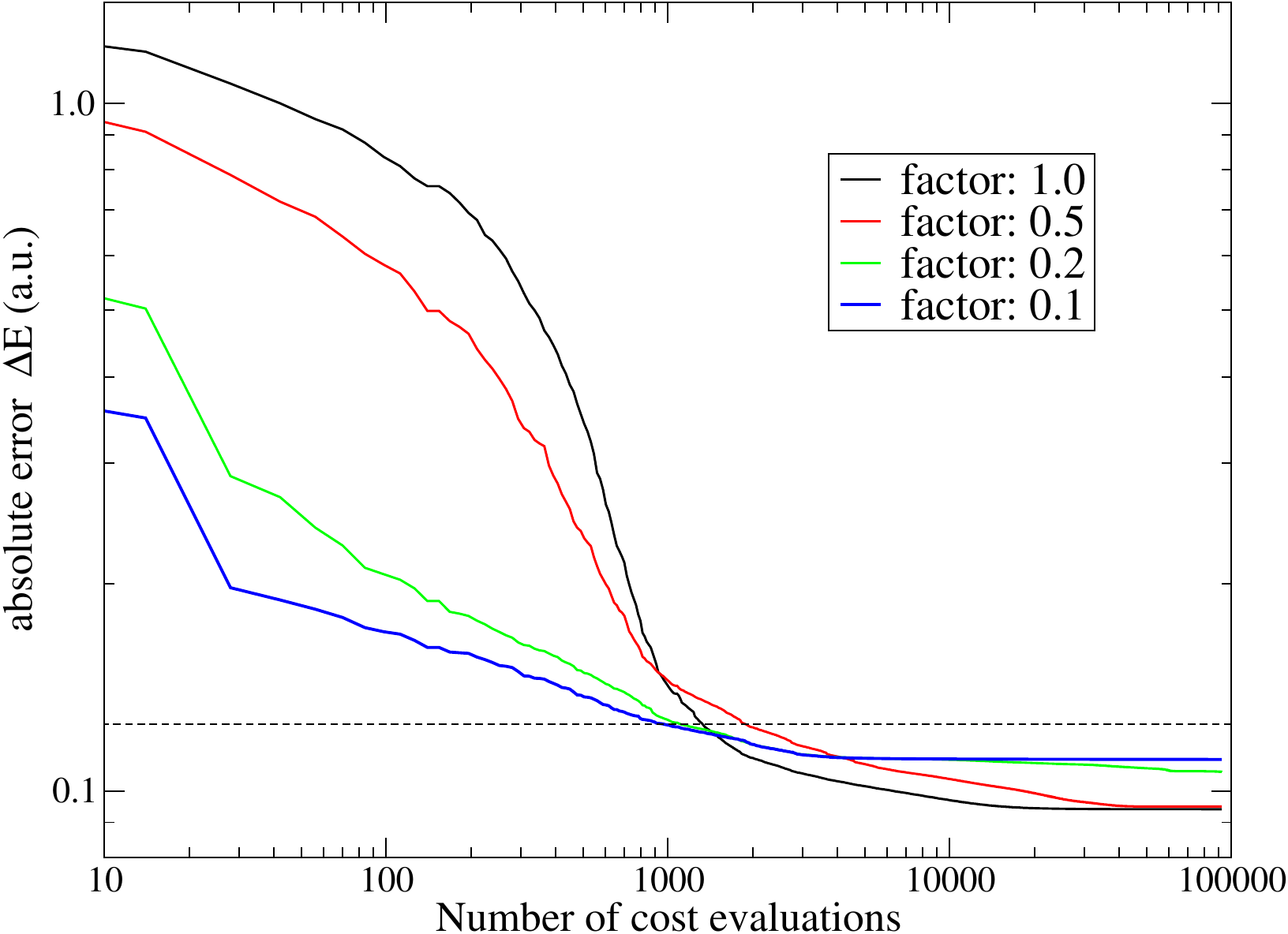}  
 \\
 (a) RBS gates with ADAM optimizer  
 &~&
 (b) Number preserving ansatz with cFQS optimizer
 \\
 \end{tabular}
 \caption{{\bf Averaged trajectories of a \ce{O2}/STO-3G molecule over 20 different optimizations with different scaling factors for random initialization.} 
 The initial states are prepared using {\it Protocol 2} with different values of a random factor $f$.
 Dashed line represents of the energy difference between the Hartree-Fock state and the exact ground state.
 }
 \label{fig:VQE_O2_sv_cfqs_factors}
\end{figure*}

\begin{figure*}[h]
 \centering
 \begin{tabular}{ccc}
 \includegraphics[scale=0.27]{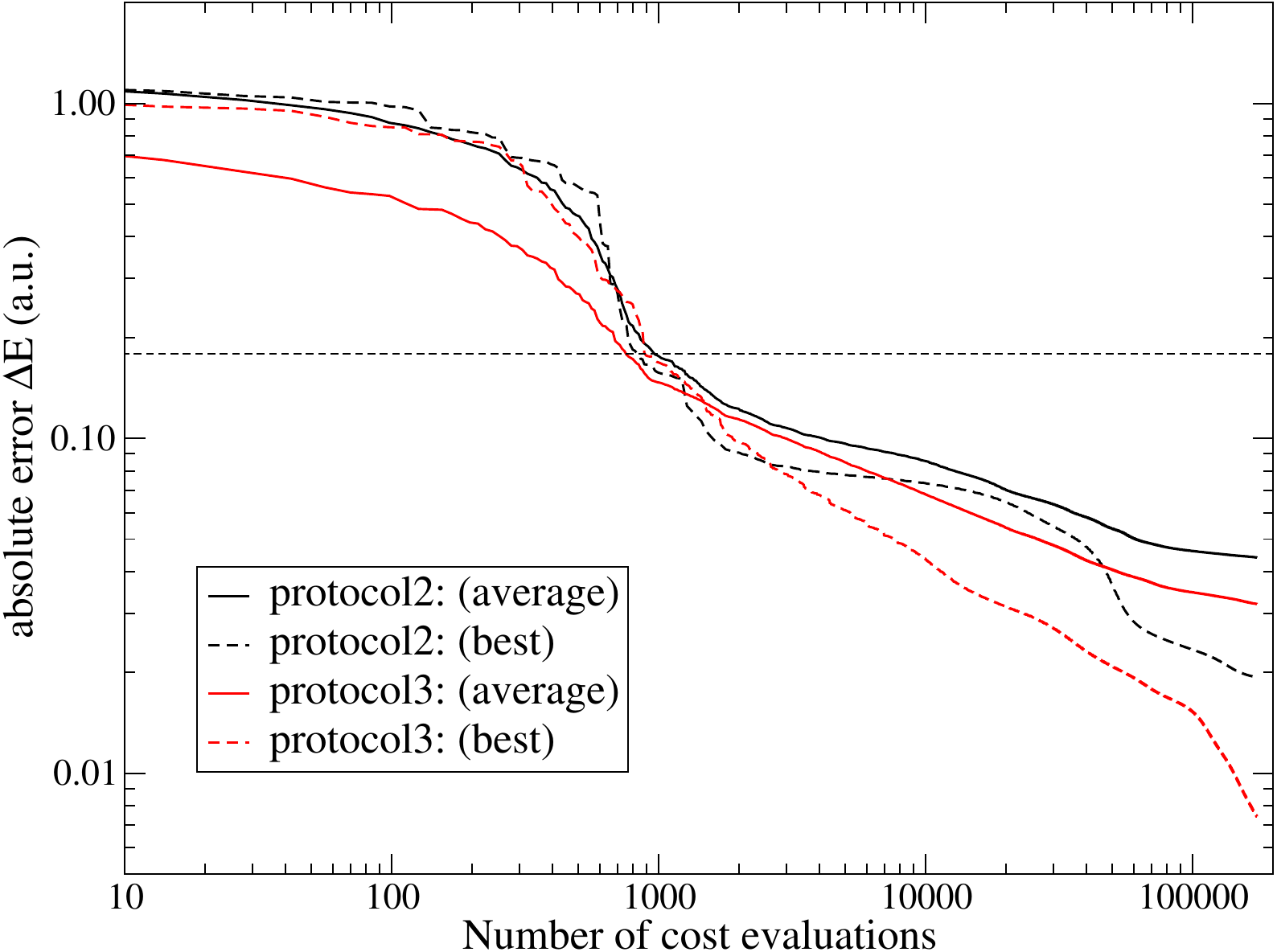}  
 &~&
 \includegraphics[scale=0.27]{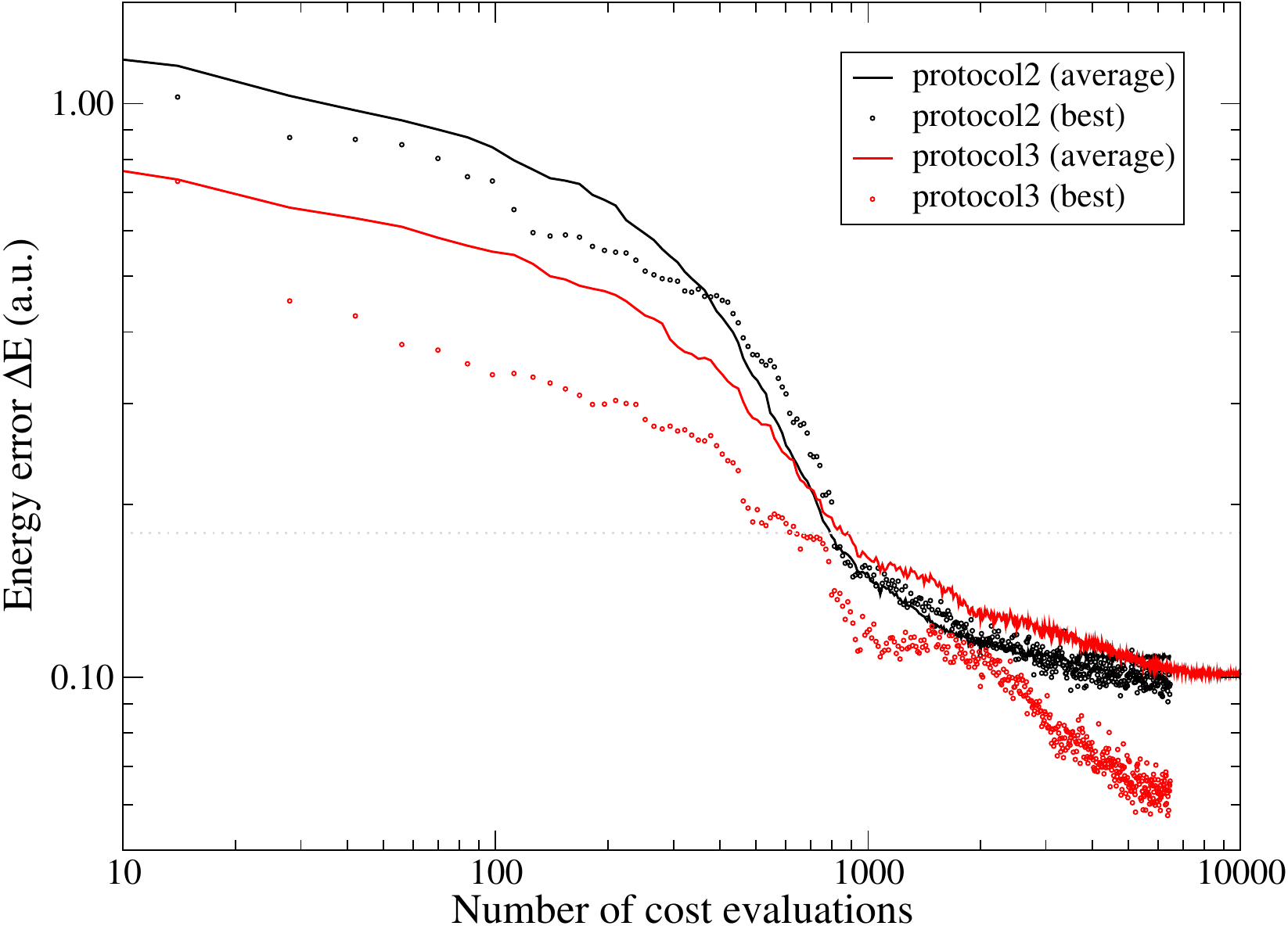}  
 \\
 (a) Statevector simulator  
 &~&
 (b) QASM simulator 
 \\
 \end{tabular}
 \caption{{\bf VQE trajectories for \ce{O2}/STO-3G with different protocols for random initialization.}  Black and red colors stand for the simulation with initial states are generated by {\it protocol 2} and {\it 3}  Solid lines represent averaged trajectories over 20 VQE optimization. Dashed lines in (a) and Circle points in (b) represent the best-case trajectories among the 20 runs.
 }
 \label{fig:VQE_O2_cfqs_initials}
\end{figure*}

\clearpage
\section{VQE for the Molecular Hamiltonian with Parity Mapping}
\begin{figure*}[h]
 \centering
 \begin{tabular}{ccc}
 \includegraphics[scale=0.27]{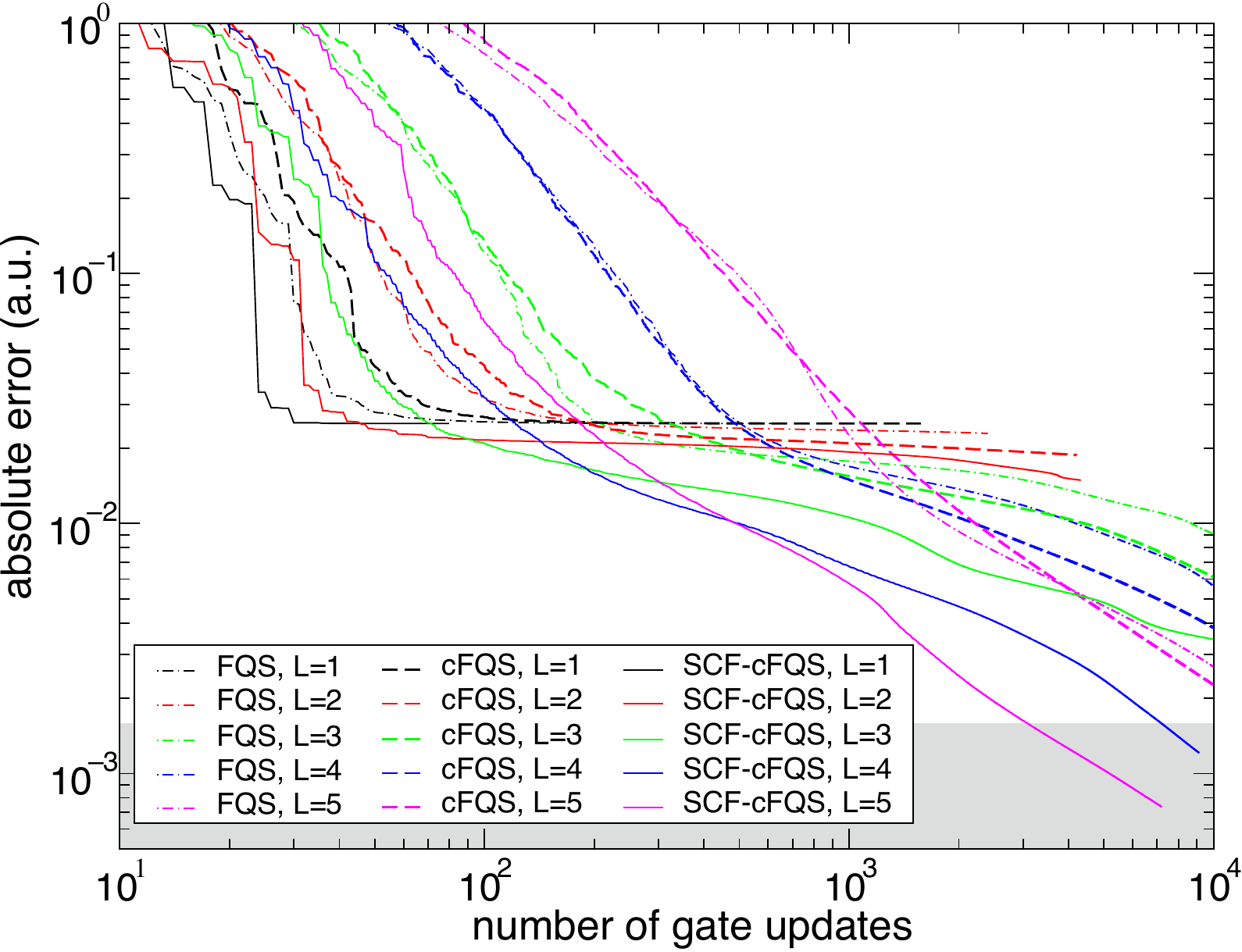}  
 &~&
 \includegraphics[scale=0.27]{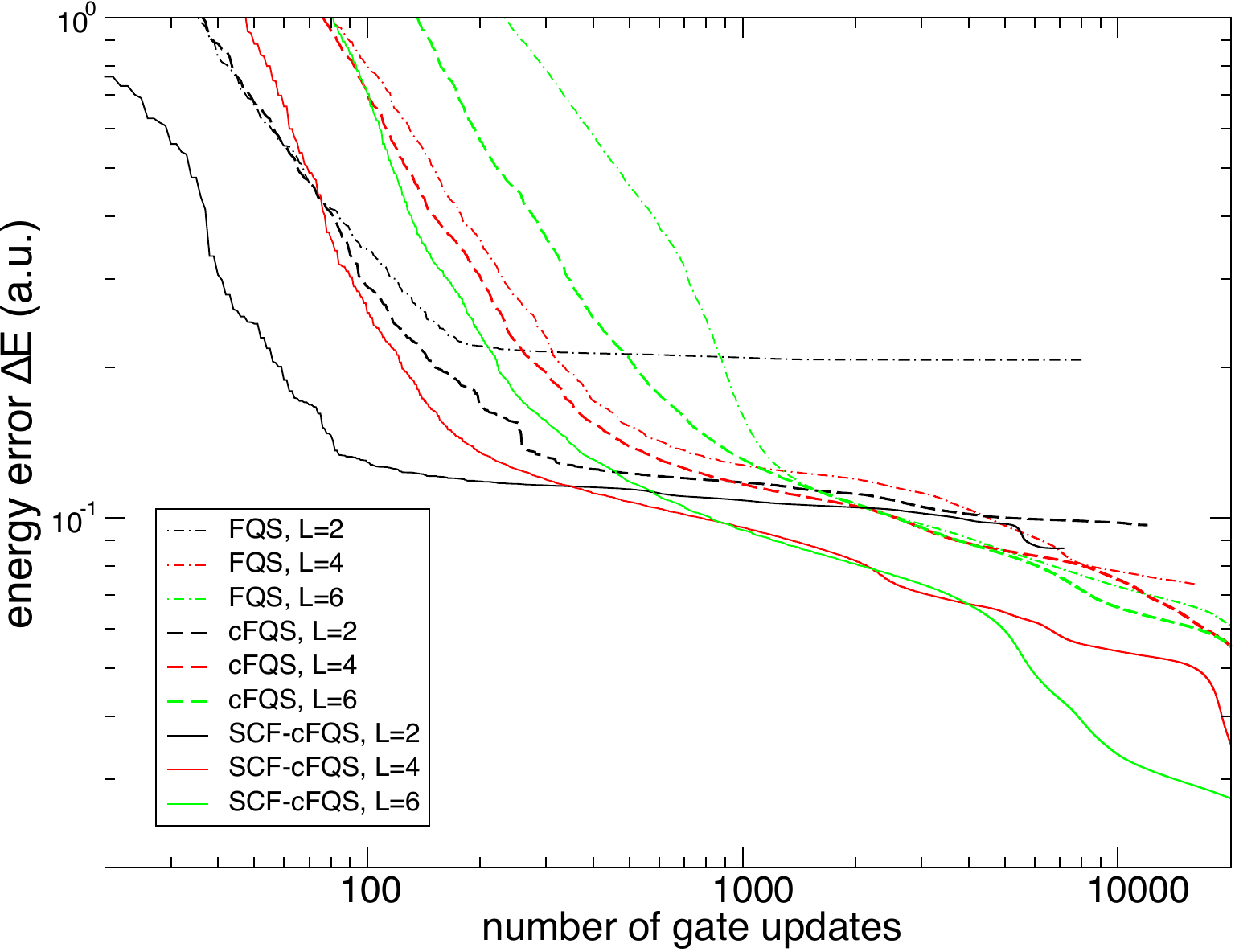}  
 \\
 (a) \ce{H2}/31G.  
 &~&
 (b) \ce{O2}/STO-3G 
 \\
 \end{tabular}
 \caption{{\bf Averaged VQE trajectories over 20 different optimizations for molecular Hamiltonian with the statevector with different number of circuit layer $L$.} The dash-dotted line represents the optimization trajectory where single-qubit gates are optimized using FQS. 
 The dashed and solid lines represent trajectories where single-qubit gates are optimized by FQS, and controlled gates are optimized by cFQS and SCF-cFQS, respectively.}
 \label{fig:VQE_molecules_sv}
\end{figure*}

\clearpage
\section{PQC structure employed for unitary compilation}
\label{apdx:PQC_structure}
\begingroup
\renewcommand{\arraystretch}{1.3}
 \begin{table}[h]
     \centering
     \caption{ {\bf Pairs of qubits used for constructing PQCs for \ce{H2}, \ce{H3}, and \ce{LiH}. } The gate block is applied to each pair of qubits and this is the structure of one layer of the PQC. For the ALT without symmetries, the gate block defined as Fig.~\ref{fig:gate_block}(b) is applied, otherwise the particle preserving gate defined as Fig. \ref{fig:PPgate} is applied. Additionally, for the ALT without symmetries, one general single-qubit gate is applied to each qubit at the last of PQC.}
     \begin{tabular}{c|c|c|c} \hline
      Molecule & Gate type & Entanglement & Qubit-pairs \\ \hline\hline
      \multirow{5}{*}{\ce{H2}}
          &Standard & Nearest neighbor   & (1,2), (3,4), (2,3), (4,1) \\ \cline{2-3}
          &$U_\mathrm{NP}$ & Nearest neighbor       & (1,2), (3,4), (2,3), (4,1) \\ \cline{2-4}
          & \multirow{2}{*}{$U_\mathrm{NP}$}&  \multirow{2}{*}{All-to-all}
            & (1,2), (3,4), (1,4) \\
            & & & (2,3), (1,3), (2,4) \\ \cline{2-4}
          &$U_\mathrm{NP}$ & Spin-preserving  & (1,2), (3,4) \\
      \hline \hline
      \multirow{9}{*}{\ce{H3}, \ce{LiH}}
      &\multirow{2}{*}{Standard}& \multirow{2}{*}{Nearest neighbor} 
          & (1,2), (3,4), (5,6), \\
          & && (2,3), (4,5), (6,1) \\ \cline{2-4}
      &\multirow{5}{*}{$U_\mathrm{NP}$}&\multirow{5}{*}{ All-to-all}
          & (1,2), (3,5), (4,6), \\
          &&& (1,3), (2,6), (4,5), \\
          &&& (1,4), (2,3), (5,6),\\
          &&& (1,5), (2,4), (3,6),\\
          &&& (1,6), (2,5), (3,4) \\ \cline{2-4}
      &\multirow{2}{*}{$U_\mathrm{NP}$} &\multirow{2}{*}{Spin-preserving}
          & (1,2), (4,5), (2,3), \\
         & && (5,6), (1,3), (4,6) \\ \hline
     \end{tabular}
 \end{table} \label{tab:qubit_connection}
\endgroup

\begin{figure*}[htb]
\centering
\begin{tabular}{ccc}
\Qcircuit @C=1.2em @R=3.8 em{
        \lstick{\ket{0}} & \multigate{1}{U_\mathrm{NP}} &\qw      &\qw & \multigate{1}{U_\mathrm{NP}} &\qw\\
        \lstick{\ket{0}} & \ghost{U_\mathrm{NP}}        &\ctrlo{1}&\qw & \ghost{U_\mathrm{NP}}        &\qw\\
        \lstick{\ket{0}} & \multigate{1}{U_\mathrm{NP}} &\gate{Z} &\qw & \multigate{1}{U_\mathrm{NP}}       &\qw \\
        \lstick{\ket{0}} & \ghost{U_\mathrm{NP}}        &\qw      &\qw & \ghost{U_\mathrm{NP}}        &\qw
        \gategroup{1}{2}{4}{4}{1.0em}{--}\\
        &&&\arrep{lll}& 
}
&
\Qcircuit @C=1.2 em @R=2.0 em{
    &\lstick{\ket{0}} & \multigate{1}{U_\mathrm{NP}} &\qw   &\sgate{U_\mathrm{NP}}{2}
        &\qw &\qw  & \multigate{1}{U_\mathrm{NP}} &\qw &\sgate{U_\mathrm{NP}}{2}&\qw\\
     &\lstick{\ket{0}} & \ghost{U_\mathrm{NP}} 
        &\multigate{1}{U_\mathrm{NP}}  &\qw   &\qw &\qw       & \ghost{U_\mathrm{NP}}        &\multigate{1}{U_\mathrm{NP}}&\qw&\qw             \\
    &\lstick{\ket{0}} & \qw 
        &\ghost{U_\mathrm{NP}}&\gate{U_\mathrm{NP}}
        &\ctrlo{1} &\qw & \qw                    &\ghost{U_\mathrm{NP}}   &\gate{U_\mathrm{NP}}&\qw\\
    &\lstick{\ket{0}} & \multigate{1}{U_\mathrm{NP}} &\qw 
        &\sgate{U_\mathrm{NP}}{2} &\gate{Z} &\qw  
        & \multigate{1}{U_\mathrm{NP}} &\qw  \qw         &\sgate{U_\mathrm{NP}}{2}&\qw\\
    &\lstick{\ket{0}} & \ghost{U_\mathrm{NP}}  
        &\multigate{1}{U_\mathrm{NP}}&\qw  &\qw &\qw 
        &\ghost{U_\mathrm{NP}}  &\multigate{1}{U_\mathrm{NP}}&\qw  &\qw      \\
    &\lstick{\ket{0}} & \qw &\ghost{U_\mathrm{NP}}       
            &\gate{U_\mathrm{NP}}
            &\qw &\qw       & \qw                    &\ghost{U_\mathrm{NP}}  &\gate{U_\mathrm{NP}}&\qw
        \gategroup{1}{3}{6}{7}{1.0em}{--}\\
    &&&&&&\arrep{llll} & 
}
    \\ \\
    4-qubit system
    &
    6-qubit system
\end{tabular}\\
\caption{{\bf Structures of the number and spin-preserving PQCs.} Two qubit blocks labeled $U_\mathrm{NP}$ are particle number-preserving gates. The structures surrounded by the dashed lines are repeated $L-1$ times, where $L$ is the number of layers. }
\label{fig:symmetric_PQC}
\end{figure*}

In the spin preserving ansatz, one additional negative controlled Z gate is appended to qubit-pairs of (2,3) for \ce{H2} and (3,4) for 6-qubit systems in order to introduce the entanglement between spin-up and spin-down states. 
We put the negative controlled Z gate in between layers. 

\clearpage
\section{Optimization trajectory of quantum compilation }
\label{apdx:traj_compliation_H2}
\begin{figure*}[h]
 \centering
 \begin{tabular}{ccc}
 \includegraphics[scale=0.27]{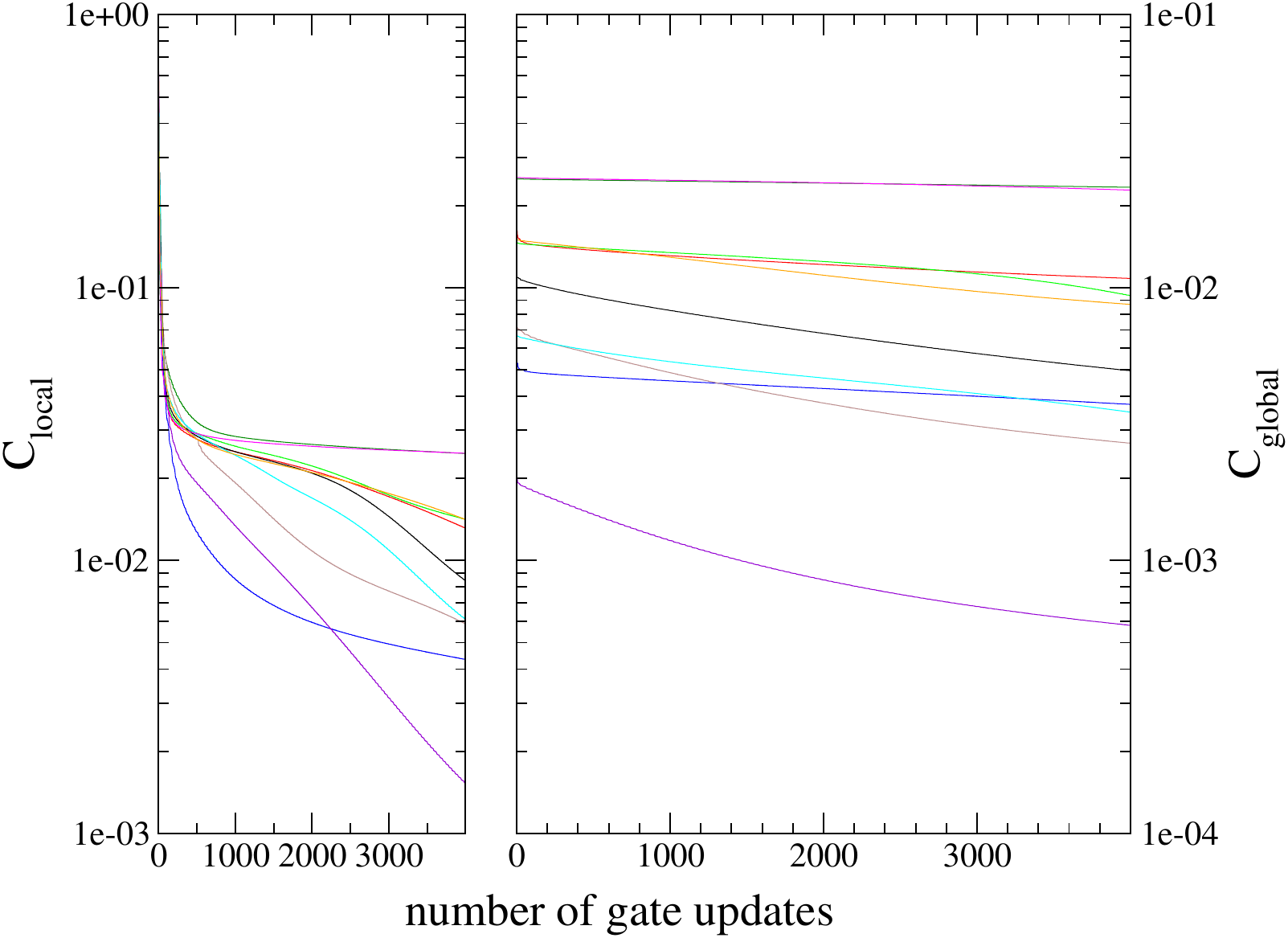}  
 &~&
 \includegraphics[scale=0.27]{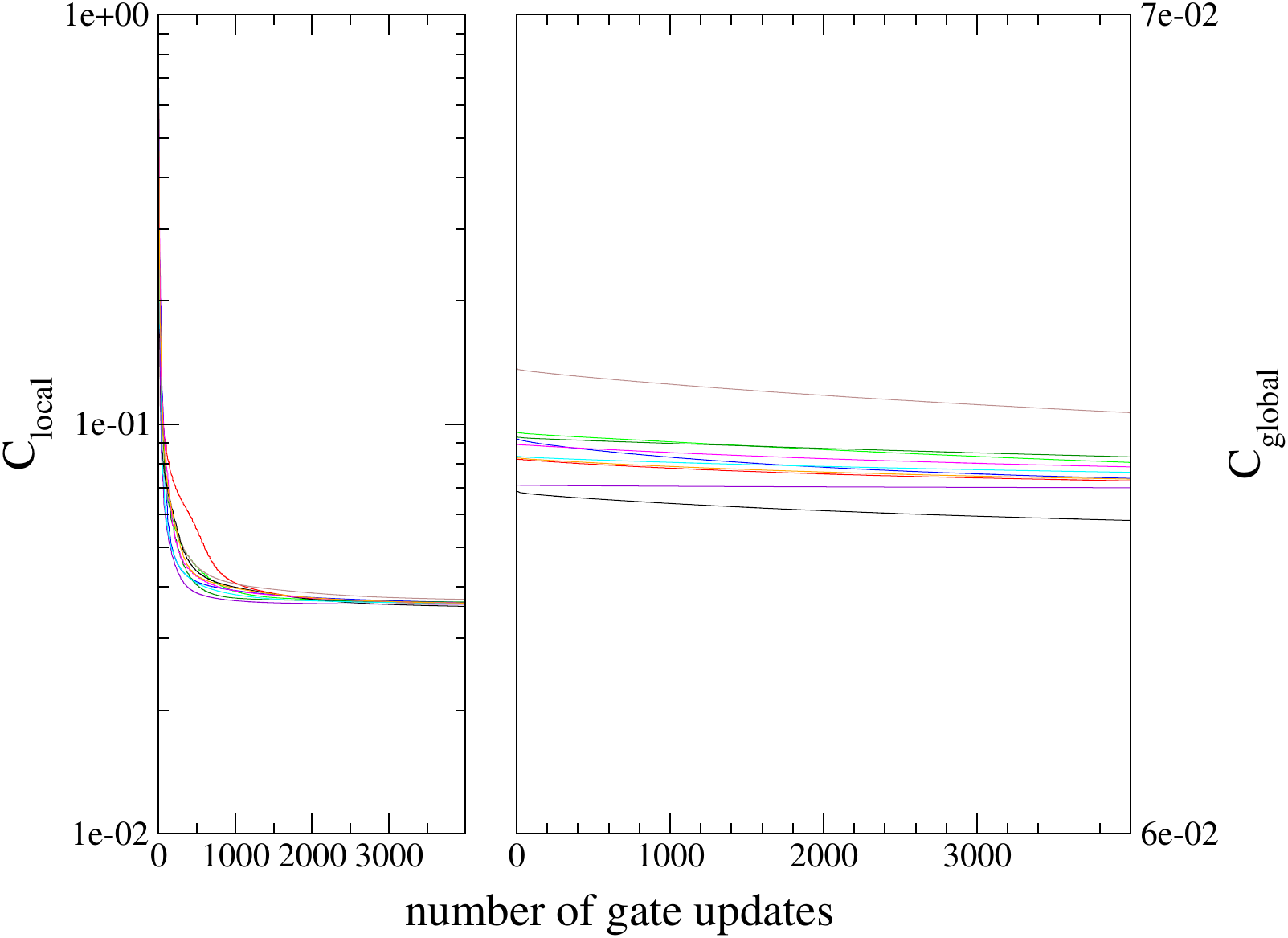} 
 \\
 (a) \ce{2-state}  
 &~&
 (b) \ce{16-state} 
 \\
 \end{tabular}
 \caption{{\bf Trajectories of unitary compilation of time evolution operator of a \ce{H2}/STO-3G molecule with input states in (a) 2- and (b) 16- dimensional subspace with the ALT.} 
 Each subspace is spanned by a basis set $W$ composed of some elements of computational basis which is defined in Table \ref{tab:set-for-subspace-h2}. }
 \label{fig:traj-2-16-subspace-H2}
\begin{tabular}{ccc}
    \includegraphics[scale=0.27]{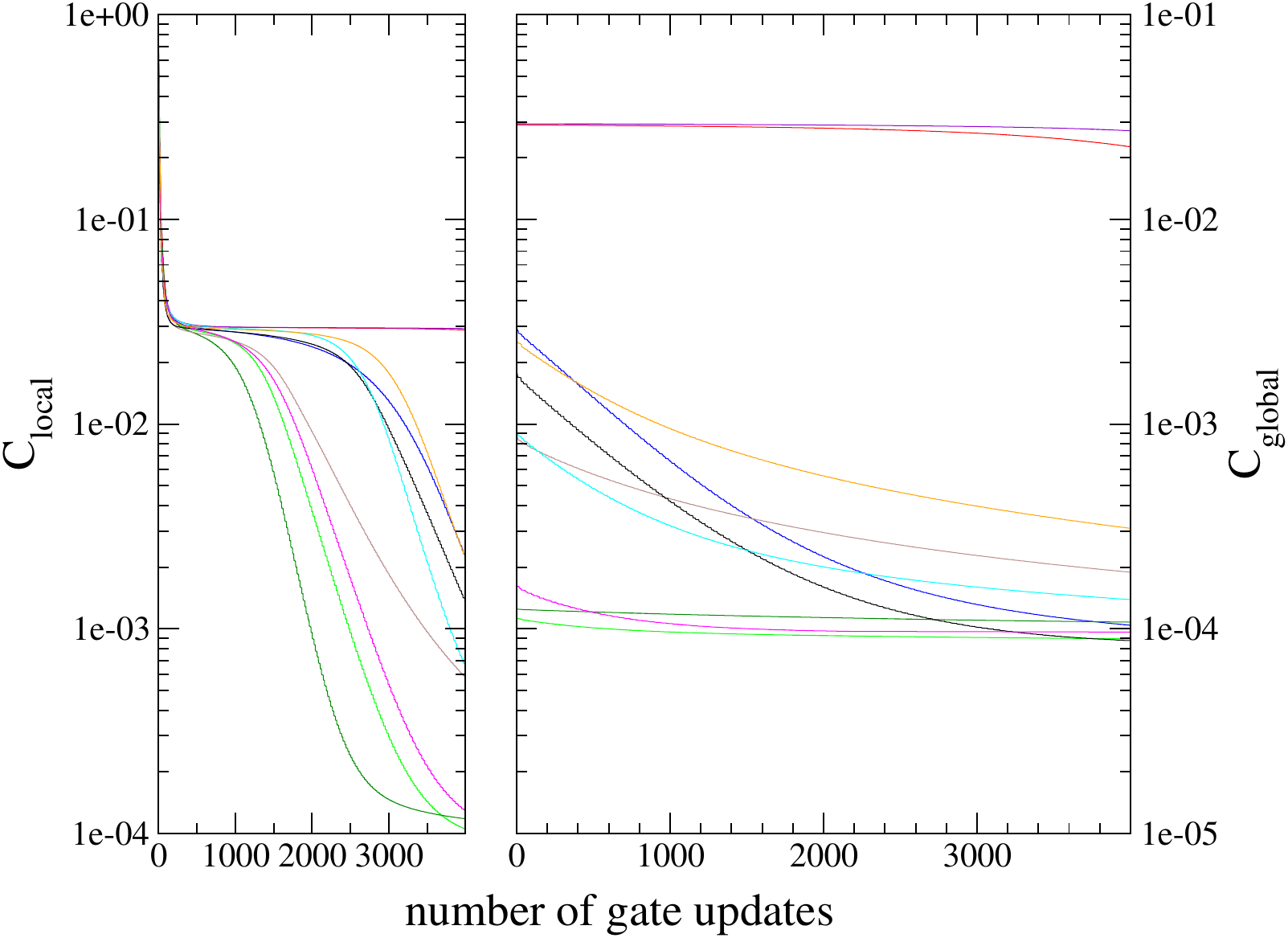}  
    &~&
    \includegraphics[scale=0.27]{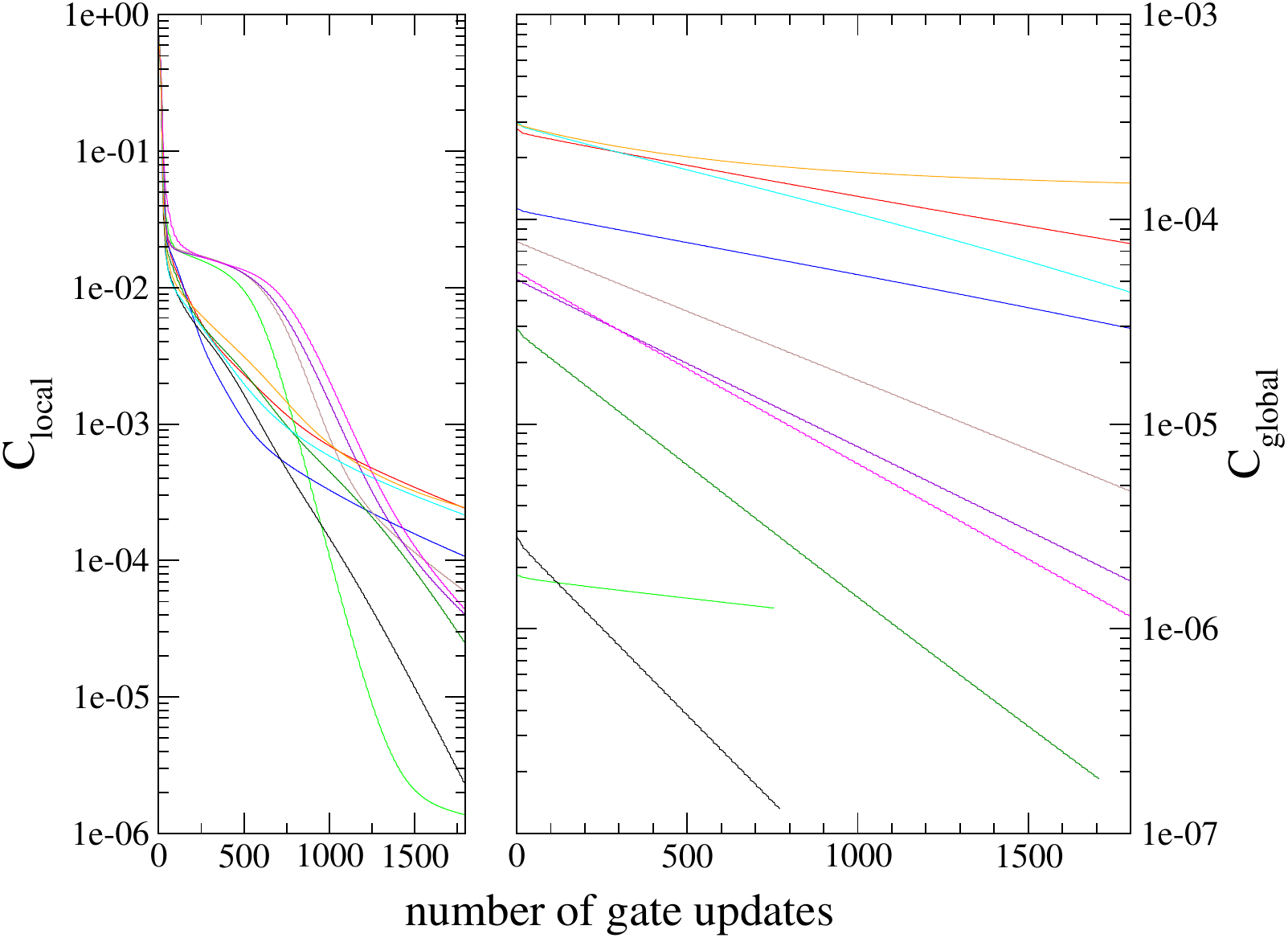}
    \\
    (a) No conserved quantity with nearest neighbor connection ($L=3$)   
    &~&
    (b) $U_\mathrm{NP}$ with all-to-all connection with $L=3$\\
    \\
    \includegraphics[scale=0.27]{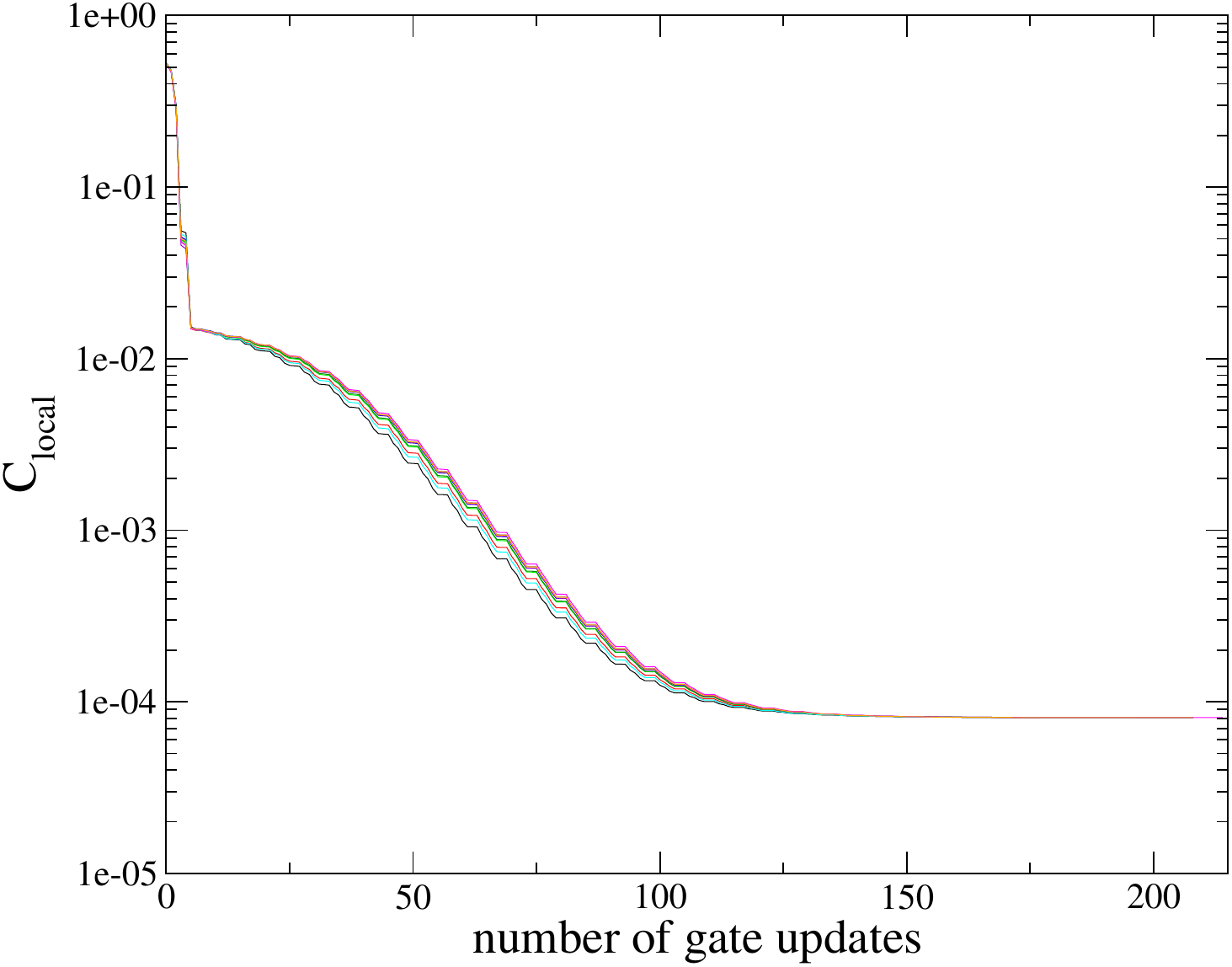}
    &~&
    \includegraphics[scale=0.27]{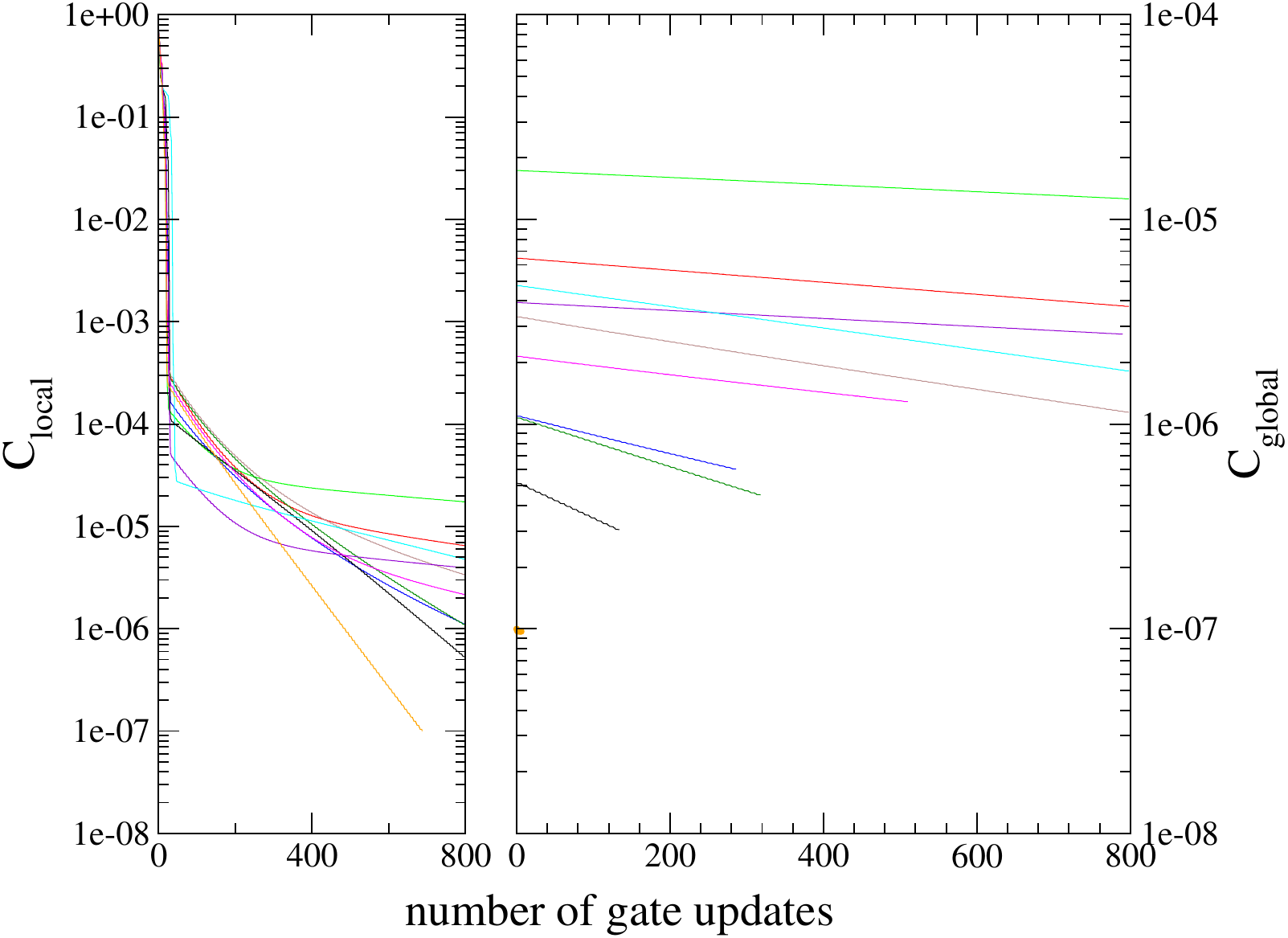}
    \\
    (c) Spin preserving ansatz with $L=3$
    &~&
    (d) Spin preserving ansatz with $L=4$
    \\ 
    \end{tabular}
    \\
    \caption{{\bf Trajectories of unitary compilation for \ce{H2}/STO-3G with input state under spin restriction and different ansaetze.} Refer to Table.~\ref{tab:qubit_connection} for qubits connection. In (c) the global cost optimization trajectory is omitted because they were not improved at all after switching from the local cost.} 
    \label{fig:traj-pqc-h2}
\end{figure*}

\clearpage
\section{The obtained cost value with different ansaetze.}
\label{apdx:compliation_final_cost}
\begin{figure}[htb]
     \centering
     \begin{tabular}{c}
     \includegraphics[width=0.45\textwidth]{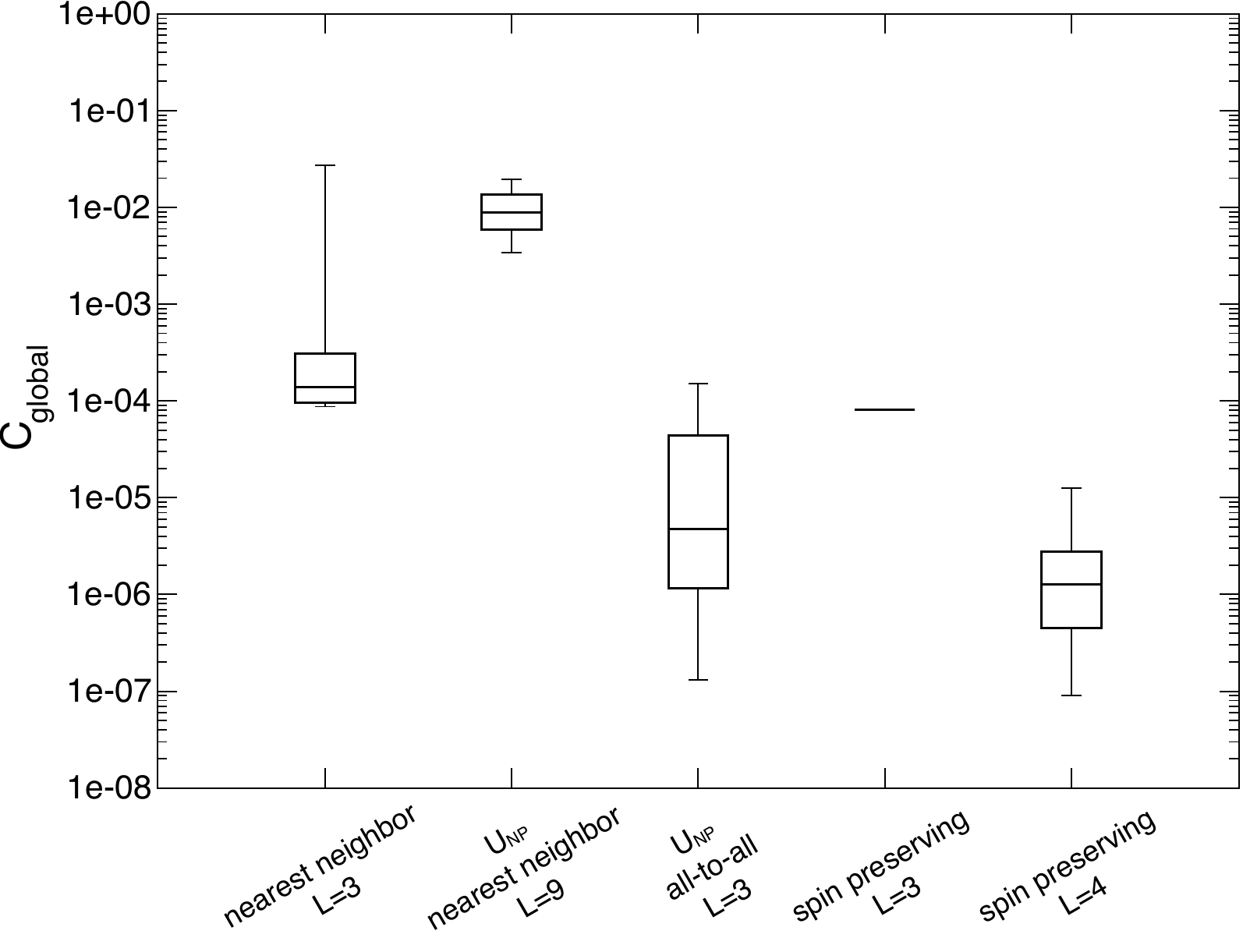} \\ 
    (a) \ce{H2}/STO-3G  \\
     \includegraphics[width=0.45\textwidth]{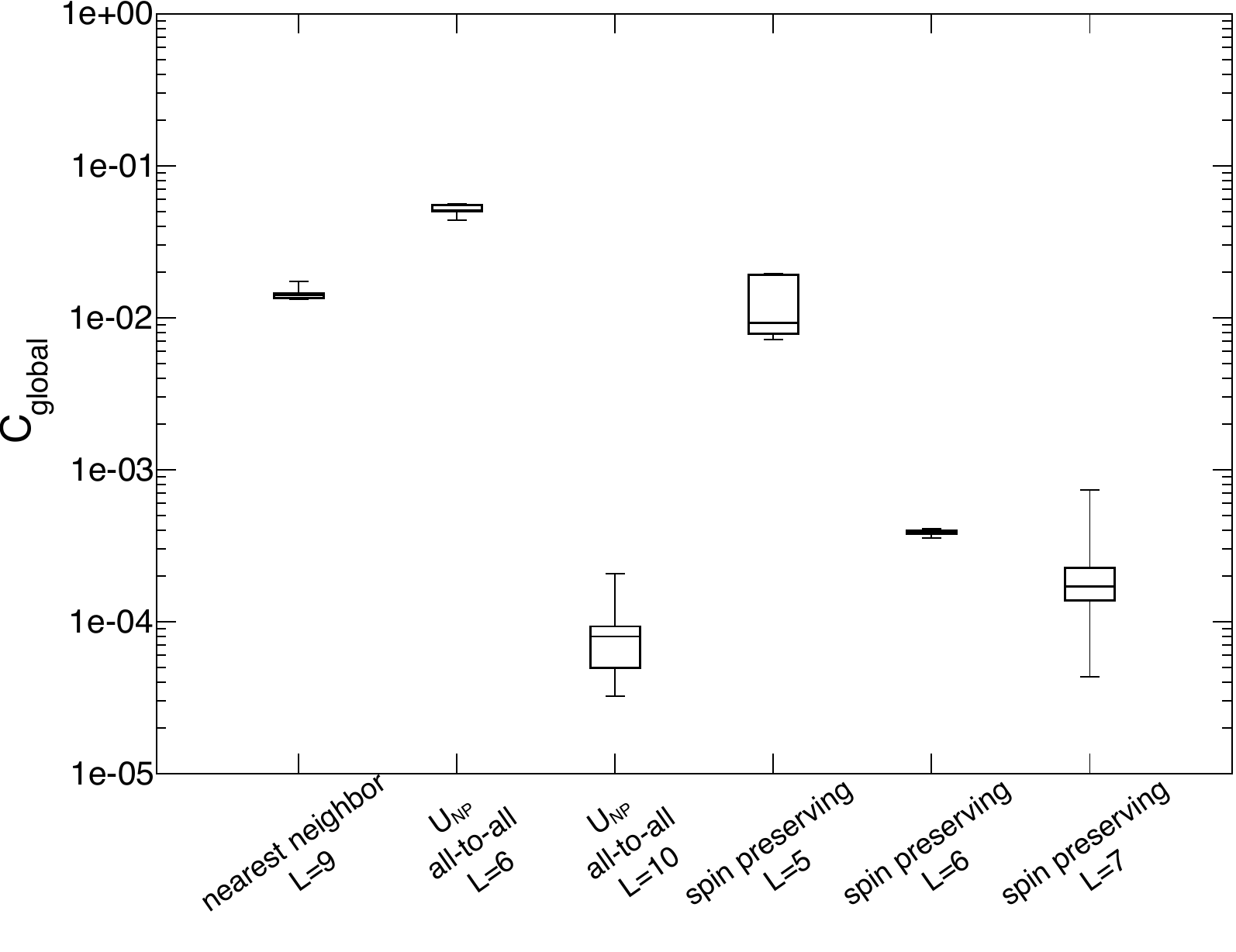}  \\
    (b) \ce{H3}/STO-3G  \\
     \includegraphics[width=0.45\textwidth]{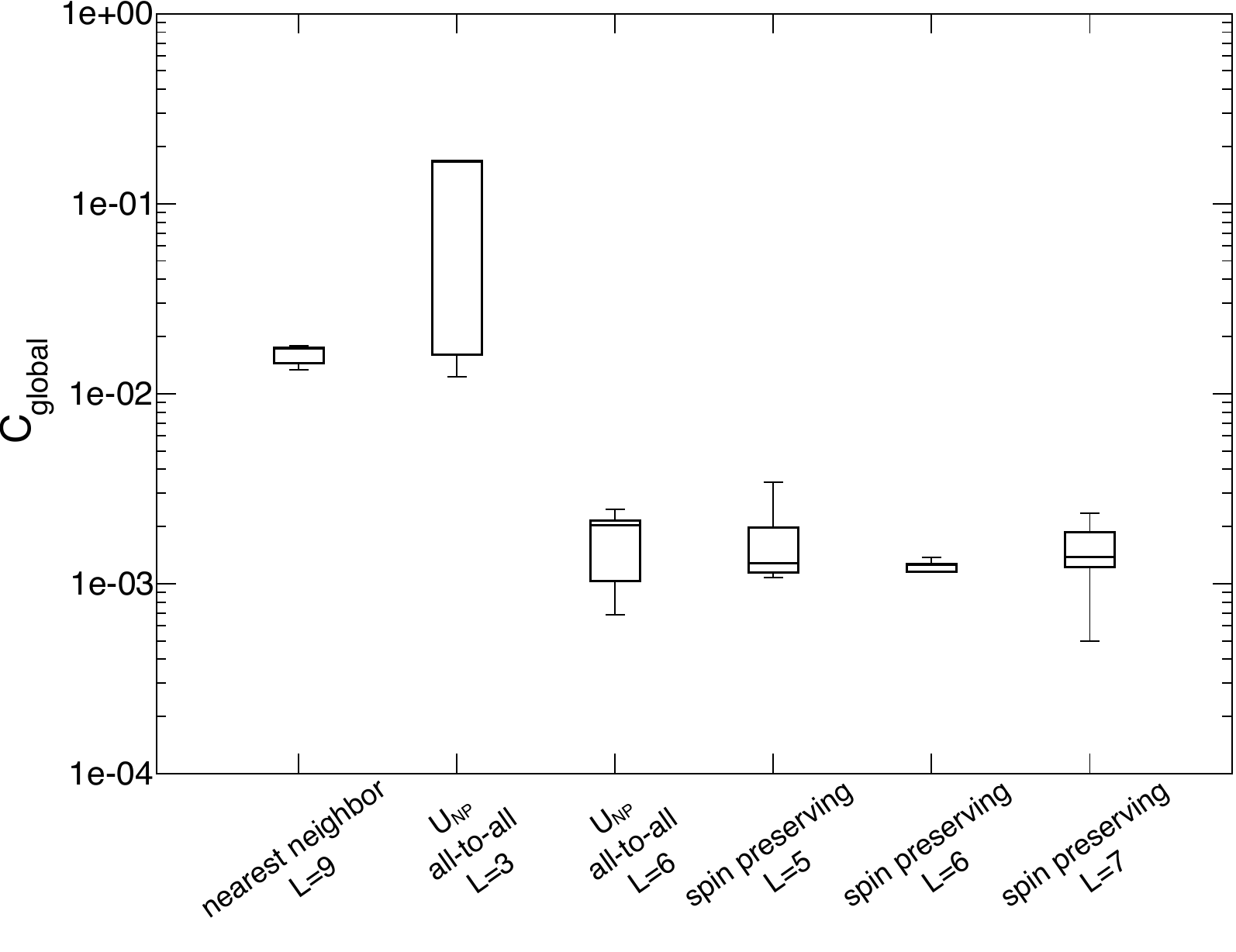} \\
    (c) \ce{LiH}/STO-3G \\
     \end{tabular}
     \caption{{\bf Distribution of the obtained cost value with various ansaetze.} 
     }
     \label{fig:pqc_restriction}
 \end{figure}

\begin{figure*}[htb]
\centering
    \begin{tabular}{ccc}
 \includegraphics[width=0.43\textwidth]{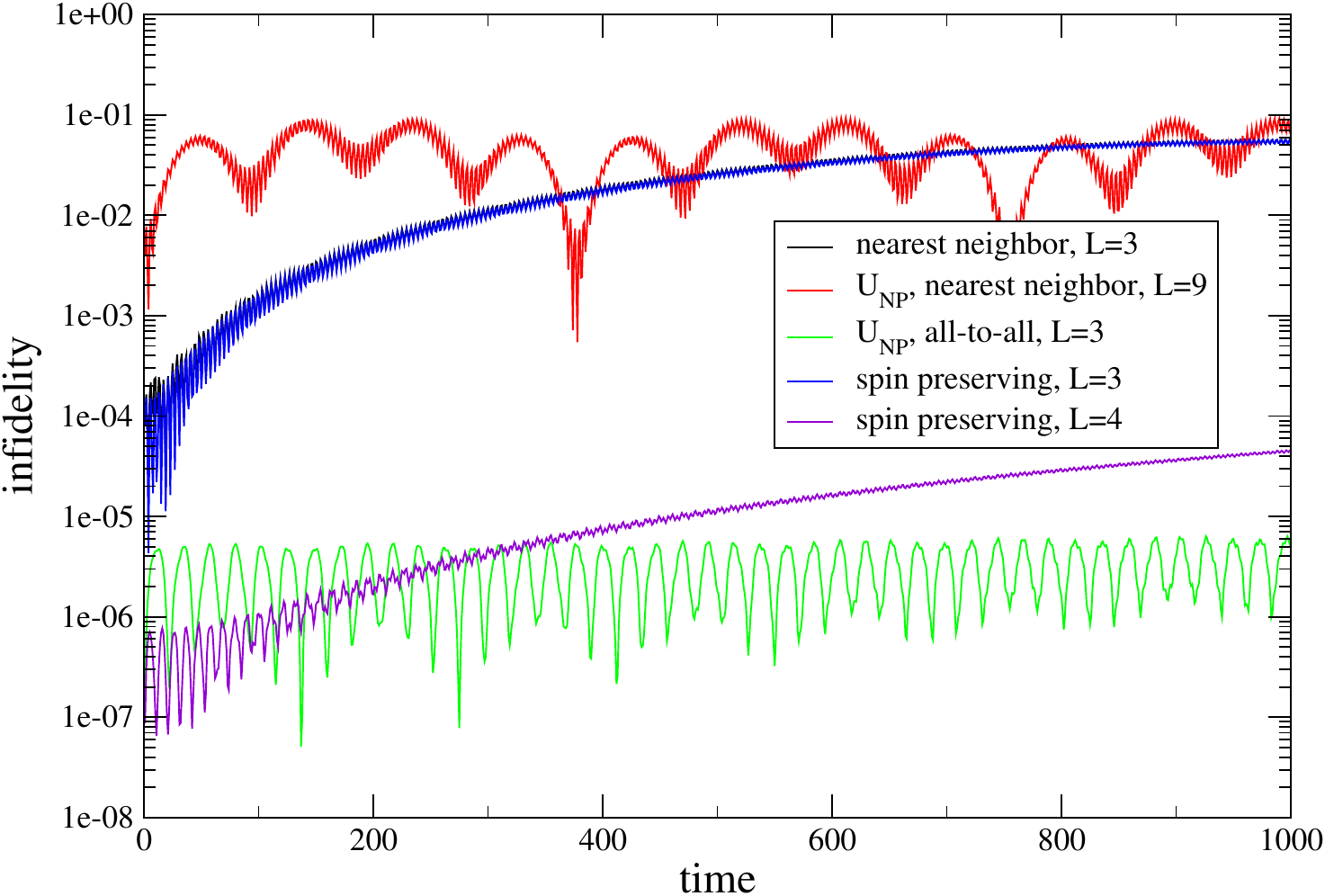}       
 &~&
 \includegraphics[scale=0.28]{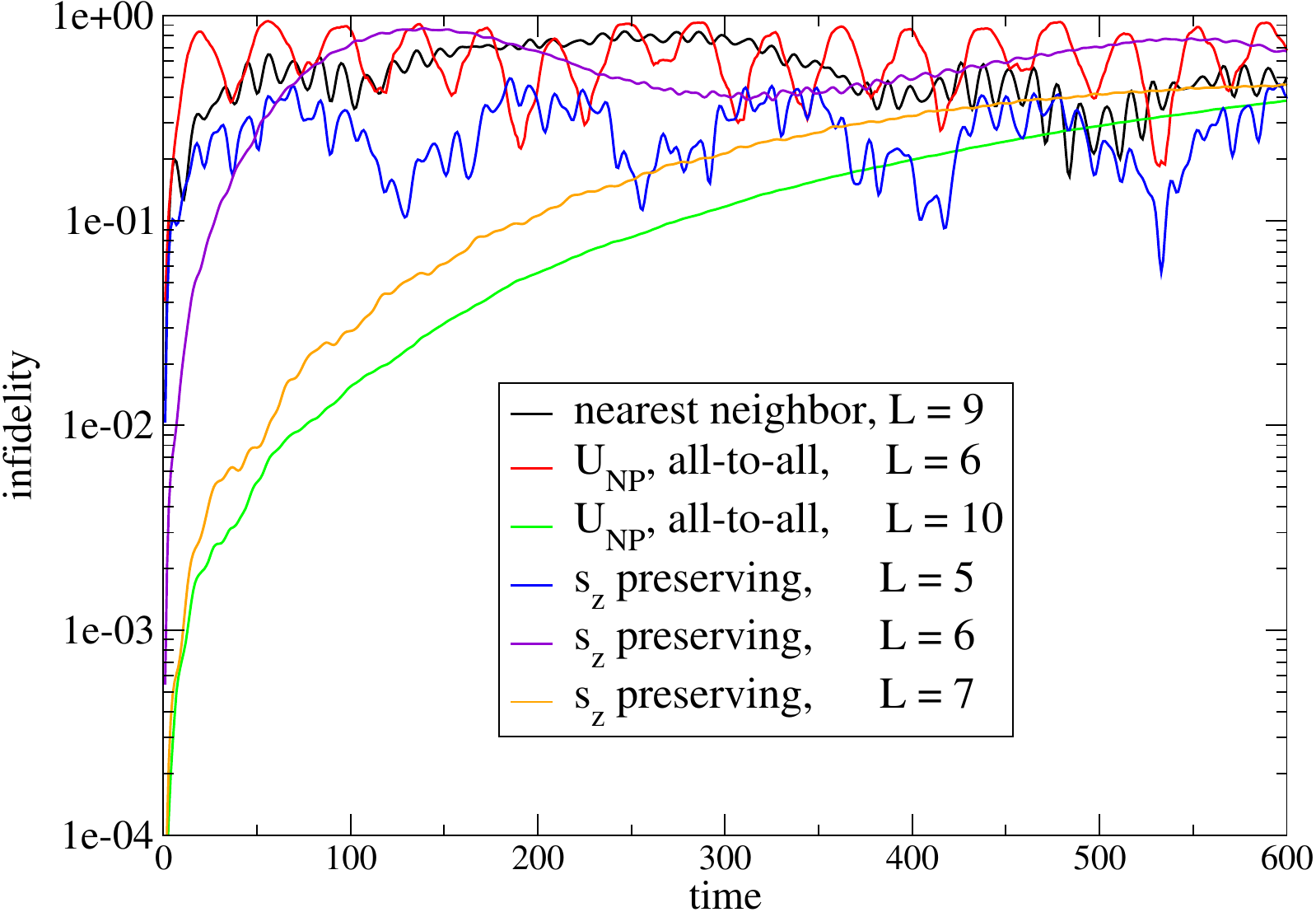} \\
 (a) \ce{H2}/STO-3G &
 ~&
 (b) \ce{H3}/STO-3G
\end{tabular}
  \caption{{\bf Infidelity between time-evolved states by the exact and approximated time evolution operators.} While the Hartree-Fock state was chosen as the initial state for \ce{H2}, the state in which two spin-up orbitals and one spin-down orbital are occupied is used as initial for \ce{H3}
 }
     \label{fig:time-evolution}
 \end{figure*}